\documentclass[a4paper,psfig,twoside,titlepage,11pt]{article}
\usepackage{epsfig}
\input{epsf}
\usepackage{amsmath}

\makeatletter \def\section{\@startsection {section}{1}{\z@}{-3.5ex
plus -1ex minus -.2ex}{2.3ex plus .2ex}{\large\bf}}
\def\subsection{\@startsection{subsection}{2}{\z@}{-3.25ex plus -1ex
minus -.2ex}{1.5ex plus .2ex}{\normalsize\bf}} \makeatother
\makeatletter 

\@addtoreset{equation}{section}
 \makeatother
\newcommand{\insertion}[3]{\begin{figure}
\noindent\fbox{\parbox{5.8in}{\centerline{Insert #1: {\bf #2}}
\smallskip #3}}
\end{figure}}

\font\cmss=cmss10 \font\cmsss=cmss10 at 7pt

\def\inbar{\vrule height1.5ex width.4pt depth0pt}
\def\IC{\relax\,\hbox{$\inbar\kern-.3em{\rm C}$}}
\def\IG{\relax\,\hbox{$\inbar\kern-.3em{\rm G}$}} \def\IB{\relax{\rm
I\kern-.18em B}} \def\ID{\relax{\rm I\kern-.18em D}}
\def\IL{\relax{\rm I\kern-.18em L}} \def\IF{\relax{\rm I\kern-.18em
F}} \def\IH{\relax{\rm I\kern-.18em H}} \def\II{\relax{\rm
I\kern-.17em I}} \def\IN{\relax{\rm I\kern-.18em N}}
\def\IP{\relax{\rm I\kern-.18em P}}
\def\IQ{\relax\,\hbox{$\inbar\kern-.3em{\rm Q}$}}
\def\bfzero{\relax\,\hbox{$\inbar\kern-.3em{\rm 0}$}}
\def\IK{\relax{\rm I\kern-.18em K}}
\def\IG{\relax\,\hbox{$\inbar\kern-.3em{\rm G}$}} \font\cmss=cmss10
\font\cmsss=cmss10 at 7pt \def\IR{\relax{\rm I\kern-.18em R}}
\def\ZZ{\relax\ifmmode\mathchoice {\hbox{\cmss Z\kern-.4em
Z}}{\hbox{\cmss Z\kern-.4em Z}} {\lower.9pt\hbox{\cmsss Z\kern-.4em
Z}} {\lower1.2pt\hbox{\cmsss Z\kern-.4em Z}}\else{\cmss Z\kern-.4em
Z}\fi} \def\bfone{\relax{\rm 1\kern-.35em 1}}   
  
\def\one{{\hbox{1\kern-.8mm l}}} \def\tilde{\widetilde}
\def\bar{\overline} 
\def\IE{\relax{{\rm I\kern-.18em E}}} 
 \def\IGam{\relax{{\rm I}\kern-.18em \Gamma}}
 
\def\beq{\begin{equation}} \def\eeq{\end{equation}}
\def\beqa{\begin{eqnarray}} \def\eeqa{\end{eqnarray}}

\begin{document}

\begin{titlepage}

\setcounter{page}{0}

\begin{flushright}

NORDITA-2003/10 HE \rightline{\hfill March 2003}
\end{flushright}

\vskip 1.6cm

\begin{center}

{\LARGE \bf Four Lectures On} \vskip 0.3cm {\LARGE \bf The
Gauge/Gravity Correspondence}

\vskip 1.4cm

{\Large Matteo Bertolini}

\vskip 0.3cm

{\large \sl NORDITA \\ Blegdamsvej 17 \\DK-2100 Copenhagen \O,
Denmark} \\ {\large \tt teobert@nbi.dk}

\vskip 1.4cm {\bf Abstract}
\end{center}
%\abstract

\noindent
We review in a pedagogical manner some of the efforts aiming to extend
the gauge/gravity correspondence to non-conformal supersymmetric gauge
theories in four dimensions. After giving a general overview, we
discuss in detail two specific examples: fractional D-branes on
orbifolds and D-branes wrapped on supersymmetric cycles of Calabi-Yau
spaces. We explore in particular which gauge theory information can be
extracted from the corresponding supergravity solutions, and what the
remaining open problems are. We also briefly explain the connection
between these and other approaches, such as fractional branes on
conifolds, branes suspended between branes, M5-branes on Riemann
surfaces and M-theory  on G2-holonomy manifolds, and discuss the
r\^ole played by geometric transitions  in all that.

\vskip 1.6cm  \centerline{\it Lectures given at SISSA/ISAS}

\centerline{\it Trieste, December 17-20 2002}

\end{titlepage}

%\input{inc.tex}

%\newpage

%\pagenumbering{roman}

%\pagenumbering{empty}

%Hope you enjoy the reading!

%\end{titlepage}

%\pagenumbering{empty}

\tableofcontents

\newpage

\addcontentsline{toc}{section}{Foreword}

%\pagenumbering{arabic}
\section*{\centerline{Foreword}} 

These lectures are based on an eight-hour course I gave at SISSA/ISAS,
December 2002, on non-conformal extensions of the gauge/gravity
correspondence. This is a vast subject and I could not discuss in
detail all approaches that have been used to pursue such a program.
The choice was to focus on cases where supersymmetry and conformal
invariance are broken by the geometry and/or D-branes, hence looking
directly for gravity duals of non-conformal theories. That  is to say,
we did not consider cases where conformal invariance is broken by mass
deformations starting from conformal theories. We analyzed in
particular two systems in some detail: fractional D-branes on orbifolds and
D-branes wrapped on supersymmetric cycles of Calabi-Yau manifolds.

This written version contains somewhat more material than I was able to
cover during the course. Some more details are provided for both
fractional and wrapped branes (lecture II and III, respectively),
while lecture IV contains a wider (but still qualitative) presentation
of other approaches used to study the gauge/gravity correspondence in
non-conformal cases, with an emphasis on the relations between them.

Throughout the text there are a few (simple) exercises that might help
in  getting a more concrete handle on the concepts and the tools
discussed. Plus, I added some inserts when some further details were
needed on topics I could not explain in detail in the main text.%, not to
%interrupt the logic flow of the lectures.

There have been hundreds of papers on this subject in the last few
years  and it would be a daunting task to give credit to all of
them. As this  is not a review but a series of lectures notes for
students, I will not  provide a complete list of references but
instead accompany the reader  into a brief tour through some of the
literature.

Hope you enjoy the reading!

\newpage

%%%%%%%%%%%%%%%%%%%%%%%%%%%%%%%%%%%%%%%%%%%%%%%%%%%%%%%%%%%%%%%%%%
%%%%%%%%%%%%%%%%%%%%%%%%%%%%%%%%%%%%%%%%%%%%%%%%%%%%%%%%%%%%%%%%%%
\section{Lecture I - Introduction And Overview}
The goal of these lectures is to review some of the efforts that have
been made in the past few years to extend the gauge/gravity
correspondence to non-conformal and less supersymmetric settings as
compared to the original duality proposed by Maldacena, which relates
type IIB string theory on ${\rm AdS}_5\times S^5$ to four-dimensional
${\cal N}=4$ Super Yang-Mills (SYM), this being a conformal and
maximally supersymmetric gauge theory. This goal has been pursued in
many different ways in the recent literature, so let me first explain
what my specific point of view will be.

The celebrated AdS/CFT correspondence is a conjectured equivalence
between two apparently very different theories
\vskip 4pt
\begin{table}[ht]
\begin{center}
\begin{tabular}{ccc}
{ Type IIB string theory} && { ${\cal N}=4$ SYM in 4D with}\\  { on
${\rm AdS}_5\times S^5$} && {gauge group $SU(N)$}
\end{tabular}
\end{center}
\end{table}
\vskip -14pt
As stated above, this correspondence does not include any kind of
D-branes. However, its seeds reside in the {\it double} nature of
D-branes, this being related to the old open/closed string duality. In
particular we know that
\begin{itemize}
\item{D-branes are hypersurfaces on which open strings can end. Their
dynamics  is described in general by a (supersymmetric) gauge theory
at low energy, this being the low energy spectrum of the corresponding
open strings.}
\item{D-branes are non-perturbative states of the closed string
spectrum (their tension going as $1/g_s$ where $g_s$ is the string
coupling) and at low energy they are described by soliton-like
solutions of the corresponding supergravity theory.}
\end{itemize}
This suggests to try and exploit the classical geometrical properties
of D-branes to study the gauge theory living on them or use the
quantum properties of the gauge theory to study the dynamics of
non-perturbative extended objects. This idea applies to any kind of
D-brane and was considered well before the work of Maldacena
(essentially it is part of the discovery by Polchinski of D-branes
being non-perturbative states of the closed string spectrum). In the
case considered by Maldacena, namely a stack of $N$ D3-branes in flat
ten-dimensional space, however, it was possible to carry this
correspondence further by taking the so-called near-horizon limit
%\begin{equation}
%\alpha' \rightarrow 0 \qquad {\rm with} \qquad
%\frac{r}{\alpha'}\,,\,g_s\,,\, N \quad {\rm held\;fixed}
%\end{equation}
and observing that in this limit the bulk (closed) and gauge (open)
degrees of freedom decouple. This decoupling is at the core of the
AdS/CFT duality.

An obvious thing one can try is to see whether a similar approach can
be used to study non-conformal and less supersymmetric theories, as
for instance ${\cal N}=1,2$ SYM (and eventually ${\cal N}=0$) in four
dimensions, starting from bound states of D-branes in less
supersymmetric backgrounds, breaking eventually conformal
invariance. There is a number of problems that one usually encounters
in pursuing this program. Let me briefly anticipate some of them.
\begin{itemize}
\item{The dual supergravity solution of a non-conformal gauge theory
does not display an AdS-like geometry. This means that, strictly
speaking, holography is not at work in these cases, in general. What
one can do, at best, is to use what one learned in the conformal case
as a guiding principle but one cannot safely rely on all theorems that
proved  so fruitful in the original AdS/CFT duality.}
\item{In general, the supergravity backgrounds one finds are
singular. To give a meaning to the solution one has of course to cure
the singularity and look for a singularity-free solution. At the same
time one should understand what the r\^ole of the singularity is from
the gauge theory point of view, as well as the meaning of its
resolution. Hence, more than being a problem, this can be a source of
interest after all. We will learn much more about this.}
\item{As already pointed out, a basic aspect of the original AdS/CFT
correspondence is the decoupling between open and closed degrees of
freedom, this being the starting point to state the exact duality. The
duality can be stated at different levels (see Insert 1) and in
general  relates a given regime of the ${\cal N}=4$ gauge theory to a
given regime  of type IIB string theory on ${\rm AdS}_5 \times
S^5$. In particular, there  exists a limit in the parameter space
(i.e. the large 't-Hooft coupling limit)
where the gauge theory is supposed just to be dual to supergravity,
without any addition of string states (this is actually the regime
where the  duality as been mostly checked). In non-conformal cases it
turns out this is not the case. Roughly speaking, if one insists in
retaining just supergravity modes, the dual gauge theory cannot be
completely decoupled from the bulk. The way this manifests is very
much case-dependent, in particular there is a qualitative difference
between the ${\cal N}=2$ and the ${\cal N}=1$ cases, as we shall see
in detail during these lectures. But it is a fact that within the
supergravity regime (which is the one we will mainly investigate) a
complete decoupling does not hold.}
\end{itemize}

\insertion{1}{Different Limits Of The AdS/CFT Correspondence
\label{insert1}}{The AdS/CFT correspondence is an equivalence
between two theories, a string theory and a gauge theory. In its more
strongest version the correspondence is supposed to hold for generic
values of the parameters defining the regime of the two theories. By
taking suitable limits the duality boils down to an equivalence
between type IIB supergravity and gauge theory at strong
coupling. This is the regime where the correspondence has been mainly
tested, so far.
\begin{center}
\begin{tabular}{cc}
\multicolumn{2}{c}{\it Exact equivalence} \\ Type IIB string theory on
${\rm AdS}^5\times {\rm S}^5$  & ${\cal N}=4$ SYM in 4D \\ string
coupling $g_s$ and string tension $T$ &  YM coupling $g_{\rm YM}$ and
number of colors $N$
\end{tabular}
\end{center}
where $T=R^2/(2 \pi \alpha')$, $R$ being the (common) radius of the
AdS space  and of the $S^5$. The dictionary is determined in terms of
the two basic relations
\begin{center}
\begin{tabular}{cc}
\multicolumn{2}{c}{$4 \pi g_s = g_{\rm YM}^2 \quad , \quad T =
\frac{1}{2\pi}\sqrt{g_{\rm YM}^2 N} = \frac{1}{2\pi}\sqrt{\lambda}$}
\end{tabular}
\end{center}
It is very difficult to test the conjecture at this level as we do not
know how to treat string theory for generic value of the string
coupling. Better to take the weak coupling limit, $g_s \rightarrow
0$. In this limit we select the sector of the gauge theory surviving
at large $N$.
\begin{center}
\begin{tabular}{cc}
\multicolumn{2}{c}{\it Equivalence in the classical limit} \\
$g_s\rightarrow 0$ with $T$-fixed & $g_{\rm YM} \rightarrow 0$ with
$\lambda = g^2_{YM}N$-fixed \\ Classical string theory
(non-interacting strings) & Large $N$ limit, planar diagrams only
\end{tabular}
\end{center}
As opposed to string theory in flat space, we do not even know how to
study classical string theory in curved backgrounds with RR
fluxes. Better to take the low energy limit. This leads to the weaker
(though more tractable) version of the conjecture, where string theory
reduces to type IIB supergravity.
\begin{center}
\begin{tabular}{cc}
\multicolumn{2}{c}{\it Equivalence at low energy} \\ $g_s \rightarrow
0$ with $T \rightarrow \infty$ & $g_{\rm YM}^2 \rightarrow 0$ with
$\lambda \rightarrow \infty$ \\ Type IIB supergravity in ${\rm AdS}_5
\times {\rm S}^5$ & ${\cal N}=4$ SYM in 4D at strong 't-Hooft coupling
\end{tabular}
\end{center}
The last equivalence means that in the original AdS/CFT correspondence
there exists a regime in the gauge theory (the large 't-Hooft coupling
regime) in which the dynamics of the gauge theory is supposed to be
captured solely by supergravity modes, without any addition of string
states.}

For these reasons we are not yet at a point where we can really state
a duality \'a la Maldacena for non-conformal theories, at least at the
supergravity level. It is believed that a proper duality holds if one
lets string states enter into the game, but as it is the case for the
original AdS/CFT correspondence, it is much harder to go beyond the
supergravity regime and check the duality at the stringy level. This
is a crucial point which makes manifest a conceptual difference
between  conformal and non-conformal dualities, but we do not discuss
this  further, for the time being. Instead, we will take a more humble
approach  and elaborate on the idea illustrated previously, namely
trying  to exploit as much as possible the power of open/closed string
duality and see what can we learn about the dynamics of non-conformal
supersymmetric gauge theories from supergravity, and vice-versa. After
all, the dynamics of these theories is richer than that of ${\cal
N}=4$ and is worth give it a try. The program is then to {\it i})
study D-branes on non-maximally supersymmetric backgrounds {\it ii})
break conformal  invariance {\it iii}) exploit open/closed string
duality {\it iv}) ... see what happens. As we shall see, we can learn
a lot and eventually shed also some light on what is the final goal,
namely to find an exact duality. There are in general two conceptually
different ways to pursue this program.
\begin{itemize}
\item{Start from a conformal theory of which we know the supergravity
dual, and deform it by means of relevant or marginal operators which
break  both supersymmetry and conformal invariance.}
\item{Start from configurations where both supersymmetry and conformal
invariance are broken from the very beginning by a specific
target-space geometry, a D-brane configuration or a combination of the
two, and search the  corresponding dual.}
\end{itemize}
In these lectures I will discuss examples belonging to the second
class only. There are many of them. One can  consider,
for instance
\begin{itemize}
\item{(Fractional) D-branes on orbifolds.}
\item{(Fractional) D-branes on conifolds.}
\item{D-branes wrapped of supersymmetric cycles of Calabi-Yau (CY)
spaces.}
\item{Branes suspended between branes.}
\item{M5-branes wrapped on Riemann surfaces.}
%\item{Geometric transitions (sometime referred to as gauge/geometry
%correspondence).}
\item{M-theory on manifolds of G2 holonomy.}
\end{itemize}
Most of these apparently different settings are related
between each other (by dualities of various kind, geometric
deformations, etc...) and in the last lecture I will outline what the
precise connections are. Still, we would like to have some
quantitative handling on all this and is the purpose of these lectures
to discuss in some detail just a couple of the above approaches. In
the next lecture we shall discuss fractional branes on orbifolds and
in the subsequent one branes wrapped on smooth CY spaces. In both
cases one can construct four-dimensional gauge theories with either
${\cal N}=1$ or ${\cal N}=2$ supersymmetry, with and without
matter. However, in order to be specific and have the time to bring
some computations until the very end, when discussing fractional
branes on orbifolds we will focus on the case with 8 supercharges,
while when discussing branes wrapped on supersymmetric cycles of CY
spaces we will focus on the case with 4 supercharges. When it comes to
write down explicit supergravity solutions and exploit them to study
the gauge theory dual,  we will use in both cases a key example, pure
(i.e. without matter)  ${\cal N}=2$ SYM for fractional branes and pure
${\cal N}=1$ SYM for  wrapped branes.

As I will be rather pedagogical and somewhat lengthly, let me
anticipate some of the features of gauge theories we will be able to
recover from the dual supergravity backgrounds.
\begin{itemize}
\item{The running of the gauge coupling and the corresponding
perturbative $\beta$-function. For the ${\cal N}=1$ case we will also
predict some (unexpected) non-perturbative contributions.}
\item{The $U(1)_R$ anomaly and its corresponding symmetry breaking
pattern.} 
\item{In the case of pure ${\cal N}=1$ the phenomenon of
gaugino condensation and the further breaking of the chiral symmetry
down to $\ZZ_2$ in the vacuum, as well as the appearance of $N$
inequivalent vacua (where $N$ is the number of colors).}
\item{The correct action for the gauge theory instantons.}
\item{For theory with moduli, as for instance ${\cal N}=2$, a general
understanding  of the moduli space in terms of D-branes dynamics.}
\item{For confining theories an estimate of some more fancy things as
the confining string tension, the domain wall tension, the glueballs
mass, etc... These computations rely heavily on tools inherited from
the original AdS/CFT correspondence and on the use of
holography. Hence, as anticipated, they should be taken with some care
and considered just as qualitative (but promising) results.}
\end{itemize}
The gauge theory information listed above belongs either to the
perturbative or to the non-perturbative regime of the gauge
theory. The  lesson one learns in studying these non-conformal gauge
theory duals is that in general the perturbative dynamics of the gauge
theory is precisely accounted for by the supergravity solutions. As
non-conformal theories changes sensibly with the scale, the
information one can get at the perturbative level are already non
trivial, and rather interesting. Also non-perturbative properties get
an holographic description and they are in general related to the way
the (would be) singular solution is deformed into a smooth one. Some
time it is possible to find quantitative predictions, some time just
qualitative, some other time it is possible just to understand what
the precise path to follow would be in order to recover them.

%%%%%%%%%%%%%%%%%%%%%%%%%%%%%%%%%%%%%%%%%%%%%%%%%%%%%%%%%%%%%%%%%%
%%%%%%%%%%%%%%%%%%%%%%%%%%%%%%%%%%%%%%%%%%%%%%%%%%%%%%%%%%%%%%%%%%
\section{Lecture II - Fractional Branes On Orbifolds}

D-branes in flat space are soliton-like states which break half the
supersymmetries of the original 32 preserved by the $\IR^{1,9}$
background. The low energy spectrum of open strings living on them is
a supersymmetric gauge theory with 16 supercharges. When considering
D-branes on orbifolds, which are backgrounds already breaking some
supersymmetry, the amount of supersymmetry on the D-brane world
volumes is also reduced. More precisely
\begin{itemize}
\item{D-branes on orbifold limits of ALE spaces ($\IC_2/\Gamma$ with
$\Gamma$ being a discrete subgroup of $SU(2)$) give rise to 1/2
supersymmetric backgrounds and their world volume theory is described
by a SYM theory with 8 supercharges.}
\item{D-branes on orbifold limits of CY spaces ($\IC_3/\Gamma$ with
$\Gamma$ being now a discrete subgroup of $SU(3)$) give rise to 1/2
supersymmetric backgrounds and their world volume theory is described
by a SYM theory with 4 supercharges.}
\end{itemize}
As we shall see, these gauge theories are in general conformal: the
vector multiplet is coupled to some matter multiplet just in the right
way to give back a conformal theory. The reason for that is that the
low energy spectrum is nothing else but the projection to lower
supersymmetry representations of the maximally supersymmetric (and
conformal) one enjoyed by D-branes in flat space. However, when
considering string theory on orbifolds, there is a specific kind of
D-branes in the spectrum, the so-called {\it fractional} branes, which
also break conformal invariance. The study of these objects and their
gravity duals is the subject of this lecture.

Playing with different orbifolds and different bound states of
fractional D-branes one can build-up SYM theories with either ${\cal
N}=1$ and ${\cal N}=2$ supersymmetry, with product gauge groups and
with matter in different representations. As anticipated, we will
focus here on the ${\cal N}=2$ case, leaving the ${\cal N}=1$ case to
the wrapped branes setting. Moreover, in order to let the students see
some concrete computations in full detail, I will mainly refer to a
key example, i.e. pure ${\cal N}=2$ SYM. Once having become
familiar with the technical tools needed and more generally with the
logic underlying the computations we will carry on for this key
example, the reader should be able to obtain the same results for any
kind of ${\cal N}=1,2$ SYM theory.

%%%%%%%%%%%%%%%%%%%%%%%%%%%%%%%%%%%%%%%%%%%%%%%%%%%%%%%%%%%%%%%%%%%%%%%
\subsection{Fractional Branes: Open String Perspective}

D-branes are defined by open strings ending on them. An open string
state is generically defined by
\begin{equation}
\label{ostate}
\lambda\, \otimes\, oscillators\; |k \rangle
\end{equation}
where $k$ is the momentum along the brane (recall that open strings
can have momentum along Neumann directions only) and $\lambda$ is the
Chan-Paton (CP) factor. For a single D-brane in flat space, $\lambda$
is a number. For a D-brane on an orbifold this is not necessarily
true: $\lambda$ is in general a matrix. This means that in this case
$\lambda$ can transform under different representations of the
orbifold group $\Gamma$. In general, for any element $h$ of the
orbifold group the CP factors transform as follows
\begin{equation}
\label{cptr}
\lambda\,\rightarrow \lambda'=\gamma(h) \,\lambda \,\gamma(h)^{-1}
\end{equation}
According to eq.~(\ref{ostate}) this leads to the possibility of
having different kinds of D-branes on orbifolds.

\underline{Def}: A {\it regular} D-brane is a D-brane whose CP factors
$\lambda$ transform under the regular representation ${\cal R}$ of the
orbifold group $\Gamma$ (this is the representation whose dimension
equals the order $|\Gamma|$ of $\Gamma$).

As the regular representation is reducible, it can be decomposed in
irreducible representations. In particular
\begin{equation}
\label{reirn}
{\cal R} = \oplus \, n_I {\cal D}_I \quad \mbox{with}  \quad
\sum_{I=0}^{m-1} n_I = |\Gamma| \quad , \quad I=0,1,\ldots,m-1
\end{equation}
where ${\cal D}_I$ are the irreducible representations of $\Gamma$ and
$n_I$ their dimension. Note that for abelian orbifolds, namely for
$\Gamma=\ZZ_N$, $n_I=1$ for any $I$. In this case the number of
irreducible representations equals the number of elements of the
orbifold group.

\underline{Def}: A {\it fractional} D-brane (of type $I$) is a D-brane
whose CP factors $\lambda$ transform in the $I$-th irreducible
representation $D_I$ of $\Gamma$.

\begin{figure}[ht]
\begin{center}
{\includegraphics{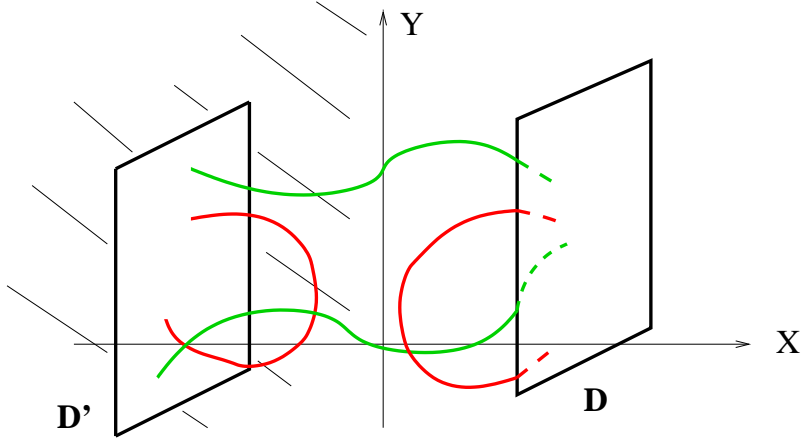}}
\caption{\small A freely moving D-brane on the most simple orbifold as it
appears in the covering space. The brane $D$ and its image $D'$ are
identified in the physical space. The $X$ axis represents the orbifold
directions (these being all transverse to the brane), the $Y$ axis the
flat longitudinal directions. To see the flat transverse
directions... we would need an extra dimension! But they are there, of
course.}
\label{reg}
\end{center}
\end{figure}

Well, here it is all the magic about fractional branes. Let us now try
to have a more physical intuition of what these definitions really
mean. When studying string theory on orbifolds the physical states are
those which are (globally) left invariant under the orbifold
projection. This applies also to D-branes, of course.  Each physical
D-brane which is free to move in the transverse space should have
images in the covering space. In figure \ref{reg} the case of the most
simple orbifold, $\Gamma=\ZZ_2$, is reported.
\vskip -15pt
From figure \ref{reg} one easily sees that in this case the CP factors
are 2 by 2 matrices
\begin{equation}
\lambda = \left(
\begin{array}{cc}
D - D  & D - D' \\ D' - D & D' - D'
\end{array}
\right)
\label{cpz2}
\end{equation}
This generalizes in a straightforward way to more complicated
orbifolds. The corresponding matrix will have a number of entries
which essentially equals the number of images in the covering space,
these being one but the same D-brane in the physical space.  The key
point now is that the set of points including the brane and its images
forms an invariant configuration of $\Gamma$ and transforms in the
{\it regular} representation! We will see this explicitly when discussing
our example.

As it is (hopefully) clear from the figure, the maximal number
of massless states arise when the D-brane is at the orbifold fixed
point $X=0$. There open strings of type, say, $D-D'$ contribute to the
massless spectrum. On the contrary, out of the orbifold fixed point the space
is essentially flat and consistently it turns out that the spectrum of a
D-brane at a generic point is equivalent to (the Coulomb branch of)
that of an usual D-brane in flat space.

What we learn then is that a regular D-brane can move in the full
transverse space, as it is a D-brane with images.

Fractional branes are D-branes whose CP factors transform in the
irreducible representations of the orbifold group $\Gamma$. This
implies they do not have images and are therefore stuck at the
orbifold fixed point. The analogue of figure \ref{reg} is now
\begin{figure}[ht]
\begin{center}
{\includegraphics{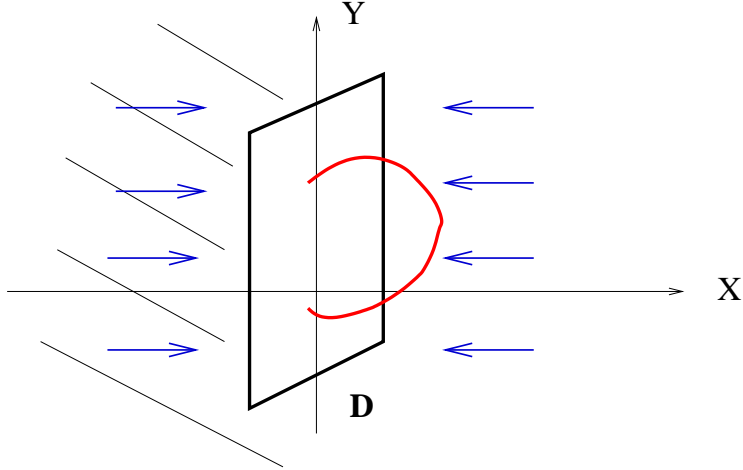}}
\caption{\small A fractional D-brane on the most simple orbifold as it
appears in the covering space. The brane does not have any image in
this  case and it is stuck at the orbifold fixed plane.}
\label{frac}
\end{center}
\end{figure}
\vskip -14pt
Fractional D-branes cannot move along the orbifold but just along the
flat transverse directions. This clearly implies their dynamics is
described in terms of less degrees of freedom as compared to regular
branes and, as we shall see, this is the basic reason why the gauge
theory living on them is a non-conformal gauge theory. Note that now
the CP factors are just numbers as there is only one possible kind  of
open strings, the $D-D$.

We would like now to compute the low energy (i.e. gauge theory)
spectrum living on regular and fractional branes. In order to work
this out explicitly, let us use our key example. Namely, we start
considering D3-branes on the orbifold $\IR^{1,5} \times
\IC^2/\ZZ_2$. The group $\ZZ_2$ consists of two generators: $g$ acting
on the coordinates of $\IC^2$ as a reflection operator, $g : X
\rightarrow - X$, and its square $g^2$ that is nothing but the
identity, $g^2 \equiv e : X \rightarrow X$. Let us suppose the
D3-brane extends along directions $x^0,...,x^3$. This means that
the transverse space consists of two flat directions, $x^4,x^5$, plus
the four orbifold directions, $x^6,...,x^9$ where the element $g$ of
the orbifold group acts non-trivially.

As already explained, to let a (regular) D3-brane be located at a
generic point on the orbifold covering space, we must include its
image (we just have one in this case) and consequently we have four
kinds of open strings. Two corresponding to open strings having both
their end-points on the brane or on its image and two other kinds
corresponding to open strings having one endpoint on the brane and the
other on its image, and vice-versa. These four kinds of open strings
are described by two by two matrices
\begin{equation}
\left(
\begin{array}{cc}
\mbox{1}  & \mbox{0} \\ \mbox{0} & \mbox{0}
\end{array}
\right) \quad , \quad  \left(
\begin{array}{cc}
\mbox{0}  & \mbox{0} \\ \mbox{1} & \mbox{0}
\end{array}
\right) \quad , \quad  \left(
\begin{array}{cc}
\mbox{0}  & \mbox{1} \\ \mbox{0} & \mbox{0}
\end{array}
\right) \quad , \quad  \left(
\begin{array}{cc}
\mbox{0}  & \mbox{0} \\ \mbox{1} & \mbox{0}
\end{array}
\right)
\label{4cp2}
\end{equation}
which altogether make-up the two by two matrix (\ref{cpz2}). The first
question we would like to answer is under which representations of the
orbifold group these CP factors transform. Using eq.~(\ref{cptr}) one
easily sees that in this case
\begin{equation}
\gamma (e) = \left( \begin{array}{cc} 1 & 0\\ 0 & 1
\end{array}
\right) ~~,~~ \gamma (g) = \left( \begin{array}{cc} 0 & 1\\ 1 & 0
\end{array}
\right)
\label{rep2}
\end{equation}
The matrix $\gamma (g)$ can be determined by requiring that it
exchanges an open string ending on the D-brane with an open string
ending on its image and vice-versa. In particular it should send the
first (third) matrix in (\ref{4cp2}) into the second (fourth) and
vice-versa.

The representation (\ref{rep2}) is a 2-dimensional representation and
is nothing but the regular representation of $\ZZ_2$. Hence we see
that D-branes having images are in fact {\it regular} branes, as
anticipated. The representation (\ref{rep2}) is reducible and
decomposes into two irreducible representations
\begin{eqnarray}
{\cal D}_0 \qquad \gamma( e ) = 1 \quad &,& \quad \gamma(g) = 1
\label{irrep2a} \\
{\cal D}_1 \qquad \gamma(e ) = 1 \quad &,& \quad \gamma(g) = -1
\label{irrep2b}
\end{eqnarray}
As it can be easily seen, the regular representation is  the direct
sum  of the above two irreducible representations
\begin{equation}
R = \oplus \, {\cal D}_I \quad , \quad I=0,1
\label{reir2}
\end{equation}
We thus have two kinds of {\it fractional} branes,  according to the
representations of their CP factors. The number of inequivalent
fractional branes coincides with the order of the orbifold group, as
the dimension of the irreducible representation is one, in this case. 
Note that the
CP factors are now numbers and, as anticipated, the corresponding
fractional branes are stuck at the orbifold fixed point as there are
not the degrees of freedom (described by open strings of the form
$D-D'$) to allow them to move along the orbifold directions.

We can now compute the massless
spectrum of open strings living on regular and fractional branes,
respectively. Let us start from regular D-branes. A massless state of
the NS sector has the following form
\begin{equation}
\lambda \;\psi_{-1/2}^{M}\, | k \rangle \quad , \quad M= 0,1 \dots 9
\label{male45}
\end{equation}
In order to keep world volume supersymmetry, $\ZZ_2 $ must act on the
fermionic coordinates in the same way as on the bosonic ones; thus the
oscillator part of the state in eq.~(\ref{male45}) transforms under
$g$  as follows
\begin{eqnarray}
\psi_{-1/2}^{\mu} | k \rangle &\rightarrow & + \,\psi_{-1/2}^{\mu} | k
\rangle~~~~~\mu =0,1,2,3 \\ \psi_{-1/2}^{i} | k \rangle &\rightarrow &
+ \,\psi_{-1/2}^{i} | k \rangle~~~~~ i= 4, 5 \\ \psi_{-1/2}^{m} | k
\rangle &\rightarrow & - \,\psi_{-1/2}^{m} | k \rangle~~~~~m=6, \dots 9
\end{eqnarray}
where we have denoted with $\mu$ the world volume directions of the
D3-brane, with $m$ the four directions along the orbifold and with $i$
the transverse ones outside the orbifold.

Taking into account the action of the orbifold group on both the
oscillators and the CP factors $\lambda$ which can be easily inferred
from eqs.~(\ref{cptr}) and (\ref{rep2}), one gets the following
globally invariant states surviving the orbifold projection
\begin{equation}
\frac{\one+ \sigma_1}{2} \otimes \psi_{-1/2}^{\mu,i} |k\rangle ~~,~~
\frac{\one - \sigma_1}{2} \otimes \psi_{-1/2}^{\mu,i} |k \rangle
\label{gau54}
\end{equation}
corresponding to two gauge fields living on the world volume of the
D3-brane represented by the index $\mu$ and four real Higgs fields
represented by the index $i$, and
\begin{equation}
\frac{ \sigma_3 + i \sigma_2}{2} \otimes \psi_{-1/2}^{m} |k\rangle
~~~,~~~ \frac{\sigma_3 - i \sigma_2}{2} \otimes \psi_{-1/2}^{m} |k
\rangle~~~m=6,7,8,9
\label{gau57}
\end{equation}
corresponding to $8$ scalars. To these bosonic states one should add
the fermionic ones which come from the R sector. At the orbifold fixed
point all these fields are massless and  are grouped together in two
${\cal{N}}=2$ vector multiplets $A_1$ and $A_2$, containing a gauge
field and two real Higgs fields each (plus fermions), and two
hypermultiplets $\Phi_1$ and $\Phi_2$, containing $4$ scalars each
(plus fermions).

The upshot of this analysis is that the gauge theory living on a
regular brane is a ${\cal{N}}=2$ supersymmetric gauge theory with
gauge group $U(1) \times U(1)$ coupled to two hypermultiplets in the
bi-fundamental (see Exercise 1 below) representation. By piling $N$
regular D3-branes on top of each other one then gets ${\cal{N}}=2$ SYM
with gauge group $U(N) \times U(N)$ plus two hypermultiplets in the
$(N,\bar N) \; ,\; (\bar N,N)$. Note that
\begin{itemize}
\item{The scalars in the vector multiplets are degrees of freedom
associated to displacement along flat directions. The scalars in the
hypermultiplets are degrees of freedom associated to displacement
along the orbifold directions. This shows that regular branes have
both {\it Coulomb} and {\it Higgs} phases.}
\item{The gauge theory is conformal, i.e. the $\beta$-functions of
both gauge groups vanish.}
\item{Regular branes on orbifolds are pretty similar to usual branes
in flat space: when moving along the orbifold directions only the
diagonal $U(N)$ survives and the low energy effective theory is
equivalent to the Coulomb branch of $U(N)$ ${\cal N}=4$ SYM, as it is
the case for D3-branes in flat space.}
\end{itemize}

\bigskip
\noindent{\bf Exercise 1 -} {\it Compute the
charges of the hypermultiplets with respect to the two gauge factors,
and show they transform in the bi-fundamental representation}. Hint:
perform a change of basis in the space of CP factors with a matrix $A
= (\one - i \sigma_2)/\sqrt{2}$.  \bigskip

We can easily repeat the above reasoning for the fractional branes,
defined through eqs.~(\ref{irrep2a}) and (\ref{irrep2b}).  In this
case, for either fractional branes of type 1 and 0, the massless open
string states surviving the orbifold projection are easily found to be
\begin{equation}
a\;\psi^{\mu,i}_{-1/2}\, |k \rangle
\label{ma43}
\end{equation}
where $a$ is just a number. These states correspond to a gauge field,
represented by the index $\mu$, and two real scalar fields,
represented by the index $i$, belonging to an ${\cal{N}} =2$ vector
multiplet (again, fermions should be added by considering the R
sector). The additional scalars belonging to the hypermultiplets are
projected out by the orbifold projection (this implying that
fractional branes are stuck on the orbifold fixed plane, as
anticipated). By piling $N$ fractional D3-branes on top of each other
one then gets pure ${\cal{N}}=2$ SYM with gauge group  $U(N)$. Note
that
\begin{itemize}
\item{Fractional branes have {\it Coulomb} phase only (the
hypermultiplets are frozen).}
\item{The gauge theory living on them is a non-conformal theory. This
is why they are so interesting objects to be studied when looking for
non-conformal extensions of the gauge/gravity correspondence.}
\end{itemize}

All what we have been saying can be extended to any orbifold of the
so-called ADE series, $\IC^2 / \Gamma$, as well as to less
supersymmetric orbifolds, i.e. orbifold limits of CY three-folds,
$\IC^3 / \Gamma$, where $\Gamma$ is now a discrete subgroup of
$SU(3)$. While for abelian orbifolds the dimension of the irreducible
representations is one, for non abelian orbifolds this is not true
anymore and the number  of different fractional branes is less
than the order of the group $\Gamma$. The analysis of the massless
open string spectrum we performed for $\ZZ_2$ can be also generalized,
of course, and the result one finds can be finally summarized as
follows

\vskip 15pt
\begin{center}
\fbox{\noindent\parbox{5.2in}{\centerline{$N$ regular D3-branes on
$\IC^2 / \Gamma$ or $\IC^3 / \Gamma$ are described by}
\begin{itemize}
\item{Conformal ${\cal N}=2$ or ${\cal N}=1$ SYM.}
\item{Gauge group $\,U(n_0 N)\times U(n_1 N) \times \, ... \,\times
U(n_{m-1}N)\,$ where $m$ is the number of irreps of $\Gamma$ and $n_I$
their dimensions.}
\item{Matter in the bi-fundamental (either hypermultiplets or chiral
multiplets). The amount of matter is such to make the theory
conformal.}
\end{itemize}}}
\end{center}

\vskip 1pt
\begin{center}
\fbox{\noindent\parbox{5.2in}{\centerline{$N$ fractional D3-branes
$\IC^2 / \Gamma$ or $\IC^3 / \Gamma$ of the $I$-th type  are described
by}
\begin{itemize}
\item{Pure ${\cal N}=2$ or ${\cal N}=1$ SYM.}
\item{Gauge group $U(n_I N)$ where $n_I$ is the dimension of the
$I$-th  irrep.}
\item{There are $m$ different types of fractional branes as
$I=0,1,...,m-1$.}
\end{itemize}}}
\end{center}
\bigskip

Note that fractional D3-branes on ${\cal N}=1$ orbifolds are
completely stuck at the orbifold apex as there is not flat transverse
space available for them to move. This is consistent with what one
learns from the gauge theory living on them, which is pure ${\cal
N}=1$ SYM  and which does not have any scalar degrees of freedom.

\bigskip
\noindent{\bf Exercise 2 -} {\it Consider the orbifold $\IC_2/
\ZZ_3$. The orbifold group $\ZZ_3$ is generated by three elements now,
$\ZZ_3 = \{g, g^2, g^3 \equiv e \}$. Compute the regular and the
irreducible  representations of $\ZZ_3$ and find the low energy gauge
theory spectrum of open strings ending on regular and fractional
D3-branes, respectively}. Hint: the group element $g$ acts on the
coordinates $z = x^6 + i x^7$ and $w = x^8 + i x^9$ as a two by two
diagonal matrix with entries $\exp(2 i \pi/3)$.  \bigskip

By considering bound states of different numbers and types of
fractional branes one can construct SYM theories with product gauge
group coupled to bi-fundamental matter (notably, in this case one can
have bound states of fractional branes which can actually move along
the orbifold). One can also introduce fundamental matter by adding
D7-branes. For instance, by considering a bound state of $N$ D3-branes
plus $M$ D7-branes on the ${\cal N}=1$ orbifold $\IC_3/\ZZ_2 \times
\ZZ_2$ one ends up with ${\cal N}=1$ SYM coupled to $M$ chiral
multiplets in the fundamental, this resembling Super QCD. In fact,
this way of brane-engineering gauge theories by means of fractional
branes has been much used recently in the context of the bottom-up
approach of string phenomenology.

\insertion{2}{Quiver Interlude \label{insert2}} {All gauge theories
one can obtain by considering D-branes on orbifolds can be organized
in terms of {\it quiver} diagrams (hence the name ``quiver gauge
theories'' often used in the literature). This originates from the
one-to-one correspondence between the gauge theories of D-branes of
$\IC_2/\Gamma$ orbifolds and the (extended) simply laced Dynkin
diagrams of the ADE series. For instance, abelian orbifolds
$\Gamma=\ZZ_m$ are classified by the extended $A_{m-1}$ Dynkin
diagrams. Each dot corresponds to a gauge group factor while lines
connecting two dots represent matter in the  bi-fundamental
representation of the two corresponding gauge groups. For instance,
the gauge theory on $N$ regular D3-branes of our key example,
$\Gamma=\ZZ_2$, is represented by
%\begin{figure}[ht]
\begin{center}
{\includegraphics{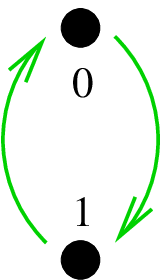}}
\end{center}
%\end{figure}
We have two gauge factors, represented by the two dots (where the
vector multiplets sit), and two bi-fundamental hypermultiplets,
represented by the arrows going from one point to the other. More 
generally, the ${\cal N}=2$ gauge  theory living on $N$ regular
D3-branes on the orbifold $\IC_2/\ZZ_m$ is  represented by the
following quiver diagram \vskip 25pt
%\begin{figure}[ht]
\begin{center}
{\includegraphics{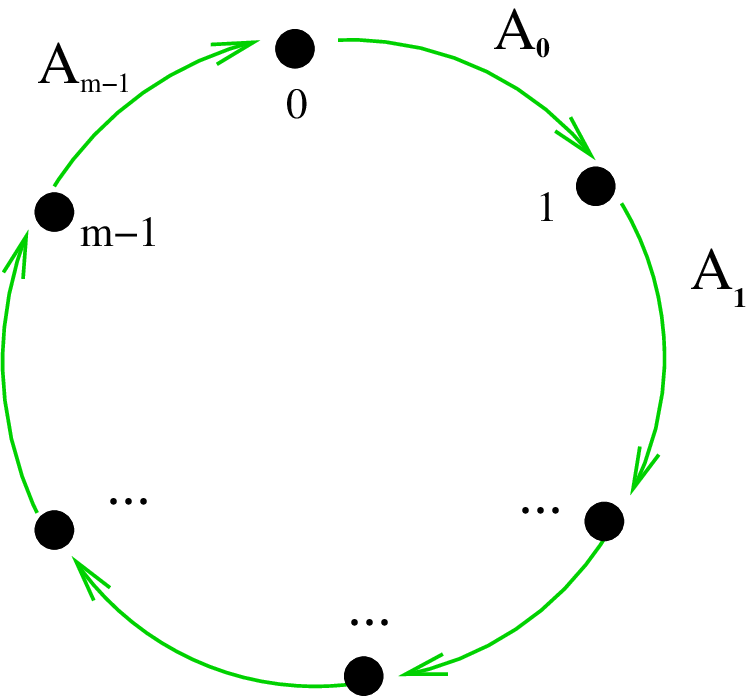}}
\end{center}
\hskip 355pt {\it Continued...}  }

\insertion{2}{\it Continued...}{Again, each dot represents a $U(N)$
gauge factor while the lines between two dots correspond to a
hypermultiplet ${A}_I$ ($I=0,...,m-1$) transforming in the
bi-fundamental representation of the two groups belonging to the two
dots. Arrows go from fundamental to anti-fundamental representations
of the corresponding gauge groups. The fermionic partner of each
bosonic field is implicit in the figure.

Although for ${\cal N}=1$ orbifolds the mathematical correspondence
just discussed does not hold, strictly speaking, in some cases all
what we have been  saying can be extended also to the ${\cal N}=1$
case. As an example, we draw the quiver diagram for the ${\cal N}=1$
quiver gauge theory on the orbifold $\IC_3/\ZZ_5\times \ZZ_3$, where
matter is now represented by chiral multiplets.
\begin{center}
{\includegraphics{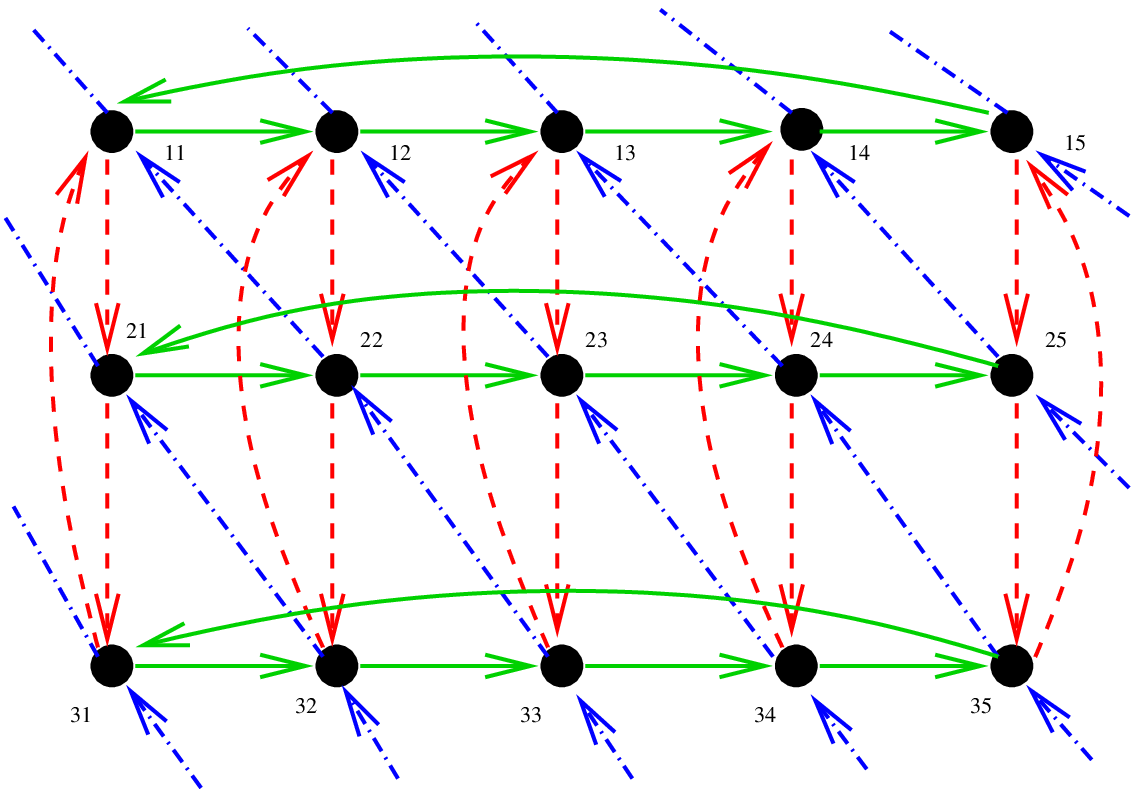}}
\end{center}
Each dot represents a $U(N)$ gauge factor, in double index notation
while the lines connecting dots correspond to bi-fundamental chiral
multiplets. Again, arrows go from fundamental to anti-fundamental
representations. We have been using different colors just to let the
picture be more readable. The blue lines starting from the first row
go back to the third one.}

All previous considerations, which we have done for D3-branes, can be
easily extended to regular and fractional D$p$-branes, with $p\not =
3$. The only essential difference, at this level, is that the low
energy  effective theory living on them is in general a
(p+1)-dimensional  gauge theory. Since we are mainly interested in
four dimensional gauge theories, we will not discuss D$p$-branes further.

As a final remark, note how eqs.(\ref{reirn}) and (\ref{reir2}) seem
to suggest that regular branes can be thought as 'made of' fractional
branes. This idea is supported also by considering the gauge theory
living on them. From what said above, it is clear that by considering
a bound state made of $N$ fractional brane of type 0, N of type 1,
etc... up to $N$ of type $m-1$ (i.e. $N_I=N$ for any $I=0,...,m-1$),
one ends up with a gauge theory which is exactly that of a bound state
of $N$ regular branes. This n\"aive identification of regular branes
in terms of fractional branes turns out to be correct, as it will
become apparent in the next section, when discussing fractional branes
from the closed string perspective.

%%%%%%%%%%%%%%%%%%%%%%%%%%%%%%%%%%%%%%%%%%%%%%%%%%%%%%%%%%%%%%%%%%%%%%%
\subsection{Fractional Branes: Closed String Perspective}

D-branes are non-perturbative states of the closed string spectrum. As
such, they couple to closed string states and to find the supergravity
solutions describing them at low energy one should know the massless 
fields to which the D-branes couple. A usual
Dp-brane in flat space couples to the graviton, the dilaton and a
RR-form potential
\begin{eqnarray}
\mbox{NSNS}: \; G_{MN}\;,\; \Phi \quad ; \quad \mbox{RR}: \; C_{(p+1)}
\nonumber
\end{eqnarray}
What do we get when considering D-branes on orbifolds? One should
remind that now there are both untwisted and twisted sectors, the
number of twisted sectors being equal to the number of non-trivial
element of the orbifold group $\Gamma$. Recall that the strings
belonging to the twisted sectors do not carry momentum along the
orbifold directions and their center of mass is stuck at the orbifold
fixed point. This means that the corresponding massless fields have a
(10-$n$)-dimensional dynamics, where $n$ is the spatial dimension of the
orbifold, as opposed to massless fields in the untwisted sector, which
of course have a ten-dimensional dynamics.

An efficient way to find out which fields couple to regular and
fractional branes on orbifolds is to implement open/closed string
duality and describe D-branes in terms of boundary states,
$|D\rangle$, and compute the overlap $\langle D| \phi \rangle$, with
$\phi$ being any possible field of the relevant supergravity
spectrum. In doing so one finds that
\begin{itemize}
\item{Regular Dp-branes couple to the untwisted sector only, right in
the same way as usual D-branes in flat space
\begin{eqnarray}
\mbox{NSNS}: \;G_{MN}\;,\;\Phi \quad ; \quad \mbox{RR}: \; C_{(p+1)}
\nonumber
\end{eqnarray}}
\item{Fractional Dp-branes couple to both untwisted and twisted
sectors. In particular, fractional Dp-branes of type $I$ (recall
previous section) couple to the fields belonging to the $I$-th twisted
sector: a scalar field $b_I$ in the NSNS sector and a (p+1)-form
potential in the RR one
\begin{eqnarray}
\mbox{NSNS}:& \; G_{MN}\;,\;\Phi \quad ; \quad \mbox{RR}:& \;
C_{(p+1)} \nonumber \\ \mbox{NSNS}_T:& \quad b_I \quad \quad \quad
\quad   ; \quad \mbox{RR}_T:&  \; A^I_{(p+1)}\nonumber
\end{eqnarray}}
\end{itemize}
The coupling of fractional branes to twisted sectors is nothing but
the closed string counterpart of fractional branes not having dynamics
along the orbifold.

Using boundary state language one can see that the analogue of
relation ~(\ref{reirn}) is now
\begin{equation}
\label{breirn}
|D\rangle = \sum_I |D\rangle_I \qquad \mbox{with}\quad I=0,1,..., m-1
\end{equation}
where $|D\rangle$ is the boundary state representing a regular brane
and $|D\rangle_I$ the boundary state representing a fractional brane
of type $I$. Consistently, one can check that by summing up the
boundary states of all kinds of fractional branes as dictated by Eq. 
(\ref{breirn}), the overall twisted coupling cancel, so
that the boundary state couples only to the untwisted sector, as it
should be the case for a regular D-brane. Note that $I$ runs on the
irreps of $\Gamma$ which equals the number of twisted sector {\it
plus} one. While this turns out to be crucial in the cancellation
procedure just mentioned, it can be rather confusing as it seems to
spoil the on-to-one correspondence between fractional branes and
twisted sectors outlined above. Actually, the number of twisted
sectors is indeed one less than the number of different fractional
branes. In order to understand this point we should make a
digression. This is worth doing though, as it will pay out a lot for
the understanding of the whole picture.

There is another point of view to look at D-branes on
orbifolds which is rather illuminating in that it helps to understand
better some of the issues raised so far in comparing regular and
fractional branes, orbifold group representations, twisted sectors and
so on. In what follows, we concentrate on ${\cal N}=2$ orbifolds,
namely those of the form $\IC_2/\Gamma$, $\Gamma$ being a discrete
subgroup of $SU(2)$. These orbifolds are singular limits of ALE
spaces, the latter being characterized by compact two-cycles ${\cal
C}_I$ which are topologically spheres and shrink to zero size in the
orbifold limit. ALE spaces are classified by the ADE Dynkin diagrams
and through a precise mathematical construction due to
McKay and Kronheimer the following one-to-one correspondences hold
\begin{eqnarray}
\label{rel1}
\alpha_I \longleftrightarrow {\cal C}_I \longleftrightarrow {\cal D}_I
\end{eqnarray}
$\alpha_I$ being the simple roots of the corresponding Dynkin diagram,
${\cal C}_I$ the 2-cycles characterizing the ALE space and ${\cal D}_I
$ the $m$ irreducible representations of the orbifold group
$\Gamma$. The trivial representation ${\cal D}_0$ is associated to the
{\it non-independent} 2-cycle ${\cal C}_0 = - \sum_{I\not = 0} d_I
{\cal C}_I$ and the corresponding root $\alpha_0$ is nothing but the
extra root in the Dynkin diagram, which is in fact an {\it extended}
Dynkin diagram. The quiver diagrams discussed in Insert 2 are just
examples of the above correspondence.

The twisted fields can be seen as the zero modes of the Kaluza-Klein
(KK) states of higher forms of the ten-dimensional supergravity
spectrum over the cycles ${\cal C}_I$ in the orbifold limit
\begin{equation}
b_I \equiv \int_{{\cal C}_I} B_{(2)} \qquad\qquad A^I_{(p+1)} \equiv
\int_{{\cal C}_I} C_{(p+3)}
\end{equation}
where $B_{(2)}$ is the NSNS 2-form potential. From this point of view
the reason why twisted fields have a (10-n=6)-dimensional dynamics is
simply due to the fact that the cycles ${\cal C}_I$ are localized at
the orbifold fixed point, in the orbifold limit.  Collecting the
following three facts

\begin{itemize}
\item{Fractional branes of type $I$ couple to twisted fields of the
$I$-th twisted sector}
\item{Fractional branes of type $I$ are associated to the irrep ${\cal
D}_I$ of $\Gamma$ and this is in one-to-one correspondence with the
cycle ${\cal C}_I$}
\item{Twisted fields of rank q are the zero modes of KK of (q+2)-forms
on ${\cal C}_I$}
\end{itemize}
one is tempted to conclude that

\vskip 8pt  \centerline{\it A fractional Dp-brane (of type $I$) is a
D(p+2)-brane}

\centerline{\it wrapped on the cycle ${\cal C}_I$ in the orbifold
limit.}  \vskip 8pt

This is indeed the case and can be proved rigorously following the
mathematical construction exemplified in the relation~(\ref{rel1}). We
are not going to do this explicitly as it is beyond the purpose of
this course, but I will soon give a consistency check for that. Before
doing so it is worth making a couple of comments.
\begin{itemize}
\item{If fractional branes branes are branes wrapped on vanishing
    cycles, how can they have a non-vanishing tension? The point is
    that in the orbifold limit the cycles ${\cal C}_I$ vanish but a
    non-zero $B_{(2)}$-flux persists on them giving a non-vanishing
    {\it background} value to the field $b_I$
\begin{eqnarray}
\label{bIbg}
b_I \equiv \int_{{\cal C}_I} B_{(2)} &=&  \frac{d_I}{|\Gamma|} \not =
0 \\ &=& \frac{1}{2} \quad \mbox{for} \quad \IC_2/ \ZZ_2 \nonumber
\end{eqnarray}
where we are using here units where $4 \pi \alpha'^2=1$.  This means
that, roughly speaking, the {\it stringy} volume of the cycle is not
vanishing. We will show this explicitly later, when discussing in
detail our usual key example. The above statement can be also
justified from a CFT point of view. One can show that a non-singular
CFT on $\IC_2/\Gamma$, as it is the string sigma model, is obtained if
and only if there is a non-vanishing background value for $b_I$.  The
precise value in the above equation turns out to be the value for
which the sigma model is free.}
\item{The fractional brane associated to ${\cal D}_0$ corresponds to a
brane wrapped on the cycle ${\cal C}_0$. This is not an independent
cycle so it could seem the corresponding fractional brane is not
independent either. In fact, this is not the case. The point is that
the fractional brane associated to ${\cal D}_0$ corresponds to a brane
wrapped on ${\cal C}_0$ with an additional background flux of the
world volume field strength ${\cal F}$ switched on, and normalized as
$\int_{{\cal C}_0}{\cal F} = 2\pi$. As it will become clear shortly,
this ensures that the fractional brane of type 0 has the same
untwisted charge of other fractional branes (so it is not an
anti-brane) and is indeed independent on them.}
\end{itemize}
The identification between fractional and wrapped branes can be
checked, for instance, by comparing the action of a fractional
Dp-brane (obtained for example using boundary state techniques) with
that of a wrapped D(p+2)-brane. If one consider a fractional
Dp-brane of type $I$, the structure of its world volume action (which 
can be derived for instance using boundary state
techniques) reads
\begin{equation}  
{\cal S}_I \;=\; - \frac{T_p}{\kappa} \int d^{p+1} \xi \;\, {\rm
e}^{\frac{p-3}{4}\, \Phi} \;\sqrt{- \det G}\; b_I \,+\,
\frac{T_p}{\kappa} \left[ \int C_{(p+1)}\; b_I \,+\, \int A_{(p+1)}
\right]
\label{pfracI}
\end{equation}
where $b_I$ is now the complete twisted scalar field (namely its
background value (\ref{bIbg}) plus the fluctuation, $\tilde b_I$),
$T_p = \sqrt{\pi}\, (2 \pi \sqrt{\alpha'})^{3-p}$ and $\kappa =  8
\pi^{7/2}\alpha'^2 g_s$. By summing up the world volume action of all
possible fractional branes in the given orbifold theory one gets the
world volume action for the corresponding regular brane, ${\cal S} =
\sum_I {\cal S}_I$. This is the analogue of relations~(\ref{reirn})
and (\ref{breirn}). From eq.~(\ref{pfracI}) we can see why the
background value of the scalar twisted field $b_I$, eq.~(\ref{bIbg}), 
is crucial to make
the fractional branes tension-full. Moreover, we also see why they are
called fractional: their RR untwisted charge (i.e. the coefficient
multiplying the WZ term for $C_{(p+1)}$) is a {\it fraction} of that
of a regular brane. What fraction this is, it is determined by the
orbifold at hand, as it can be seen from eq.~(\ref{bIbg}).

Let us now see how one can get eq.~(\ref{pfracI}) following the wrapped
brane program previously outlined. We concentrate for simplicity on
our key example, i.e. the orbifold $\IC_2/\ZZ_2$. According to what we
have learnt, in this case we have two different kinds of fractional
D-branes, as we have two irreducible representations of the orbifold
group, and just one shrinking cycle, ${\cal C}_1$.  The fractional
Dp-brane of type 1 should correspond to a D(p+2)-brane wrapped  on
${\cal C}_1$. The fractional Dp-brane of type 0 to a D(p+2)-brane
wrapped on ${\cal C}_0 = - {\cal C}_1$ with a non vanishing ${\cal
F}$-flux on it. In the following we will be pedantic with all
factors of $\pi$, $\alpha'$, etc... In the  Einstein frame the world
volume action of D(p+2)-brane has in  general the form
\begin{eqnarray}
{\cal S} =  &-& \frac{T_{p+2}}{\kappa} \int d^{p+3} \xi  \,\,{\rm
e}^{\frac{p-1}{4} \Phi} \sqrt{- \det \left[ G +  {\rm e}^{- \phi/2}
\left(B + 2 \pi \alpha'  {\cal F}\right)\right]} \, + \nonumber \\
&+&\frac{T_{p+2}}{\kappa} \int \left[ C \wedge {\rm e}^{B + 2 \pi \alpha'
{\cal F}} \right]_{(p+3)}
\label{bp+2}
\end{eqnarray}
Our D(p+2)-brane is not generic, but actually a very specific one. The
smooth limit of the $\ZZ_2$ orbifold is the well known  Eguchi-Hanson
space, which has an antiself-dual two-form $\omega_2$  which is
associated to the compact 2-sphere ${\cal C}_1$. The $\omega_2$
satisfies the following properties
\begin{equation}
\omega_2 = - {}^{*} \omega_2~~,~~ \int_{{\cal C}_1} \omega_2 = 1~~,~~
\int {}^* \omega_2 \wedge \omega_2 = \frac{1}{2}
\label{omega289}
\end{equation}
The compact cycle vanishes in the orbifold limit but, as already said,
a non-zero $B_{(2)}$-flux persists on it. In order to obtain from the
action in eq.(\ref{bp+2}) the world volume actions of the two
fractional Dp-branes, we should start from an action  with no world
volume fields switched-on along the p+1 non-compact directions of the
world volume. That is to say,  both $B_{(2)}$ and ${\cal F}$ are
non-vanishing only on the cycle ${\cal C}_1$. Moreover, to make the
action in eq.~(\ref{bp+2}) describing a brane wrapped on ${\cal C}_1$,
we  should consider its (p+3)-dimensional world volume ${\cal
V}_{p+3}$ as a product of a flat (p+1)-dimensional volume ${\cal
V}_{p+1}$ times the volume of the cycle ${\cal C}_1$ and keep only
those fields that are left in  the orbifold limit
\begin{equation}
{\cal V}_{p+3} = {\cal V}_{p+1} \times {\cal C}_1~~,~~ B_{(2)} = b_1
\,\omega_2~~,~~C_{(p+3)} = A_{(p+1)} \wedge \omega_2
\label{wra98}
\end{equation}
By noticing that the metric has no support on the vanishing cycle, one
can easily factorize the matrix in the determinant in the action
(\ref{bp+2}) as a direct product of a  $(p+1) \times (p+1)$ matrix $G$
times a $2\times 2$ matrix where only $B_{(2)}$ and ${\cal F}$ are
present
\begin{eqnarray}
&&\sqrt{- \det \left[ G +  {\rm e}^{- \Phi/2} \left(B + 2 \pi \alpha'
{\cal F}\right)\right]_{p+3\times p+3} } = \nonumber \\
&=& \sqrt{- \det
\left[G\right]_{p+1\times p+1} } \; {\rm e}^{- \Phi/2}  \int_{{\cal
C}_1} \left(B_{(2)} + 2 \pi \alpha' {\cal F}\right)
\end{eqnarray}
We now have all the ingredients to get the desired result. Let us consider
first  the case of a fractional brane of type 1. We want to show that
it corresponds to a D(p+2)-brane wrapped on ${\cal C}_1$ with no
${\cal F}$-flux. Inserting the expressions~(\ref{wra98}) into
eq.~(\ref{bp+2}) one gets
\begin{eqnarray}
{\cal S}_1 = &-& \frac{T_{p+2}}{\kappa} \int d^{p+1} \xi \,\,{\rm
e}^{\frac{p-3}{4}\Phi} \sqrt{- \det G }  \int_{{\cal C}_1} B_{(2)} +
\nonumber \\ &+& \frac{T_{p+2}}{\kappa} \left[  \int
C_{(p+1)}\int_{{\cal C}_1}  B_{(2)}  + \int A_{(p+1)} \right]
\end{eqnarray}
Recalling now that in this case
\begin{equation}
\label{bfl1}
\int_{{\cal C}_1} B_{(2)} = 4 \pi \alpha'^2 \left(\frac{1}{2} +
\frac{1}{4 \pi \alpha'^2} \,\tilde b_1\right)
\end{equation}
and $T_{p} = 4 \pi \alpha'^2 \, T_{p+2}$ we finally get
\begin{eqnarray}
{\cal S}_1 = &-& \frac{T_{p}}{\kappa} \int d^{p+1} \xi  \,\,{\rm
e}^{\frac{p-3}{4}\Phi} \sqrt{- \det G} \, \left(\frac{1}{2} +
\frac{1}{4\, \pi^2 \alpha'} \,\tilde{b}_1 \right) + \nonumber \\  &+&
\frac{T_{p}}{\kappa} \left[\int C_{(p+1)} \left(\frac{1}{2}+
\frac{1}{4\,  \pi^2 \alpha'}\, \tilde{b}_1 \right) +  \frac{1}{4\,
\pi^2 \alpha'} \int A_{(p+1)} \right]
\label{bpf1}
\end{eqnarray}
which is the desired result, see eq.~(\ref{pfracI}). By repeating the 
same reasoning for a
D(p+2)-brane which is now wrapped on ${\cal C}_0 = - {\cal C}_1$ but
with an additional ${\cal F}$-flux normalized as $\int_{{\cal C}_0}
{\cal F} = 2 \pi$ one gets
\begin{eqnarray}
{\cal S}_0 = &-& \frac{T_{p+2}}{\kappa} \int d^{p+1} \xi \,\,{\rm
e}^{\frac{p-3}{4}\Phi} \sqrt{- \det G } \int_{{\cal C}_0}
\left(B_{(2)} + 2 \pi \alpha' {\cal F}\right) + \nonumber \\  &+&
\frac{T_{p+2}}{\kappa} \left[ \int C_{(p+1)}\int_{{\cal C}_0}  \left(
B_{(2)} + 2 \pi \alpha' {\cal F}\right) - \int A_{(p+1)}  \right]
\end{eqnarray}
Using now that
\begin{equation}
\int_{{\cal C}_0} \left(B_{(2)} \,+\, 2 \pi \alpha'  {\cal F}\right) =
- 4 \pi^2 \alpha'\,\left(\frac{1}{2}  + \frac{1}{4 \pi^2\alpha'}
\,\tilde{b}_1 \right) + 4 \pi^2 \alpha' = 4 \pi^2 \alpha'
\,\left(\frac{1}{2} - \frac{1}{4 \pi^2\alpha'} \, \tilde{b}_1 \right)
\label{for968}
\end{equation}
we finally get
\begin{eqnarray}
{\cal S}_0 = &-& \frac{T_{p}}{\kappa} \int d^{p+1} \xi  \,\,{\rm
e}^{\frac{p-3}{4}\Phi} \sqrt{- \det G} \, \left(\frac{1}{2} -
\frac{1}{4 \pi^2 \alpha'}\, \tilde{b}_1 \right) +  \nonumber \\  &+&
\frac{T_{p}}{\kappa} \left[ \int C_{(p+1)} \left(\frac{1}{2} -
\frac{1}{4  \pi^2 \alpha'} \,\tilde{b}_1 \right) - \frac{1}{4 \pi^2
\alpha'} \int A_{(p+1)} \right]
\label{bpf2}
\end{eqnarray}
which represents the world volume action of the fractional Dp-brane of
type 0.  We can see from eq.~(\ref{for968}) how the presence of the
${\cal F}$-flux makes the asymptotic value of the {\em untwisted}
charge to be unchanged, as anticipated. The difference between the
actions  (\ref{bpf1}) and (\ref{bpf2}) amounts to a different sign in
the couplings to the fields of the twisted sector. This is a trivial
case since we have just one twisted sector. In general, the fractional
brane of type 0 couples to all twisted sectors with a minus sign with
respect to the other fractional branes. Still, its coupling to the
untwisted sector is the same as the other branes. This ensures it is a
brane and not an anti-brane. Note that by summing up the two actions
(\ref{bpf1}) and (\ref{bpf2}) the coupling to the twisted sector
cancels out and one gets back the world volume action of a regular
brane, as expected. This agrees with  the idea that regular branes can
be thought of as bound states of fractional branes of different
kinds. The same holds, of course, for a more general orbifold: the sum
of the world volume actions of all fractional branes give the action
of the regular brane.

\bigskip
\noindent{\bf Exercise 3 -} {\it Consider the orbifold $\IC_2/
\ZZ_3$. There are two non trivial cycles now, ${\cal C}_1\,,\,{\cal
C}_2$, two twisted sectors and three different kinds of fractional
branes. Derive the world volume  action of the three different
D3-branes and show they sum-up giving the world volume action of the
corresponding regular D3-brane, the latter not coupling to the twisted
sectors}. Hint: the background value of the $b_I$ fields ($I=1,2$) is
now $1/3$.

%%%%%%%%%%%%%%%%%%%%%%%%%%%%%%%%%%%%%%%%%%%%%%%%%%%%%%%%%%%%%%%
\subsection{Gauge Theory From Gravity: Example 1}

We have now all the information we need to derive the supergravity
solutions  describing (bound states of) fractional branes of various
kind as we know what fields of the supergravity spectrum the branes
couple to. Moreover, for any case we like to study, we know how to 
determine the
corresponding dual supersymmetric gauge theory should be. Once we find
the supergravity solutions, we can then exploit them to see what can
we learn about the corresponding dual gauge theories.

To be specific, we will go on focusing on our key example. This
amounts to study the supergravity solution describing a bound state of
$N$ fractional D3-branes on the orbifold $\IC_2/\ZZ_2$ whose low
energy effective dynamics is four dimensional ${\cal N}=2$ pure SYM.
Still, the underlying philosophy is all the same, and the logical
procedure and technical tools developed in this case should allow the
reader to understand (and work out) any other case without much effort.

From what we learned in the previous section fractional D3-branes  on
the orbifold $\IC_2/\ZZ_2$ couple to the graviton, a twisted scalar
$b$ (as we have just one twisted sector we omit the index $I$) and two
RR potentials, one in the untwisted sector, $C_{(4)}$, and one in the
twisted sector, $A_{(4)}$. Hence we expect that the other fields of
the supergravity spectrum are trivial in the solution.

Let us start from the type IIB supergravity action in 10 dimensions
which reads
\[
{\cal S}_{\rm IIB} = \frac{1}{2 \kappa^2} \Bigg\{ \int d^{10} x~
\sqrt{-\det G}~ {\cal R} - \frac{1}{2} \int \Big[ d \Phi \wedge {}^* d
\Phi \,+\, {\rm e}^{- \Phi} H_{(3)}  \wedge {}^* H_{(3)}\,+\, {\rm
e}^{2 \Phi}\, F_{(1)} \wedge {}^* F_{(1)}
\]
\begin{equation}
+ \,\,{\rm e}^{\Phi} \,{\widetilde{F}}_{(3)} \wedge {}^*
{\widetilde{F}}_{(3)} \,+\, \frac{1}{2}\, {\widetilde{F}}_{(5)} \wedge
{}^* {\widetilde{F}}_{(5)}  \, + \,  C_{(4)} \wedge H_{(3)} \wedge
F_{(3)} \Big] \Bigg\}
\label{SIIb}
\end{equation}
where
\begin{equation}
H_{(3)} = d B_{(2)}~,~F_{(1)}=d C_{(0)}~,~ F_{(3)} = d
C_{(2)}~,~F_{(5)} = d C_{(4)}
\end{equation}
are, respectively, the field strengths of  the NS-NS 2-form and the
0-, 2- and 4-form potentials of the R-R sector, and
\begin{equation}
{\widetilde{F}}_{(3)} = F_{(3)} - C_{(0)} \wedge H_{(3)}~
,~{\widetilde{F}}_{(5)} = F_{(5)} - C_{(2)} \wedge H_{(3)}
\end{equation}
As usual, the self-duality constraint ${}^* {\widetilde{F}}_{(5)} =
{\widetilde{F}}_{(5)}$  has to be implemented on shell. In order to
find the classical solution corresponding to a bound state of $N$
fractional  D3-branes of, say, type 1, we have to add to the previous
bulk action ($N$ times) the corresponding world volume action ${\cal
S}_1$ previously found, eq.~(\ref{bpf1}). By varying the sum of the 
bulk and boundary
actions one can derive the  equations of motion for the various fields 
and eventually find the desired solution. As
we are choosing a static gauge and the D3-branes extend along
directions $x^0,...,x^3$, the fields in the solution will depend only
on the transverse space coordinates, the flat ones, $x^4, x^5$, and the
four orbifold directions $x^6,..., x^9$.

On the action (\ref{SIIb}) we have of course to implement the suitable
ansatz for the relevant fields, in particular (recall the previous
section) we have
\begin{equation}
B_{(2)} = b \, \omega_2 \quad , \qquad C_{(6)} = A_{(4)} \wedge
\omega_2
\end{equation}
where $b$ and $A_{(4)}$ only depend on the flat transverse
coordinates, $x^4, x^5$. Actually, we are going to write the solution
in terms of the metric, the 5-form field strength and a twisted {\it
complex scalar}, $\gamma$, defined as $\gamma = c + {\rm i}\, b$ where
$c$ is nothing but the Hodge dual (in the six dimensional sense) of
$A_{(4)}$. We are forced to do that. Indeed, this Hodge duality is
inherited from that between $C_{(6)}$ and $C_{(2)}$ in ten dimensions
(hint: to prove this use the relations (\ref{omega289}) and the ansatz 
(\ref{wra98})). As well
known, the supergravity action (\ref{SIIb}) is expressed in terms of
lower form potentials, hence $C_{(6)}$ does not enter there but rather
its dual $C_{(2)}$ on which we can then implement the fractional brane
ansatz, $C_{(2)} = c \,\omega_2$. As the fractional D3-branes couple
electrically to $A_{(4)}$, we expect a magnetic-like solution for the
field $c$.

By working pretty hard with the equations obtained following the
outlined procedure (this is rather lengthly!) one finally gets the
following solution
\begin{eqnarray} 
\label{metsol} ds^2 &=& H^{-1/2} \, \eta_{\mu\nu}
dx^\mu dx^\nu + H^{1/2}\,  \left[ d\rho^2 + \rho^2 d\theta^2 +
\delta_{mn}\, dx^m dx^n \right] \\
\label{f5sol}{\widetilde{F}}_{(5)}&=& d \left(H^{-1} dx^0 
\wedge ... dx^3 \right) + {}^* d \left(H^{-1} dx^0 \wedge ... dx^3
\right)\\
\label{tssol} 
\gamma &=& c + {\rm i}\, b \;=\; 4 \pi \alpha' g_s N \, \log
\frac{z}{\rho_E}
\end{eqnarray} 
where $\rho = (x_4^2 + x_5^2)^{1/2}$, $z = \rho\, e^{{\rm i}\,
\theta}$, $r = (x_6^2 + ... + x_9^2)^{1/2}$, $\rho_E$ a short distance
regulator for the logarithm, and $H$ a function of the radial
coordinates $\rho$ and $r$, whose explicit expression we do not need,
for the time being.

Now that we have the supergravity solution, we should exploit it and
see what properties of the gauge theory we can reproduce. As this is the
core of the gauge/gravity correspondence, let pause a bit and make a
couple of comments. The basic  point of the duality consists in providing
a precise mapping between supergravity quantities and gauge theories
operators. If one were able to give a prove of the duality, this
should come out for free. In fact, this is not the case. As far as our
present understanding of the gauge/gravity correspondence is
concerned, what has been done, so far, was to check its validity (both
in conformal as well as in non-conformal cases),  but to prove it from
first principles is something different. This is  indeed a very
challenging task which is far from being reached, yet. Rather than
giving a prove, what we can do, at best, is to make a proposal, based
on some reasonable assumptions, and check it. What follows is then
nothing but a (well based) recipe to get such a dictionary. The
dictionary can then be put under test. The way we derive the
dictionary is ultimately related with open/closed string duality but
there could possibly be other equivalent ways to arrive to the same
result. The other one we are aware of is based on purely geometric
reasoning and it would be interesting to find out the relations (which
are there, of course) between these two apparently different
approaches. This is part of present research and could actually also
shed some light on the actual prove of the correspondence.
 
Let us summarize our recipe. The logic behind it is rather simple.
Essentially, what one is doing when dealing with D-branes is studying
their dynamics, which is given in terms of open strings, in the
background of the closed strings. At low energy this amounts to study
the back-reaction of the supergravity background on the D-branes  and
this, in the end, let one express various gauge theory quantities,
which are those governing the dynamics of D-branes at low energy, in
terms of supergravity fields.

The dynamics of D-branes is described by the non-abelian Born-Infeld
(BI)  action, and this we do not really know. What we know is that at
low energy  this is nothing but the action of a given non-abelian
gauge theory. The procedure is then to start considering the abelian
BI action (that we know) with abelian world volume fields switched-on, 
take the $\alpha' \rightarrow 0$ limit and
eventually promote these fields to the adjoint representation of the
gauge group and replace all derivatives with  covariant ones.

\insertion{3}{Some Facts About $\mathbf{{\cal N}=2}$ Super Yang-Mills
\label{insert3}} {Let us briefly recall some features of 
${\cal N}=2$ Super Yang-Mills with gauge group $SU(N)$. This is  a
supersymmetric gauge theory whose field content (auxiliary fields are
not included) is described by the ${\cal N}=2$ vector supermultiplet
\begin{eqnarray}
\left( A_\mu\; , \; \psi^\pm \;, \; \phi \right) \nonumber
\end{eqnarray}
corresponding to a vector field, two Majorana spinors and a complex
scalar, all transforming in the adjoint representation of the gauge
group. On shell this corresponds to 4 bosonic and 4 fermionic degrees
of freedom.

The theory has both a scale anomaly and a $U(1)_R$ anomaly, at the
quantum level. The scale anomaly is accounted for by the
$\beta$-function which is one-loop exact perturbatively and  reads
\begin{eqnarray}
\beta = \frac{g_{\rm YM}^3}{16 \pi^2} \left( - \frac{11}{3}\, c_v +
\frac{1}{6}\, n_s\, c_s + \frac{2}{3} \,n_f \,c_f \right) = - \frac{2
N}{16 \pi^2} \,g_{\rm YM}^3 \nonumber
\end{eqnarray}
where $c_v, c_s$ and $c_f$ are the quadratic Casimir of the adjoint
representation, $c_v = c_s = c_f = N$, $n_s=2$ is the
number of real scalars and $n_f = 2$ that of Majorana fermions. The
corresponding running coupling constant reads
\begin{eqnarray}
\frac{1}{g_{\rm YM}^2(\mu)} = \frac{1}{g_{\rm YM}^2(\Lambda_0)} +
\frac{2 N}{8 \pi^2} \log \frac{\mu}{\Lambda_0}
\label{gymn2}\nonumber
\end{eqnarray}
where $\Lambda_0$ is some UV cut-off and $\mu$ the subtraction scale.
The theory possesses a $U(2) \simeq SU(2) \times U(1)_R$
R-symmetry. The $U(1)_R$ is anomalous and gets broken to $\ZZ_{4 N}$
at the quantum level (this can be seen by computing the triangular
one-loop diagram with one global current and two gauge currents). We have
\begin{eqnarray}
\partial_\mu J^\mu_R \rightarrow \partial_\mu J^\mu_R = n_f\, c_f\, R
\,\frac{1}{16 \pi^2} \, F^A_{\mu\nu}\, \tilde F_A^{\mu\nu} = 4\, N
\,q(x)\nonumber
\end{eqnarray}
where $q(x) \equiv \frac{1}{32 \pi^2} \, F^A_{\mu\nu} \tilde
F_A^{\mu\nu}$ while $R=1$ is  the R-charge of the gauginos (this means
that $\psi$ transform as $e^{{\rm i} \,\epsilon} \psi$ under a
$U(1)_R$ transformation with parameter $\epsilon$).

\hskip 355pt  {\it Continued...}}

\insertion{3}{\it Continued...}{The effect  of the anomaly is
equivalent to assigning the $\theta_{\rm YM}$-angle transformation
properties under $U(1)_R$ as
\begin{eqnarray}
\theta_{\rm YM} \rightarrow \theta_{\rm YM} - 4 N \epsilon \nonumber
\end{eqnarray}
The theory is invariant under shifts $\theta_{\rm YM} \rightarrow
\theta_{\rm YM} + 2 \pi k$. So, if $\epsilon = \frac{\pi k}{2 N}$ the
theory is unchanged even at the quantum level. This shows that 
the full quantum theory is invariant
under $\ZZ_{4 N}$ transformations only. In the
${\cal N}=2$ theory the scale and the $U(1)_R$ anomaly sit in the same
supermultiplet and one can construct a complex coupling $\tau$ out of
$g_{\rm YM}$ and $\theta_{\rm YM}$
\begin{eqnarray}
\label{tau}
\tau = \frac{\theta_{\rm YM}}{2 \pi} + \frac{4 \pi {\rm i}}{g_{\rm
YM}^2} =  {\rm i} \, \frac{2 N}{4 \pi} \, \log
\frac{\phi^2}{\Lambda^2} \nonumber
\end{eqnarray}
where the last equality can be proved to hold exactly, at the
perturbative level ($\Lambda$ is the dynamically generated scale).
This last equation implies that the anomalies can be read from the
response of $\tau$ to a re-scaling of the energy by a parameter $s$
and to a $U(1)_R$ transformation with parameter $\epsilon$. Indeed
under these re-scalings the complex scalar $\phi$ (which has dimension of
energy and R-charge $R=2$) transforms as
\begin{eqnarray}
\phi \longrightarrow s \; e^{2\, {\rm i}\, \epsilon}\, \phi \nonumber
\end{eqnarray}
and one gets
\begin{eqnarray}
\tau \longrightarrow \tau + {\rm i} \, \frac{2 N}{2 \pi} \, \left(
\log s + 2 {\rm i} \epsilon \right)\;\; \;\; \Bigg\{\begin{matrix}
\;\;\quad\frac{1}{g_{\rm YM}^2} \rightarrow \frac{1}{g_{\rm YM}^2} +
\frac{2 N}{8 \pi^2} \log s \\ \theta_{\rm YM} \rightarrow \theta_{\rm
YM} - 4 N \epsilon \end{matrix} \nonumber
\end{eqnarray}
These are all perturbative features of the theory. However, ${\cal
N}=2$ SYM possesses a moduli space and the metric of this moduli
space, which is proportional to the inverse of the gauge coupling
squared, is always positive definite. This means we do not expect a
Landau pole as instead predicted by the perturbative running of the
gauge coupling. As shown by Seiberg and Witten time ago, at scales of
order $\Lambda$ instantons become relevant and change the (otherwise
divergent) perturbative running of the gauge coupling.}

It is not difficult to show that the world volume action of a
fractional  D3-brane on our orbifold with world volume fields
switched-on reads (we are  again pedantic about numerical factors)
\begin{eqnarray}
{\cal S} = &-& \frac{T_{3}}{\kappa} \int d^{p+1} \xi  \, \sqrt{- \det
\left[G + 2 \pi \alpha' F\right]} \;  \frac{b}{4\, \pi^2 \alpha'}
\nonumber \\   &+& \frac{T_{3}}{\kappa} \left[\int C_{(4)} \;
\frac{b}{4\, \pi^2 \alpha'} +  \frac{1}{4\, \pi^2 \alpha'}  \int
A_{(4)} + \frac{\alpha'}{8} \int d^4\xi \;F_{\mu\nu}  \, \tilde
F^{\mu\nu}\right]
\label{bid3f}
\end{eqnarray}
Starting from this action one should
\begin{enumerate}
\item{Choose the static gauge, i.e. $X^\mu = x^\mu\,,\,X^i =
X^i(x^\mu)$ with $\mu = 0,...,3$ and $i=4,...,9$.}
\item{Expand the above action up to terms quadratic in the derivatives
of the fields.}
\item{Take the $\alpha' \rightarrow 0$ limit keeping the combination
$\phi = (2 \pi \alpha')^{-1} z$ fixed.}
\item{Promote $F \rightarrow F_A T^A$ where $T^A$ are the generators
of $SU(N)$ normalized as $Tr\left(T^A T^B\right) = \frac{1}{2}
\,\delta_{AB}$, $A,B$  being adjoint indexes.}
\item{Evaluate the action on the classical solution, call it  ${\cal
S}_{YM}$ ... and see what happens.}
\end{enumerate}
If doing so, after some algebra (this is left as a non-trivial
exercise) one gets
\begin{equation}
{\cal S}_{YM} = -\,\frac{1}{g_{\rm YM}^2} \int d^4 \xi  \left[
\frac{1}{4} F^A_{\mu\nu} F_A^{\mu\nu} + \frac{1}{2}\, D_{\mu} \bar
\phi^A D^{\mu} \phi_A  \right]  + \frac{\theta_{\rm YM}}{32 \pi^2}
\int d^4 \xi\; F^A_{\mu\nu} {\tilde{F}}_A^{\mu\nu} \; + \; \mbox{ferm.}
\label{ymfn2}
\end{equation} 
where
\begin{eqnarray}
\frac{1}{g_{\rm YM}^2 } &=& \frac{1}{16 \pi^3 \alpha' g_s}\int B_{(2)}
= \frac{N}{4 \pi^2} \log \frac{\rho}{\rho_E}
\label{gyn2} \\ 
\theta_{\rm YM} &=& \frac{1}{2 \pi \alpha' g_s} \int C_{(2)} = - 2 N
\theta
\label{thn2}
\end{eqnarray}
Equations (\ref{gyn2}) and (\ref{thn2}) are the supergravity
predictions for the running coupling constant and the $\theta_{\rm
YM}$-angle of the dual ${\cal N}=2$ SYM theory. Let us  see what we
learn from them.

%%%%%%%%%%%%%%%%%%%%%%%%%%%%%%%%%%%%%%%%%%%%%%%%%%%%%%%%%%%%%%%
\vskip 10pt {\it -- The perturbative predictions} \vskip 5pt

From the identification $\phi = (2 \pi \alpha')^{-1} z$ and recalling
that $z = \rho\, e^{{\rm i}\, \theta}$ and that $\phi$ has R-charge 2
we get that under a chiral transformation the {\it physical} angle
$\theta$ shifts as
\begin{equation}
\theta \rightarrow \theta + 2 \pi
\end{equation}
Since $(1/4 \pi^2 \alpha' g_s) \int C_{(2)}$ is allowed to change by
integer values we get that
\begin{equation}
\theta \rightarrow \theta + \frac{\pi}{N}\, k
\end{equation}
are true symmetries of the supergravity solution. On the gauge theory
side this corresponds, see the holographic relation (\ref{thn2}), to a
R-transformation with parameter
\begin{equation}
\epsilon = \frac{\pi\, k}{2\, N}
\end{equation}
which means that every time $\epsilon$ changes by these discrete
values the corresponding gauge theory, as determined from the
supergravity dual, is not changed. This is exactly the expected
$\ZZ_{4 N}$ symmetry of ${\cal N}=2$ SYM. So, the classical
supergravity solution knows about a quantum phenomenon, the breaking
of the $U(1)_R$ down to $\ZZ_{4 N}$! This is an encouraging result.

Let us now consider the other holographic relation,
eq.~(\ref{gyn2}). To this end, there is in fact a lacking element. We
need to establish a precise energy/radius relation which is crucial to
interpret correctly (\ref{gyn2}). What is the relation between the
radial distance $\rho$, which is the unique dimensional quantities in
the supergravity solution, and the energy $\mu$ where the gauge theory
is defined? This is a very deep and important issue in any well
established gauge/gravity correspondence and for non-conformal
theories there is not an unique answer. We will appreciate this point
much more in the next lecture, when discussing the case of ${\cal
N}=1$ SYM, as in the present case the answer will be rather
simple. Still, we want to emphasize this point, already.

The general recipe is to find the gravity dual of some protected
operator of the gauge theory and exploit the corresponding equivalence
to extract a relation between $\rho$ and $\mu$. Here things are pretty
simple as the protected operator is nothing but $\phi$, the complex
scalar. Under a scale transformation with parameter $s$ the mass
scales as $m \rightarrow s m$ and then we get $\phi \rightarrow s 
\phi$. Recalling now that $\phi = (2 \pi \alpha')^{-1} z$ and that $z
= \rho\,e^{{\rm i}\, \theta}$, we easily extract the energy/radius
relation
\begin{equation}
\rho = 2\, \pi \alpha' \mu \quad (\rho_E = 2\, \pi \alpha' \Lambda)
\quad \rightarrow \frac{\rho}{\rho_E} = \frac{\mu}{\Lambda}
\label{idn2}
\end{equation}
Inserting this result into eq. (\ref{gyn2}) we get
\begin{equation}
\label{rung23}
\frac{1}{g_{\rm YM}^2} = \frac{2 \, N}{8 \, \pi^2}\;
\log \frac{\mu}{\Lambda}
\end{equation}
which is indeed the correct result for the (perturbative) running
coupling constant of ${\cal N}=2$ SYM. From the above equation one
obtains also the expected perturbative $\beta$-function, of
course. From the identification (\ref{idn2}) we also understand what
the meaning of the regulator $\rho_E$ is at the gauge theory level:
this is nothing but the dynamically generated scale, where the
perturbative running coupling is expected to diverge. Note that we can
rewrite eq.~(\ref{gyn2}) by implementing the value of the $B_{(2)}$
background flux (use eq.~(\ref{bfl1})) and get
\begin{eqnarray}
\frac{1}{g_{\rm YM}^2(\mu)} &=& \frac{N}{4 \pi^2} \log
\frac{\rho}{\rho_E} \equiv \frac{1}{16 \pi^3 \alpha' g_s} \int B_{(2)}
=\nonumber \\ &=& \frac{1}{4 \pi g_s} \left(\frac{1}{2} + \frac{N
g_s}{\pi} \log \frac{\rho}{\rho_0} \right) = \frac{1}{g_{\rm
YM}^2(\Lambda_0)} + \frac{N}{4 \pi^2} \log \frac{\mu}{\Lambda_0}
\end{eqnarray}
where $\rho_0$ is now a long distance regulator for the logarithm
related to $\rho_E$ by $\rho_E = \rho_0 e^{- \pi/ N g_s}$. Upon the
corresponding identification $\rho_0 = 2 \pi \alpha' \Lambda_0$, with
$\Lambda_0$ being the scale where the bare coupling is defined,  we
finally get the expected relation between the dynamically generated
scale and the UV cut-off in the gauge theory, $\Lambda = \Lambda_0 \,
e^{- 8\pi^2 / N g_{\rm YM}^2}$. Putting together all what we have 
learnt, we can summarize our findings saying that
the twisted complex scalar $\gamma$ is nothing but the holographic
dual of the complex gauge coupling of the gauge theory, $\tau$ (modulo
$4 \pi^2 \alpha' g_s$).

All the results we have been found for our working example have in
fact a  much wider validity. In particular
\begin{itemize}
\item{Adding D7-branes amounts to add fundamental matter. Considering 
the bound state of $N$ fractional D3-branes
with $M$ D7-branes, one ends up with ${\cal N}=2$ SYM with gauge group
$SU(N)$ coupled to $M$ fundamental hypermultiplets. The solution is
more complicated as the D7-branes couple to the dilaton $\Phi$ and the
axion $C_0$ but also in this case the predicted running coupling
constant and $\theta_{\rm YM}$-angle turn out to be the correct ones
\begin{eqnarray}
\frac{1}{g_{\rm YM}^2} &=& \frac{1}{16 \pi^3 \alpha' g_s} \, e^{-\Phi}
\int B_{(2)} = \frac{2 N -M}{8 \pi^2} \log \frac{\rho}{\rho_E}
\nonumber \\ \theta_{\rm YM} &=& \frac{1}{2 \pi \alpha' g_s}
\left[\int C_{(2)} + C_{(0)} \int B_{(2)} \right] = (2N -M) \; \theta
\nonumber
\end{eqnarray}
}
\item{Considering more complicated orbifolds of the ADE series one
gets product gauge groups $U(N) \rightarrow U(N_0) \times U(N_1)
\times ... \times U(N_{m-1}) +$ bi-fundamental matter. Also in this
case the perturbative features recovered for the pure case are
reproduced correctly.}
\item{Considering orbifold limit of CY three-folds one gets theories
with ${\cal N}=1$ supersymmetry. As explained previously, playing with
different orbifolds and different bound states of branes, one can
construct a zoology of gauge theories with matter, study the
corresponding supergravity solutions and finally recover gauge theory
information. A particularly interesting case is to consider a bound
state of $N$ fractional D3-branes with $M$ D7-branes. This corresponds
to ${\cal N}=1$ SYM coupled to chiral  matter, resembling very much
Super QCD. Again, also for these cases  all perturbative gauge theory 
properties are correctly reproduced by the dual supergravity backgrounds.}
\end{itemize}

%%%%%%%%%%%%%%%%%%%%%%%%%%%%%%%%%%%%%%%%%%%%%%%%%%%%%%%%%%%%%%%
\vskip 10pt {\it -- The enhan\c{c}on and the non-perturbative
corrections} \vskip 5pt

All what we have been found so far are perturbative information. A
question naturally arises at this point: what about non-perturbative
contributions? Are these supergravity duals able to recover them, too?

To start with, a detailed analysis of the supergravity solution
(\ref{metsol})-(\ref{tssol}) shows that it is actually singular: the
explicit form of the warp factor $H=H(\rho,r)$ shows that the metric
has a {\it naked} singularity for some $r = r_s$ (i.e. $H(r_s)=0$).
This singularity should be cured, somehow, for the all story to make
sense.

Let us start from the end. There is a general a mechanism, known as
the {\it enhan\c{c}on} mechanism, which excises the singularity rendering
a singularity-free solution: at a distance $\rho=\rho_E$, the 
{\it enhan\c{c}on}, new light
degrees of freedom not accomplished by the supergravity approximation
come into play and modify the background.  This means that the
physical description of the system at distances $\rho < \rho_E$ should
include these states which change the form of the solution in the
interior. The important point is that the singularity is cloaked
behind $\rho_E$ and is therefore excised from the solution.  This
means that the singularity we have
found was a fake one. No matter what the ``true'' solution looks like,
the one we have found has a meaning as a low  energy description of
our system of fractional branes only at distances bigger than the
enhan\c{c}on, and so the singularity was never really there, after all.

A number of questions naturally arise at this point. The first one is
of course: what is the enhan\c{c}on, in more quantitative terms? This
is simple to answer. The enhan\c{c}on is the locus where the
$B_{(2)}$-flux vanishes, namely where the fluctuations of the $b$
fields cancels its background value
\begin{equation}
\int B_{(2)} = b \,\sim \, 4 \pi^2 \alpha' \, \frac{1}{2} \;+\; \tilde
b \,|_{\rho = \rho_E} \equiv 0
\end{equation}
Given the above definition, we know what is happening at the
enhan\c{c}on: fractional brane probes become tensionless as  their
world volume action (recall what we learned in section 2.2)  is
proportional to the $B_{(2)}$-flux. To be more precise, by probing the
geometry generated by our stack of fractional D3-branes by a
fractional Dq-brane probe one finds that the probe tension is
$\rho$-dependent and vanishes at $\rho = \rho_E$.

Recalling the discussion after eq.(\ref{rung23}), the enhan\c{c}on is 
then just the scale where the dual  gauge theory becomes strongly
coupled since
\begin{equation}
\frac{1}{g_{\rm YM}^2} \sim \int B_{(2)}
\end{equation}
The final lesson is then that in order to incorporate the
non-perturbative contributions in the gauge theory we should go beyond
the supergravity approximation. This looks quite unfortunate but we
could already have guessed it from eq.~(\ref{gyn2}). It is well known
that the metric of the moduli  space of ${\cal N}=2$ SYM, which is
proportional to the inverse gauge coupling squared, is positive
definite. This means the gauge coupling can never diverge in the exact
theory, in disagreement with the perturbative prediction (\ref{gyn2}).

Now that we have learned what the enhan\c{c}on is, the second question
we should try to answer is: what are these new light degrees of
freedom? This depends on the supergravity solution (or equivalently
the dual gauge theory) one is considering. Again, let us start from
the end. The bound state of fractional Dp-branes generating the
background (let us be general here, and let $p$ be also different from
3) couples to
\begin{eqnarray}
\mbox{NSNS}: \; G_{MN}\;,\;\Phi\;,\;b \quad ; \quad \mbox{RR}:  \;
C_{(p+1)}\;,\;A_{(p+1)} \quad (\leftrightarrow \;A_{(3-p)}) \nonumber
\end{eqnarray}
where the solution is written in general in terms of the magnetic dual
of the RR twisted potential $A_{(p+1)}$, i.e. $A_{(3-p)}$, to which
the branes couple magnetically. The light degrees of freedom
becoming relevant at the enhan\c{c}on come from fractional D-branes
carrying an {\it electric} charge with respect to $A_{(3-p)}$.

As this could seem just an abstract recipe, let us see how it works in
our example. In the fractional D3-brane solution we have discussed,
the r\^ole of  $A_{(p+1)}$ is played by $A_{(4)}$ while that of
$A_{(3-p)}$ by the RR scalar $c = \int C_{(2)}$. The corresponding
electric state is then a fractional D(-1)-brane whose world volume
action looks like
\begin{equation}
{\cal S}_{D(-1)} =  \frac{T_{-1}}{\kappa} \left[ \left( C_{(0)} +
e^{-\phi}\right) \int B_{(2)} + {\rm i} \int C_{(2)}  \right]
\end{equation}
By evaluating it in the background of our solution we get
\begin{equation}
\label{insgr}
{\cal S}_{D(-1)} =  \frac{1}{2 \pi \alpha' g_s} \left(\int B_{(2)} +
{\rm i} \int C_{(2)} \right) = - \, \frac{{\rm i}}{2 \pi \alpha' g_s}
\gamma = - \, 2 \pi {\rm i} \, \tau
\end{equation}
where in the last step we have used the holographic identification
between $\gamma$ and $\tau$.

From the ${\cal N}=2$ point of view we expect instantons to become
relevant at strong coupling and de-singularize the moduli space. The
instanton contribution enters the partition function with an action as
\begin{equation}
{\cal S}_{ins} = \frac{8 \pi^2}{g_{\rm YM}^2} - {\rm i}\, \theta_{\rm
YM} \equiv - 2 \pi {\rm i} \, \tau
\end{equation}
in perfect agreement with equation~(\ref{insgr}). Hence, we are on the
right path. Can we do better? In order to completely solve the theory
one should include these extra states (the fractional D(-1)-branes) in
the analysis and look for a solution which is asymptotically identical
to the one we have found but changing sensibly at short
distances. This should let one reproduce the full moduli space of
${\cal N}=2$ from the supergravity (or better the string)
dual. Unfortunately, this goal has  not been achieved, yet.

As we just said, the enhan\c{c}on phenomenon is not specific to the
D3-brane case, but has a much wider validity. Let us consider for
instance a gauge/gravity correspondence testing SYM with 8
supercharges in 3 space-time dimensions (in fact, this has been the
first case where the enhan\c{c}on phenomenon was observed). The 
fractional branes
one has to consider in this case are the fractional D2-branes of type IIA
string theory. The all story goes on pretty much the same. The branes
couple to the graviton, the dilaton (this is non-trivial now) and the
RR 3-form potential $C_{(3)}$ in the untwisted sector, and to the
scalar field $b$ and the RR 3-form potential $A_{(3)}$ in the twisted
sector (we take for simplicity the orbifold $\IC_2/\ZZ_2$ so to
have just one twisted sector).  The solution is given in terms of the
magnetic dual of $A_{(3)}$, i.e. the 1-from potential $A_1$, to which
the D2-branes couple magnetically. Following the recipe outlined
above, we expect the degrees of freedom associated to fractional
D0-branes to become relevant at the enhan\c{c}on. The world volume
action of fractional D0-branes is
\begin{equation}
{\cal S}_{D0} \sim \mp \frac{T_0}{\kappa} \int d^3 \xi \sqrt{- {\rm
det}\, G}  \; e^{-\,\frac{1}{4}\,\Phi} \; \int_{{\cal C}_1} B_{(2)} \,
\pm \, \frac{T_0}{\kappa}  \left[ \int C_{(1)} \int_{{\cal C}_1}
B_{(2)} +  \int A_{(1)} \right]
\end{equation}
and at the enhan\c{c}on they become tensionless. The $\pm$ sign is
related to which type of fractional brane we are speaking about (they
both contribute). The r\^ole of these states is similar to that of
$W^\pm$ boson as they couple electrically to the "gauge" field
$A_{(1)}$, enhancing the $U(1)$ symmetry to $SU(2)$ at the
enhan\c{c}on, where they become massless (hence the name enhan\c{c}on
for this phenomenon, which however applies, as we have seen, also to
other situations)
\begin{equation}
\label{encd2}
A_{(1)} \longrightarrow   \begin{pmatrix} A^+_{(1)} \\ A_{(1)} \\
A^-_{(1)}
\end{pmatrix}   \qquad \quad U(1) \longrightarrow SU(2)
\end{equation}
The upshot is that in the interior, namely at distances smaller than
the enhan\c{c}on radius, the solution should be described by an
$N$-charge $SU(2)$ monopole. On the contrary, at large distances the
deformations due to these extra states is negligible and the solution
is the same as the singular one. It turns out this perfectly agrees
with gauge theory expectations. Indeed it is known that the moduli
space of  $SU(2)$ monopole is smooth and is equivalent to that of
$2+1$ SYM with 8 supercharges, automatically including the instanton
corrections. Again, the supergravity/string background gives the
correct prediction. We will come back to this issue in the last
lecture, when discussing dualities, as we will have a different (but
dual-equivalent) description of the enhan\c{c}on phenomenon.

Before ending this section a last comment is in order. There is in
fact another interesting (and related) feature of our solution,
eqs.~(\ref{metsol})-(\ref{tssol}). If one computes the flux of the
5-form ${\widetilde{F}}_{(5)}$ along a closed surface intersecting the
$z$-plane on some curve $\Sigma$ one gets
\begin{equation}
\label{f5}
\Phi({\widetilde{F}}_{(5)}) = 4 \pi^2 g_s N \left( \frac{1}{2} +
\frac{N g_s}{\pi} \log \frac{\rho}{\rho_0}\right)
\end{equation}
Differently from what happens in the original AdS/CFT correspondence
where this integral gives just $N$ (the number of constituent branes
which represents the number of colors of the dual gauge theory), we get
here a scale dependent result! Well, this is correct. In a
non-conformal theory the number of effective degrees of freedom
decreases through the infrared and the above formula predicts that
(recall the relation between $\rho$ and $\mu$). This is just
qualitative agreement and one would like to have a more quantitative
one. To start with, note that the value of the flux goes to zero at
the enhan\c{c}on (this is due to the fact that the
${\widetilde{F}}_{(5)}$-flux turns out to be proportional to the
$B_{(2)}$-flux). What is the meaning of all that? There has been some
debate in the literature about the precise field theory interpretation
of this phenomenon and its relation with the enhan\c{c}on.

The safer thing we can say, so far, is as follows. One can think to
build-up the configuration giving raise to the supergravity solution
we have been discussing as a step-by-step process. Starting from a
single (fractional) D3-brane positioned at the origin, one can take a
second one far at infinity and move it toward the origin (this can be
done at no cost as the system is BPS), then a third one and so on. In
doing so, for the reasons we have been just discussing, one sees it is
not possible to build up a source made of fractional D-branes located
at the origin, $\rho =0$: at the enhan\c{c}on, whose radius is
proportional to the number of constituent branes, the branes one is
taking from infinity becomes tensionless. It evaporates, in a
sense. Moreover, one can see that the Newton-like force experienced by
a brane  probe in the supergravity background under consideration  ($F
\sim - \nabla G_{00} = \nabla H^{-1/2}$) becomes repulsive at short
distances, thus indicating that the branes tends to repel each other,
gravitationally. The final picture is then that the branes making-up
the configuration are not at the origin but rather expand to form a
ring-like shell (the enhan\c{c}on in fact) around a region which
appears singular in the original supergravity solution. The
enhan\c{c}on is just the distance where the tension of the branes
drops to zero and that is the last radius where there is a meaning to
the constituent branes as localized sources. In this way it is clear
that, while  the exterior solution, due to Gauss' theorem, is of
course unchanged (and it is nothing but the supergravity solution we
have found), the interior one could look completely different, and is
actually flat at leading order in $1/N$, as there is no point-like
source in the inside. This is the classical stable solution one should
start with, and which reproduces (is dual to) the perturbative part of
the gauge theory moduli space. To take into account non-perturbative
corrections, one should include the already mentioned new light
degrees of freedom which become relevant at the enhan\c{c}on. These
would take care of higher order corrections in $1/N$ and would modify
the interior region. This is what string theory tell us, at large $N$.
This picture has been perfectly confirmed by a dual field theory
analysis: studying the gauge theory side of the duality by means  of
the corresponding Seiberg-Witten curve, it has been shown the results
agree quantitatively with the above picture at large $N$. Indeed, for
large $N$, instanton corrections are  implemented by assuming there is
no running of the gauge coupling below the scale $\Lambda$ and this is
automatically taken into account by imaging the fractional branes
being smeared at the enhan\c{c}on.

Still, it has not been possible to find the complete answer including
higher order corrections. This would imply one was able to ``resolve''
the enhan\c{c}on by really including the new light degrees of freedom
in the low energy closed string analysis. But, as already stressed,
this has not been done, yet.

%\newpage
%%%%%%%%%%%%%%%%%%%%%%%%%%%%%%%%%%%%%%%%%%%%%%%%%%%%%%%%%%%%%%%%%%
%%%%%%%%%%%%%%%%%%%%%%%%%%%%%%%%%%%%%%%%%%%%%%%%%%%%%%%%%%%%%%%%%%
\section{Lecture III - Branes wrapped on Calabi-Yau spaces}

Another possibility to reduce    the number of supersymmetries  is  to
consider D-branes whose  world    volume is (partially)     wrapped on
topologically   non-trivial   supersymmetric   cycles   of  a  CY
manifold.   This  is the   natural  counterpart  in  smooth  spaces of
fractional   branes on orbifolds   (in fact,  these configurations are
related by  T-duality,  as we will see   later). The logic behind  the
construction  of  the  gauge  theory  out   of  these D-branes  can be
summarized as follows. Consider a Dp-brane wrapped on a q-cycle inside
a CY space (when we say CY space we really mean a {\it non-compact} CY
space; this will always be understood in the following).
\begin{itemize}
\item{The unwrapped part of the brane world volume remains flat and
supports a (p+1-q)-dimensional effective theory.}
\item{In order to preserve some supersymmetry the q-cycle normal 
bundle has to be partially {\it twisted}. As we shall see, this
implies that some (would be massless) fields become massive and
decouple, at low-energy.}
\item{Taking a low energy limit where both the string states and the
KK excitations on the q-cycle decouple, one ends up with a
supersymmetric gauge theory in (p+1-q)-dimensions. The amount of
preserved supersymmetry depends on the way the q-cycle is embedded in
the ambient space, a CY three-fold or a CY two-fold.}
\end{itemize}
This is how one engineers the gauge theory. The general idea is that 
the supergravity solution generated
by the bound state of D-branes one is considering should be dual, in
some limit, to the (p+1-q)-dimensional supersymmetric gauge theory
living  on them.

This procedure was first used by Maldacena and Nu\`nez (MN) to study
pure ${\cal N}=1$ SYM in four dimensions and later generalized to
other cases with different space-time dimensions and/or amount of
preserved supersymmetry. We will focus in what follows on the case
discussed by MN as we want to study the basic case of pure ${\cal
N}=1$ SYM. As it was the case for fractional branes, the logic
procedure and technical tools we will develop should enable the reader
to address any other case.

The MN setting corresponds to consider $N$ D5-branes wrapped  on a
supersymmetric two-cycle, which is topologically a two-sphere,
inside a CY three-fold. A CY three-fold preserves 1/4 supersymmetries
and  the D-branes break 1/2 more so the gauge theory living on the
D-branes preserves 4 supercharges. We have in this case p=5 and q=2
so, according to the discussion above, we end up with a
four-dimensional ${\cal N}=1$ gauge theory with gauge group $SU(N)$,
at low energy. As we are going to show, it turns out there is no
matter coupled to it so the theory at hand is described by vector
multiplets only, i.e. is pure. Once the supergravity solution
generated by this bound state of D-branes is derived, one can exploit
it and study the perturbative and the non-perturbative properties of
${\cal N}=1$ SYM theory.

%%%%%%%%%%%%%%%%%%%%%%%%%%%%%%%%%%%%%%%%%%%%%%%%%%%%%%%%%%%%%%%%%%%
\subsection{Wrapping Branes: The Topological Twist}

The first thing we have to understand is to what extent we can wrap a
D-brane on a topologically non-trivial cycle of a CY space and at the
same time preserve supersymmetry. In fact, supersymmetry gets all
broken, in general. This is a non-trivial geometric problem which
would require a careful treatment. In what follows we are just going
to give a rough idea on how things work, this being sufficient for our
purpose. In order to preserve some  supersymmetry on the D-brane world
volume one should solve an equation like
\begin{equation}
\label{cov1}
D_M \;\epsilon = \left( \partial_M + \omega_M \right)\; \epsilon = 0
 \qquad  \quad M=0,1,...,p
\end{equation}
where $\omega_M$ is a short cut for $\omega_M^{NP} \gamma_{NP}$,
$\omega_M^{NP}$ being the spin connection on the q-cycle under
consideration, and $\epsilon$ is a killing spinor. This equation does
not admit any solution as in general there are no covariantly constant
spinors on a topologically non-trivial cycle.

Here it is where the twist comes about. If the theory has some global
R-symmetry group, we can couple the theory to an external ``gauge''
field $A_M$ that couples to the R-symmetry current. In doing so,
eq.~(\ref{cov1}) becomes
\begin{equation}
\label{cov2}
\left( \partial_M + \omega_M - A_M \right) \epsilon = 0
\end{equation}
This is not done by hand as it could seem from this discussion, and
has a simple geometrical explanation. The coupling to the external
field $A_M$ is just what takes into account the fact that the q-cycle
is non-trivially fibered within the CY space: the directions normal to
the q-cycle form a non-trivial bundle, the so-called normal bundle,
and $A_M$ is nothing but the connection on this normal bundle.

Upon the identification $\omega_M = A_M$ one can easily preserve
supersymmetry as now eq.~(\ref{cov2}) becomes
\begin{equation}
\label{cov3}
\partial_M \; \epsilon = 0
\end{equation}
and this amounts to have just a constant spinor, which we certainly
have. This is of course not the only way to solve eq.~(\ref{cov2}),
but it turns out it is the way things work when dealing with D-branes.

The  above  operation has a  striking  effect: the Lorentz assignment
(i.e. the spin) of the various fields gets  changed (for instance, as
it  can be seen from eq.~(\ref{cov3}), the supersymmetry parameter
becomes a  scalar!). This is why  we say the  resulting theory is {\it
twisted}. The crucial point now is that although the theory on the
(p+1)-dimensional world volume is indeed twisted, that on the  flat
(p+1-q)-dimensional part, which is the one we are finally interested
in, is not.  That is just an ordinary field theory.

Let us see how this works in the case we want to study, a bunch of
D5-branes wrapped on a two-sphere inside a CY space. Let us start
considering a D5-brane in flat space. The presence of the brane breaks
the ten-dimensional Lorentz group as follows
\begin{equation}
D5 \quad : \quad SO(1,9) \longrightarrow  SO(1,5) \times SO(4)
\end{equation}
where SO(4) is nothing but the R-symmetry group of the six-dimensional
gauge theory living on the world volume of the D5-brane. If
considering a  D5-brane wrapped on an $S^2$ we have instead
\begin{equation}
\label{d5s2}
D5_{S^2} \quad : \quad SO(1,9) \longrightarrow  SO(1,3) \times SO(2)
\times SO(4)
\end{equation}
where $SO(2) \simeq U(1)_J$ is the tangent bundle of the
two-sphere. The twist is introduced by identifying $U(1)_J$ with some
$U(1) \subset SO(4) \simeq SU(2)_L \times SU(2)_R$. Let us see what
does this imply for the world volume fields living on the
D5-brane. First we have to see how the fields transform with respect
to the broken ten-dimensional Lorentz group $SO(1,3) \times U(1)_J
\times SU(2)_L \times SU(2)_R$. Then we impose the twist, i.e. the
identification between $U(1)_J$ and a $U(1)\subset SU(2)_L  \times
SU(2)_R$. This is just an operative procedure since we are actually
breaking supersymmetry  on the way, when we perform the twist. This is
done for pedagogical  purpose, but one should bare in mind that the
actual situation for a  wrapped brane world volume theory is directly
what we will get in the  end, of course.

The (flat) D5-brane field content amounts to a vector $V_M$
($M=0,1,...,5$),  four scalars $\phi^A$ ($A=1,...,4$), and two complex
Weyl spinors with opposite chirality $\psi^{\pm}$. Playing  a  bit
with representation theory we easily get
\begin{table} [ht] 
\begin{center}
\begin{tabular}{c|c|c}
& $SO(1,5) \times SO(4)$ & $SO(1,3) \times U(1)_J \times SU(2)_L
\times SU(2)_R$ \\ \hline && \\ $V_M$ & $(6,1)$ &  $(4_0,1,1) \oplus
(1_\pm,1,1)$ \\&& \\$\phi^A$  & $(1,4)$  & $ (1_0,2,2)$ \\
&&\\$\psi^{+}$ & $(4,2)$ & $ (2_+,2,1) \oplus  (\bar 2_-,2,1)$ \\ &&
\\$\psi^{-}$ & $(4',2')$ & $ (\bar 2_+,1,2) \oplus (2_-,1,2)$
\end{tabular}
\end{center}
\end{table}

where the subscript $0,\pm$ in the second column represents the
$U(1)_J$  charge. This is the field content in six dimensions as seen
from a four-dimensional point of view. We have a vector, six scalars
and four fermions. So far this is just representation theory, the
supersymmetry is all there (16 supercharges).

Let us now implement the twist. This can be done in different ways and
the number of preserved supersymmetries on the D-brane world volume
changes accordingly. To our purpose we have two inequivalent choices.

\vskip 7pt  $\;\bullet\;$ $\,U(1)_J \equiv U(1)_D \subset SU(2)_D
\equiv D\left( SU(2)_R \times SU(2)_L\right )$

Let us rewrite the table above taking into account the identification
between $U(1)_J$ and $U(1)_D$ and let us call the corresponding group
$U(1)$. We have to consider the charge of the various fields with
respect to this $U(1)$ since the massless fields in four dimensions
are those fields which are singlets under $U(1)$ (this can be
understood as a charge under the overall $U(1)$ acts as a mass term
from the four-dimensional point of view). Moreover, recall we are at
energies such that both string states as well as KK modes on the $S^2$
are decoupled so we should care only of the zero modes in the
expansion of six-dimensional massless fields into $S^2$ harmonics. 
Implementing the twist we easily get the following
decomposition
\begin{table} [ht] 
\begin{center}
\begin{tabular}{c|c}
& $SO(1,3) \times U(1)$ \\ \hline \\ $V_M$ & $4_{0} \oplus 1_{\pm}$ \\
\\$\phi^A$  &  $ 2 \times 1_{0} \oplus  1_{+} \oplus 1_{-}$ \\
\\$\psi^{\pm}$ & $\begin{matrix} 2\times 2_{0}\oplus 2\times 2_{++} \\
2 \times \bar 2_{0} \oplus 2 \times \bar 2_{--}\end{matrix}$
\end{tabular}
\end{center}
\end{table}

From the above table we see that the {\it massless} field content in
four dimensions amounts to a vector $A_\mu \simeq 4_0$, two scalars
$\phi^a \simeq 2 \times 1_0$  with $a=1,2$, and two Majorana spinors,
$\psi^a \simeq 2 \times \left( 2_0 \,\oplus\,\bar 2_0\right)$ (one
could equivalently arrange the fermionic degrees of freedom into two
Weyl spinors of opposite chirality). This is nothing but the field
content of a ${\cal  N}=2$ vector multiplet in four dimensions. Hence
the twist we have been considering leaves a theory with 8 supercharges
only (of which we consider just those supermultiplets which appear
massless from the four dimensional point of view). Note that all these
four dimensional fields have ordinary statistic,  as anticipated: the
four-dimensional theory living on the flat part of the world volume of
the D5-brane is not twisted. The supergravity solution generated by a
bound state of $N$ such D5-branes is thus relevant to study pure
${\cal N}=2$ SYM theory with gauge group $SU(N)$ in four dimensions.

One can show that the geometric counterpart of this operation amounts
to make the $S^2$ non-trivially fibered on a 2-dimensional
sub-manifold of the overall transverse space. In other words, the
geometry of the CY three-fold is $K3 \otimes \IR^2$. This is a
degenerate CY manifold as the number of preserved supersymmetries is
1/2 rather than 1/4 as for an ordinary CY with $SU(3)$ holonomy (the
holonomy in this case is just $SU(2)$, in fact). As the brane breaks
further 1/2 supersymmetry, one would expect a world volume theory with
8 preserved supercharges, i.e. ${\cal N}=2$ in four dimensions,
consistently with what we have just found.

\vskip 7pt  $\;\bullet\;$ $\,U(1)_J \equiv U(1)_L \subset SU(2)_L$

This choice is of course equivalent with the identification of
$\,U(1)_J$ with $U(1)_R \subset SU(2)_R$ (there is a left-right
symmetry here). We should proceed right in
the  same way as we did before. Decomposing the six-dimensional fields
with respect to the twisted Lorentz group we have in this case
\begin{table} [ht] 
\begin{center}
\begin{tabular}{c|c}
& $SO(1,3) \times U(1)$ \\ \hline  \\ $V_M$ & $4_{0} \oplus 1_{\pm}$
\\  \\$\phi^A$  &  $ 2 \times 1_{+} \oplus 2 \times 1_{-}$ \\
\\$\psi^{\pm}$ & $\begin{matrix} 2_{0}\oplus  2_{++} \oplus 2\times
2_{-} \\ \bar 2_{0} \oplus \bar 2_{--} \oplus 2 \times \bar 2_{+}
\end{matrix}$
\end{tabular}
\end{center}
\end{table}

We see that the {\it massless} field content in four dimensions
amounts now to a vector $A_\mu \simeq 4_0$ and a Majorana spinor,
$\psi \simeq  2_0 \, \oplus \,\bar 2_0$.  This is nothing but the
field content of a ${\cal  N}=1$ vector multiplet in four dimensions.
Hence the twist leaves now a theory with 4 supercharges. As in the
previous case, all the fields we are considering in four dimensions
are ordinary ones, i.e. the four-dimensional field theory is not
twisted. The supergravity solution generated by a bound state of $N$
such D5-branes is thus relevant to study pure ${\cal N}=1$ SYM theory
with gauge group $SU(N)$ in four dimensions.

Also in this case there is a geometric counterpart of all that. The
$S^2$ turns out to be non-trivially fibered on the full transverse
space. In other words, the target space geometry is now that of an
ordinary CY three-fold with $SU(3)$ holonomy which preserves 1/4
supersymmetry. As the brane breaks further 1/2 supersymmetry, one
would expect a world volume theory with 4 preserved supercharges,
i.e. ${\cal N}=1$ SYM in four dimensions, consistently with what we
have just found.

This is the case we are going to deal with in the reminder of this
lecture. Our task will then be to find the supergravity solution
describing a bound state of $N$ D5-branes wrapped on a two-sphere
inside a CY three-fold. As just seen, this is the configuration  to
exploit in order to study pure ${\cal N}=1$ SYM in four dimensions.

All what we have been doing so far can of course be extended to other
cases. Considering for instance D6-branes wrapped on non-trivial
three-cycles of a CY manifold, one would again obtain ${\cal N}=1$ SYM
in four dimensions. Alternatively, one could consider D5-branes
wrapped on three-cycles or D4-branes wrapped on two-cycles, getting in
this case three-dimensional supersymmetric gauge theories at low
energy. In the latter case, if the CY space has $SU(2)$ holonomy, one
would get three-dimensional SYM with 8 supercharges.  The logic in
getting the spectrum, starting from the flat brane case  and then
performing the twist, goes on pretty much the same.

\bigskip
\noindent{\bf Exercise 4 -} {\it Consider ${\cal N}=2$ SYM theory in 
four dimensions. This theory admits an internal symmetry $SU(2)_I$
while the Lorentz group is $L = SU(2)_L \times SU(2)_R$. Define a  new
Lorentz group $L' =  SU(2)_L \times SU(2)'_R$ where $SU(2)'_R \equiv D
\left(SU(2)_R \times SU(2)_I \right)$ (this is a twist) and  compute
how the field statistic (i.e. the spin) changes.}  Hint: recall that the
(original) field content consists of a gauge field $A_\mu$, two 
Majorana spinors $\psi^{1,2}$ and two real scalars $\phi^{1,2}$.   
%\bigskip

%%%%%%%%%%%%%%%%%%%%%%%%%%%%%%%%%%%%%%%%%%%%%%%%%%%%%%%%%%%%%%
\subsection{Wrapping Branes: The R\^ole Of Gauged Supergravity}

To find the supergravity solution generated by a bound state of
wrapped Dp-branes is not in general an easy task. In principle, one
should just come out with the right ansatz in ten dimensions and
solve the relevant equations of motion. But this turns out to be hard
to do, as compared to the flat brane case.

As originally proposed by MN, an efficient way to circumvent these
difficulties is to find a (wrapped) domain wall solution in the lower
dimensional (p+2)-dimensional gauged supergravity and then lift the
solution up in ten dimensions on a (8-p)-sphere. In order to
understand how and why this works, let us consider flat branes, to
start with. The near horizon geometry of a bound state of D3-branes in
flat space is ${\rm AdS}_5 \times {\rm S}^5$. This is the geometry
which is expected to be dual to four-dimensional ${\cal N}=4$ SYM,
the theory describing D3-brane dynamics, at low
energy. This ten-dimensional geometry can be thought of as a domain
wall solution of five-dimensional gauged supergravity, which is then
up-lifted to ten dimensions on a five-sphere. This is a simple case
though, since in the solution none of the gauge fields of the
five-dimensional gauged supergravity are switched-on, nor any scalar.
Notice that the isometry group  of $S^5$ is $SO(6)$, this being the
R-symmetry group of ${\cal N}=4$ SYM theory, the theory living on  the
D3-branes world volume, as well as the group one could gauge, in fact,
in the five-dimensional effective supergravity theory.

Let us now consider the case of (still flat) Dp-branes with  p$\not =
3$. In this case one can show there always exists a frame, the
so-called dual frame, where the near horizon geometry looks (locally)
like ${\rm AdS}_{p+2} \times {\rm S}^{8-p}$. As it is the case for the
D3-branes, this solution can be thought of as a domain wall in the
(p+2)-dimensional supergravity theory up-lifted to ten dimensions on a
(8-p)-sphere (in fact this is consistent in ten dimensions only for
p=2,3,5,7, but this is enough for us). In this case, however, we do not
really have an ${\rm AdS}$-like geometry, since  the metric is warped
now, i.e. there is an $r$-dependent function  multiplying it: the
space is only conformally equivalent to ${\rm AdS}$. This makes the
presence of some non-trivial scalar necessary for the consistency of
the solution, while also in this case all gauge fields of the
(p+2)-dimensional effective theory are trivial in the solution. The
isometry group of $S^{8-p}$ is now $SO(9-p)$,  this being the
R-symmetry group of the corresponding maximally supersymmetric gauge
theory living on the Dp-branes world volume, as well as the group one
could gauge in the lower dimensional supergravity theory.

What we have learnt in the previous section is that in order to
describe a curved Dp-brane we should have some non-trivial external
field $A$ coupled to the world volume theory. This field, which is the
one responsible for the twist, gauges the corresponding generator of
the R-symmetry group. It is clear then that if we follow the same
logical  procedure as before, we should just look now for a slightly
more involved (p+2)-dimensional domain wall solution where the
corresponding gauge field $A$ of the (p+2)-dimensional gauged
supergravity has a non-trivial profile. Note that once the coordinates
parameterizing the (p+2)-dimensional space are chosen, the
identification  between the spin connection of the q-cycle and the
gauge field of the  R-symmetry group also suggests what the suitable
ansatz for this gauge field should be.

The lesson is then that what is twisting in the world volume theory
(open string side) translates into having some non-trivial gauge
fields in the (p+2)-dimensional supergravity solution (closed string
side). These gauge fields are precisely those fields gauging  the
R-symmetry group of the world volume theory of the branes. This is the
guiding principle one should follow in searching for any kind of
supergravity solution describing (the near-horizon geometry of) 
wrapped branes.

Let us see how all this works in the case we are interested in,
i.e. that of $N$ D5-branes wrapped on a two-cycle inside a CY
three-fold. We have
in this case a domain wall solution of seven-dimensional gauged
supergravity which should be up-lifted to ten dimensions on a
three-sphere. The isometry group of the three-sphere is $SO(4)$, this
being the R-symmetry group of  the six-dimensional gauge theory living
on the D5-brane. What we have learnt  in the previous section is that
in order to describe such brane we should  have some external field
gauging a $U(1)  \subset SO(4)$. This has its natural seven-dimensional
gauged supergravity counterpart: this same gauge field is just the one
we should make a non-trivial ansatz for, in searching for the
seven-dimensional domain wall solution corresponding to our wrapped
brane. To say it in other words, a non-trivial gauge field $A \subset
SU(2)_D$ or $A \subset SU(2)_L$ should  be present in the
seven-dimensional supergravity ansatz in order to find a solution
corresponding  to a D5-brane wrapped on an $S^2$ inside a K3 or a CY
manifold and being dual to pure ${\cal N}=2$ and ${\cal N}=1$ SYM,
respectively.

As the ${\cal N}=2$ case has already been discussed in the previous
lecture, in the following we will concentrate on ${\cal N}=1$. We will
derive the (near-horizon limit of the) supergravity solution describing 
$N$ D5-branes wrapped on an $S^2$ inside a CY three-fold and exploit 
it to learn something about pure ${\cal N}=1$ SYM.

%%%%%%%%%%%%%%%%%%%%%%%%%%%%%%%%%%%%%%%%%%%%%%%%%%%%%%%%%%%%%%%%%%%%%%
\subsection{Gauge Theory From Gravity: Example 2}

Let us start by briefly mentioning the structure of the domain-wall
solution of the seven-dimensional $SO(4)$ gauged supergravity theory
we should start with. The fields present in this theory are the
metric, six gauge fields transforming in the adjoint of $SO(4)$, ten
scalar fields organized in a symmetric matrix $T_{ij}$ ($i,j$ being
vector indexes of $SO(4)$), and a two-form potential. In order to
obtain the solution we are looking for, we must first truncate the
$SO(4)$ gauge group to its $SU(2)_L$ part and then identify a
$U(1)\subset SU(2)_L$ with the spin connection, as dictated by the
${\cal N}=1$ twist discussed in section 3.1. One can show this
truncation requires the scalar matrix $T_{ij}$ be proportional to a
$\delta$-function, while the two-form can be consistently put to
zero. By choosing the seven-dimensional coordinates as
$x_0,\ldots,x_3,\rho,\theta_1,\phi_1$, where the angles $\theta_1$ and
$\phi_1$ parameterize the two-sphere, the solution looks like
\begin{eqnarray} 
\label{mn7d} 
ds^2 &\sim& f(r) \left( dx^2_{1,3} + dr^2 \right) +
\frac{1}{\lambda^2} \, g(r) \left( d\theta_1^2 + \sin^2  \theta_1
d\phi_1^2\right) \\
\label{mns7d}
T_{ij} & \sim& e^{y(r)} \,\delta_{ij} \\  A &\sim & \cos \theta_1 \,
d\phi_1
\end{eqnarray} 
where $\lambda$ is some dimension-full parameter and $f(r)$, $g(r)$
and $y(r)$ are functions of the radial coordinate $r$ (we have not
being precise with numerical coefficients as this is not important for
our purposes, for the time being). Note how the actual form of $A$
makes explicit its identification with the $S^2$ spin  connection.

It turns out that this solution is singular (as it was the case for 
fractional branes), but remarkably the singularity, which is at $r=0$, 
can be easily removed, this time! This can be done starting from a 
more general ansatz where all the three one-forms belonging to $SU(2)_L$ 
are switched-on
\begin{equation}
\label{ga7d}
A^1 \sim a(r) \,d \theta_1 \quad , \quad A^2 \sim a(r)  \sin \theta_1
\, d \phi_1 \quad , \quad  A^3 \sim \cos \theta_1\, d \phi_1
\end{equation}
where $a(r)$ is again a function of the radial coordinate $r$.
Inserting the ansatz (\ref{mn7d}),(\ref{mns7d}) and (\ref{ga7d})  in
the effective seven-dimensional supergravity lagrangian one can
determine the functions $f(r) \,, \,g(r)$ and $a(r)$ and verify the
new solution is indeed free of singularity.

As we are not going to discuss the ${\cal N}=2$ case explicitly, it is
probably worth making a comment at this point. If repeating the above
reasoning for the ${\cal N}=2$ case, considering now a domain wall
solution with a $U(1)$ field belonging to $SU(2)_D$ (see section 3.1),
one ends up again with a singular solution. However, differently from
the present case, it turns out there is no way to smooth out the
singularity. The solution is singular in seven dimensions and remains
so also after the up-lifting to ten dimensions. The nature of the
singularity is the same as the one discussed in the previous lecture,
namely is a repulson-like singularity, and can be cured. This goes
along the same lines as for the fractional brane case, i.e. the
enhan\c{c}on phenomenon is at work here, too. There is a distance
where new light degrees of freedom come into play and change the low
energy effective theory, and correspondingly the form of the solution,
at short distances. How this can be done, we all know by now, but
explicit results have not been really achieved, yet. This same thing
happens for all situations in which one is studying supergravity duals
of supersymmetric gauge theories with eight supercharges. No matter
the theory under consideration is pure or with matter, no matter if it
is in four, three or any number of dimensions. The enhan\c{c}on
phenomenon and its gauge theory meaning is a common feature of
non-conformal supersymmetric gauge theories with eight supercharges
and does not depend on the way (fractional branes, wrapped branes or
anything else) one is using to study the supergravity dual. In fact,
the r\^ole of the singularity looks rather different between the case
of four supercharges and that of eight. In both cases the singularity
can be cured. However, in the first case this can be done within the
supergravity framework, in the second case one needs to include string
states. This difference resides in the different properties of the
corresponding gauge theories, of course.

Let us now come back to our solution.  The non-singular
seven-dimensional solution (\ref{mn7d})-(\ref{ga7d}) has to be lifted
up in ten dimensions on a three-sphere which we parameterize with
coordinates $(\psi, \theta_2, \phi_2)$. An important point to notice
is that the effect of the twist is to make the seven-dimensional
geometry   non-trivially embedded in the ten-dimensional one. This
means that there is a non-trivial mixing between the three-sphere used
for the up-lift and the two-sphere of the seven-dimensional
solution. This is related to the two-sphere being non-trivially
fibered in the CY space. As we shall see, this fact has important
consequences.

By following the inverse path which led to the consistent Kaluza-Klein
truncation of ten-dimensional supergravity theory down to the $SO(4)$
gauged seven-dimensional supergravity, one can easily obtain the
ten-dimensional solution from the seven-dimensional one (there are
simple up-lift formul\ae$\,$ one can use). The ten-dimensional
solution is expected to be described by a metric, a dilaton $\Phi$,
and a RR three-form magnetic flux  $F_{(3)}$ (this is because
D5-branes couple magnetically to the corresponding RR two-form
potential $C_{(2)}$). This is exactly what one gets  performing the
up-lift. The final result (we are now precise with all numerical
factors) turns out to be, in the string frame
\begin{eqnarray} 
\label{mnsol} ds^2 &=& e^\Phi dx^2_{1,3} + e^{\Phi} \alpha' g_s N 
\left[ e^{2h} \left( d\theta_1^2 + \sin^2 \theta_1 \,d\phi_1^2 \right)
+ d\rho^2 + \sum_{a=1}^3\left(\omega^a - A^a \right)^2 \right] \\
\label{mnsol1} e^{2 \Phi} &=& \frac{\sinh 2\rho}{2\, e^{h}} 
\\
\label{mnf3} F_{(3)} &=& 2 \, \alpha' g_s N \, \prod_{a=1}^3 
\left( \omega^a - A^a \right) - \alpha' g_s N \, \sum_{a=1}^3 F^a
\wedge \omega^a
\end{eqnarray} 
where
\begin{eqnarray} 
\label{mnsol2} A^1 &=& - \frac{1}{2} \,a(\rho) \,d\theta_1 \quad , 
\quad A^2 =  \frac{1}{2} \,a(\rho) \,\sin\theta_1 \,d \phi_1 \quad ,
\quad A^3 = - \frac{1}{2} \,\cos \theta_1 \,d\phi_1 \\
\label{mnsol3} e^{2h} &=&  \rho \coth 2 \rho - 
\frac{\rho^2}{\sinh^2 2 \rho} -\frac{1}{4} \quad , \quad a(\rho) =
\frac{2 \rho}{\sinh 2 \rho}
\end{eqnarray} 
and $F^a = \nabla A^a \equiv d A^a + \epsilon^{abc} A^b \wedge A^c$.
The $\omega^a$ are the left-invariant one-forms parameterizing  the
three-sphere
\begin{eqnarray} 
\label{sigmas} 
\omega^1 &=& \frac{1}{2} \left(\cos \psi \,d\theta_2 + \sin \psi
\sin\theta_2 d\phi_2\right) \quad , \quad  \omega^2 = - \frac{1}{2}
\left(\sin \psi \,d\theta_2 - \cos \psi \sin\theta_2 d\phi_2\right)
\nonumber \\ \omega^3 &=& \frac{1}{2} \left(d\psi + \cos\theta_2
d\phi_2 \right)
\end{eqnarray}
and the five angles are defined as $0 \leq \theta_{1,2}\leq \pi$ , $0
\leq \phi_{1,2}\leq 2 \pi$ and $0 \leq \psi \leq 4 \pi$. Finally,
$\rho$ is a dimensionless quantity and is related to the actual radial
distance $r$  as $\rho = r / \sqrt{\alpha' g_s N}$. For later
convenience, let us also report the asymptotic behavior of the
functions $a(\rho)$, $h(\rho)$ and of the dilaton
\begin{eqnarray} 
\label{asym1} 
a(\rho) & \underset{\rho\to\infty}\sim & \rho \; e^{- 2 \rho}
\rightarrow 0 \quad , \quad \;\; a(\rho) \underset{\rho\to 0}\sim \; 1
\\   e^{2 h(\rho)} &\underset{\rho\to\infty}\sim & \rho \qquad \quad
\qquad \, , \quad  e^{2 h(\rho)} \underset{\rho\to 0}\sim \; \rho^2  \\
\label{asymdil} e^{2 \Phi}  &\underset{\rho\to\infty}\sim & \rho^{- 
1/2} \,e^{2 \rho} \quad \quad , \quad  \;\; \; e^{2 \Phi}
\underset{\rho\to 0}\sim \; 1
\end{eqnarray}
as well as the explicit expression for the RR two-form potential
$C_{(2)}$ one can get out of $F_{(3)}$. A straightforward
computation gives
\begin{eqnarray}
C_{(2)} &=& \frac{1}{4}\, \alpha' g_s N \left[ \left( \psi +
\psi_0\right) \left( \sin \theta_2\, d \theta_2 \wedge d \phi_2 - \sin
\theta_1 \, d \theta_1 \wedge d \phi_1\right) + \cos \theta_1 \cos
\theta_2 \,d\phi_1 \wedge d\phi_2 \right] \nonumber \\  &+&
\frac{1}{2} \,\alpha' g_s N a(\rho) \left[ d \theta_1 \wedge \omega^1
- \sin \theta_1 \, d \phi_1 \wedge \omega^2 \right]
\label{mnc2}
\end{eqnarray}
Note that integrating
$F_{(3)}$ the angular variable $\psi$ does not get fixed and is
defined modulo an arbitrary constant, $\psi_0$. However, the specific
value of this constant is not really relevant as the physics sees
$F_{(3)}$ rather than $C_{(2)}$, of course.

A couple of comments are in order at this point.
\begin{itemize}
\item{As it can be seen from eq.~(\ref{ga7d}), the function $a(\rho)$
is responsible for the non-abelian structure of the seven-dimensional
solution and, as already pointed out, this is what makes the solution
smooth and free of singularities. From the asymptotic behavior
(\ref{asym1}) one sees that $a(\rho)$ plays a relevant r\^ole at short
distances only: at large  distances the solution is not sensibly
different from the singular  one, as the function $a(\rho)$ vanishes
and so do $A^{1}$ and $A^{2}$, while it displays its non-abelian
structure at short distances. Another important point to notice is
that  the function  $a(\rho)$ does not preserve the twist. This is not
a  problem though, since the twist should be preserved only at large
distances. Indeed (this will be shown later), this is the  region
corresponding to the UV of the gauge theory, where the perturbative
D5-brane spectrum has been computed and the twist performed. As soon
as  the theory becomes strongly coupled, and eventually confines, it
looks  completely different and the relevant degrees of freedom do not
coincide  with those in the UV. As we shall see, the
function $a(\rho)$ plays a crucial r\^ole in obtaining the precise 
gauge/gravity dictionary.}
\item{We said the solution is smooth, i.e. no singularity at $\rho=0$
is  present. Let us see this. The singularity could arise from the 
transverse part of
the  metric only, as the dilaton smoothly goes to zero for small
$\rho$  (see eq.~(\ref{asymdil})), so no problem there. By computing
the  six-dimensional transverse part of the metric at $\rho=0$ one
easily gets
\begin{eqnarray} 
\label{met0}
&&ds^2_5 \sim \frac{1}{4} \,(\cos\psi\sin\theta_2 d\phi_2 -\sin\psi
d\theta_2 -\sin\theta_1 d\phi_1)^2 + \\  &&+\,\frac{1}{4} \,
(\sin\psi\sin\theta_2 d\phi_2 +\cos\psi d\theta_2 + d\theta_1)^2  +
\frac{1}{4} (d\psi + \cos\theta_1d\phi_1+\cos\theta_2d\phi_2)^2
\nonumber
\end{eqnarray} 
which is finite (in fact, this is topologically a three-sphere with
constant radius, as it will become apparent shortly). Hence there is
indeed  no singularity.}
\item{As we will explicitly show later, what was the r\^ole of the
$B_{(2)}$-flux over the shrinking cycles in the fractional brane case,
is now played by the volume of the cycle the branes are wrapped on. It
is then crucial to identify exactly what the actual cycle is, in the
ten-dimensional geometry (\ref{mnsol}). This is not as trivial as it
could seem at first sight. Naively, one would say that this cycle is
the cycle parameterized by the two coordinates ($\theta_1 , \phi_1$).
This is the original cycle characterizing the seven-dimensional
solution one has started with. In fact, the seven-dimensional solution
is non-trivially embedded in ten dimensions. As a result of this,
as we have already pointed out, there is a non-trivial mixing  between
the three coordinates of the three-sphere along which one up-lifts the
solution ($\theta_2 , \phi_2, \psi$) and those of the two-sphere along
which the original seven-dimensional domain wall is wrapped ($\theta_1
, \phi_1$). This mixing can be seen explicitly by the appearance of
the seven-dimensional gauge connection in the ten-dimensional metric
(\ref{mnsol}). We can say that the seven-dimensional domain wall
already knows about the ten-dimensional geometry via the twist.  From
a seven-dimensional point of view this actually mixes space-time
degrees of freedom with internal ones (note that in ten dimensions all
these degrees of freedom are relative to space-time). For this reason,
it turns out that the correct two-cycle is different from that
suggested by the naive intuition.}
\end{itemize}
To identify the relevant two-cycle we should focus on the
five-dimensional angular part of the metric (\ref{mnsol}). Let us
consider the limit $\rho\to\infty$. At large $\rho$, from the solution
(\ref{mnsol}) we easily get
\begin{equation} 
\label{metinf}
ds^2_5 ~ \sim~ \rho\, (d\theta_1^2 + \sin ^2\theta_1\, d\phi_1^2) +
\frac{1}{4}  (d\theta_2^2+\sin^2\theta_2\, d\phi_2^2) + \frac{1}{4}
(d\psi + \cos\theta_1\, d\phi_1 + \cos\theta_2 \, d\phi_2)^2
\end{equation} 
It is easy to see that this is the metric of a $T^{1,1}$ manifold.
This is a coset manifold, $T^{1,1} = \left[SU(2) \times SU(2)\right]/
U(1)$, which can be seen as a $U(1)$ bundle over two $S^2$ (the
explicit form of the metric above makes this structure
manifest). Still, topologically $T^{1,1}$ can be seen as a two-sphere
fibered over a three-sphere. This is the more convenient way we should
think about it, in the present context. Indeed, together with the
coordinate $\rho$, this manifold makes-up the CY three-fold
characterizing our target space, and this is characterized by
topologically non-trivial two and three-cycles. Even if the metric
(\ref{metinf}) differs from the standard $T^{1,1}$ one, as it is
re-scaled in a way it is no longer an Einstein space, we can anyhow
determine the non-trivial cycles. They are those of the standard
$T^{1,1}$, since the only difference with the above manifold is just a
metric difference. In our set of coordinates the (topologically
non-trivial) two-cycle is not
uniquely defined and one can show there are two possible choices
\begin{eqnarray} 
\label{ciclo} 
S^2 &:& \theta_1 = \pm \theta_2 \quad , \quad \phi_1=-\phi_2 \quad ,
\quad \psi = \mbox{any}
\end{eqnarray}
where the value of $\psi$ does not get fixed to any specific value, by
topological arguments. The three-cycle is instead parameterized by
\begin{eqnarray} 
\label{ciclo3}
S^3 &:& \theta_1 = \phi_1=0
\end{eqnarray}
The angle $\psi$ in eq.~(\ref{ciclo}) gets fixed by the physical
requirement that the cycle is that of minimal volume, which means
minimal energy, since the volume of the cycle is proportional to the
tension of the wrapped D5-branes and we are looking for classically 
stable configurations. By computing  the volume of the
$S^2$ using eqs.~(\ref{ciclo}) and (\ref{mnsol}),  it is easy to show
(this is left as an exercise) that the following  holds
\begin{eqnarray} 
\label{ciclo0} 
S^2 &:& \theta_1=-\theta_2 \quad , \quad \phi_1=-\phi_2 \quad , \quad
\psi=0 \;\; \mbox{mod}\;\; 2\pi\\
\label{ciclop} S^2 &:& \theta_1=\,\theta_2 \; \;\quad , \quad \phi_1=-\phi_2 
\quad , \quad \psi=\pi \;\, \mbox{mod}\;\; 2\pi
\end{eqnarray} 
The two cycles (\ref{ciclo0}) and (\ref{ciclop}) are physically
equivalent from the dual gauge theory point of view. This is due to
the fact that the corresponding volumes are equal, namely
\begin{equation} 
\label{vol} \mbox{Vol}(S^2) = \int_{S^2}{\rm  e}^{- \Phi}\,\sqrt{\mbox{det}G} 
~\sim~ 4 \,{\rm e}^{2\,h(\rho)} + \,\left(a(\rho)-1\right)^2~
\end{equation}
where we set for simplicity $\theta \equiv \theta_1$ and $\phi \equiv
\phi_1$ in both cases. Note that in eqs.~(\ref{ciclo0}) and
(\ref{ciclop}) $\psi$ is defined modulo $2 \pi$, this {\it not} being
its period, which is instead $4 \pi$. This has a precise (and very
nice) gauge theory interpretation, as we shall see.

Let us now consider the metric at the origin. This is just
eq.(\ref{met0}) and is nothing but the metric of a deformed conifold
(a cone in the radial coordinate $\rho$ with base $T^{1,1}$) at the
apex. The parameterization of the topologically non-trivial two and
three-cycle is known for this metric, and is consistent with the one
found above. By implementing eq.~(\ref{ciclo0}) (or equivalently
eq.~(\ref{ciclop})) and eq.~(\ref{ciclo3}) in the metric (\ref{met0})
one finds a vanishing radius for the two-sphere while a finite one for
the three-sphere, as expected for a {\it deformed} conifold at
$\rho=0$.  In fact, the size of the basis of an ordinary cone shrinks
to zero at the apex. This cone is deformed as a non-vanishing volume
for the three-sphere persists at $\rho=0$. The  blown-up volume is
what makes the metric (\ref{mnsol}) non-singular. Note that this is
due to the presence of the function $a(\rho)$, which goes to one at
$\rho=0$, see eq.(\ref{asym1}). For future reference, remind that a
{\it resolved} conifold would be instead a conifold where the
singularity at the apex is removed by the two-sphere being
blown-up. As our manifold is not topologically a conifold for any
$\rho$ but just looks like a conifold in the two limits $\rho
\rightarrow 0$ and $\rho \rightarrow \infty$, the above terminology
gets extended to deformed CY and resolved CY, meaning a smooth CY
three-fold which is topologically a three-sphere or a two-sphere at
$\rho=0$,  respectively.

Let us summarize the lesson. We start with a CY manifold being
characterized by a topologically non-trivial two-cycle where the
D5-branes are wrapped and the gauge theory is being engineered. The
back-reaction  of the D-branes  deforms the original background and
changes its topology: in the resulting geometry the two-cycle shrinks
to zero size  at $\rho=0$ while a three-cycle has blown-up. The
manifold has undergone a {\it geometric transition}. We will further
discuss this point in the last lecture.

\insertion{4}{Some Facts About $\mathbf{{\cal N}=1}$ Super Yang-Mills
\label{insert4}} {Let us briefly recall some features of 
${\cal N}=1$ Super Yang-Mills with gauge group $SU(N)$. This is  a
supersymmetric gauge theory whose field content (auxiliary fields are
not included) is described by the ${\cal N}=1$ vector supermultiplet
\begin{eqnarray}
\left( A_\mu\; , \; \lambda \right) \nonumber
\end{eqnarray}
corresponding to a vector field and a Majorana spinor, the gaugino,
both transforming in the adjoint representation of the gauge group. On
shell this corresponds to 2 bosonic and 2 fermionic degrees of freedom.

This theory has many similarities with QCD, the only difference, at a
superficial level, being that fermions transform in the adjoint
representation rather than in the fundamental representation. In fact,
it shares with QCD many physical properties. The theory is asymptotically 
free, only
colorless asymptotic states exist, it is expected to confine and that
a mass gap is dynamically generated, i.e. all particles in the
spectrum are massive.

The theory has both a scale anomaly and a $U(1)_R$ anomaly, at the
quantum level. The gaugino has R-charge $R=1$, i.e. under a $U(1)_R$
transformation it transforms as $\lambda \rightarrow e^{{\rm i}
\epsilon} \lambda$, while the vector field  has R-charge $R=0$. The
scale anomaly is accounted for by the $\beta$-function which in
general receives corrections at all loops and in the Pauli-Villars
renormalization scheme reads
\begin{eqnarray}
\beta = - 3 \, \frac{N g_{\rm YM}^3}{16 \pi^2}  \left(1 - \frac{N
g_{\rm YM}^2}{8 \pi^2} \right)^{-1} \nonumber
\end{eqnarray}
This is known as the Novikov-Shifman-Vainstein-Zakharov (NSVZ)
$\beta$-function.  By analytic transformations in the gauge coupling
the above result changes, corresponding to a different renormalization
scheme, but universality of the  two-loop coefficient is maintained
(i.e. changes enter beyond two-loops only). In  the Wilsonian scheme,
which is related to the above by a {\it singular} transformation, the
$\beta$-function becomes exact at one-loop. This is useful, as this is
the scheme where holomorphicity is made manifest.

\hskip 355pt  {\it Continued...}}

\insertion{4}{\it Continued...}{ The $U(1)_R$ symmetry which is
conserved at the classical level, is in fact anomalous and gets broken
to $\ZZ_{2 N}$ by quantum effects  (this can be seen by computing the
triangular one-loop diagram with one global current and two gauge
currents). The effect of the anomaly is equivalent to assigning the
$\theta_{\rm YM}$-angle transformation properties under a global
$U(1)_R$ transformation with parameter $\epsilon$ as
\begin{eqnarray}
\theta_{\rm YM} \rightarrow \theta_{\rm YM} - 2 N \epsilon \nonumber
\end{eqnarray}
The theory is invariant under shifts $\theta_{\rm YM} \rightarrow
\theta_{\rm YM} + 2 \pi k$. So if $\epsilon = \frac{\pi k}{ N}$ the
theory is unchanged even at the quantum level. This shows that 
the full quantum theory is invariant under $\ZZ_{2 N}$ transformations 
only.

This is not the full story. The $\ZZ_{2N}$ is spontaneously broken down to
$\ZZ_2$ in the IR and there exist $N$ degenerate vacua where only a
$\ZZ_2$ invariance is conserved. This further breaking is accompanied
by the phenomenon of gaugino condensation, i.e. the fermion bilinear
$\, {\rm Tr} \lambda^2\,$ acquires a vacuum expectation value
\begin{eqnarray}
S \equiv \langle \,{\rm Tr} \lambda^2 \,\rangle =  \Lambda^3 \; e^{2 \pi
{\rm i} k/N} \quad , \quad k=0,1,..., N-1 \nonumber
\end{eqnarray}
where $\Lambda$ is the dynamically generated scale and the vacuum
angle $\theta_{\rm YM}$ has been set equal to zero. Under a chiral
transformation (we choose the gaugino to have R-charge equal 1) we
have $\langle {\rm Tr} \lambda^2 \rangle \rightarrow e^{2 {\rm
i}\alpha} \langle {\rm Tr} \lambda^2 \rangle$ so in the vacuum we see
that only a $\ZZ_2$ invariance is preserved, as anticipated. Choosing
$\theta_{\rm YM}=0$, as we did above, the phases of the gaugino
condensate label the different vacua. However, one can show that the
$\theta_{\rm YM}$ dependence of the gaugino condensate is
\begin{eqnarray}
\langle \,{\rm Tr} \lambda^2 \,\rangle_{\theta_{\rm YM}} \;=\; \langle
\,{\rm Tr} \lambda^2 \,\rangle_{\theta_{\rm YM}=0}\,\; e^{{\rm i}
\,\theta_{\rm YM}/N} \nonumber
\end{eqnarray}
showing that the $N$ vacua are intertwined as far as the $\theta_{\rm
  YM}$ evolution is concerned: since $\theta_{\rm YM} \simeq
  \theta_{\rm YM} + 2 \pi k$ for physical purpose, doing chiral
  transformations out of $\ZZ_2$ one can generate all vacua starting
  from a given one. In particular it is possible to make the gaugino
  condensate to be real in each vacuum, the different vacua being
  equivalently well labeled by the compensating value of the
  $\theta_{\rm YM}$-angle.}

We have now all the necessary ingredients to finally investigate the
gauge/gravity correspondence for the system we have been studying. The
procedure to get the proper gauge/gravity dictionary is essentially
the same as the one we pursued for the fractional brane case. By
expanding the action of the (wrapped) D5-branes up to terms quadratic
in the world volume fields (which, in the low energy limit we are
considering, we let depend on the flat space directions only) and
evaluating it in the background (\ref{mnsol})-(\ref{mnf3}) we get in
this case the action for pure ${\cal N}=1$ SYM (this is left as an exercise). 
The gauge coupling and the $\theta_{\rm YM}$-angle are expressed in 
terms of supergravity quantities as
\begin{eqnarray} 
\label{gym1} \frac{1}{g_{\rm YM}^2} &=& 
\frac{1}{2(2\pi)^3 \alpha'g_s} \int_{S^2}{\rm  e}^{- \Phi}\,\sqrt{
\mbox{det}G} \,= \,\frac{N}{16 \pi^2} \,Y(\rho) \sim {\rm Vol}(S^2)\\
\label{theta1} \theta_{\rm YM} &=& - \,\frac{1}{2\pi \alpha'g_s} \int_{S^2} 
C_{(2)} = - N \,\psi_0
\end{eqnarray} 
where
\begin{equation} 
Y(\rho) = 4 \,e^{2 h(\rho)} + \left( a(\rho) -1\right)^2 = 4 \,\rho
\,\tanh \rho \quad
\end{equation} 
and where we have considered, for definitiveness, the two-cycle
(\ref{ciclo0}). The gauge coupling is related to the volume of the
two-cycle, as  anticipated. From eq.~(\ref{gym1}) we easily see that
\begin{eqnarray} 
\label{gyminf} 
\frac{1}{g_{\rm YM}^2} &\simeq & \frac{N\rho}{4 \pi^2} \qquad {\rm
for} \quad \rho \rightarrow \infty \\
\label{gym0} \frac{1}{g_{\rm YM}^2} &\simeq & \,\,0
\;\;\; \qquad {\rm for} \quad  \rho \rightarrow 0
\end{eqnarray} 
Therefore large distances in supergravity correspond
to the UV of the gauge theory, since for large $\rho $ the gauge
coupling happens to become smaller and smaller (this is what expected
at high energy since ${\cal N}=1$ SYM is an asymptotically-free
theory). At short distances the gauge coupling becomes instead bigger
and bigger (in fact, it seems there is a Landau pole at
$\rho=0$). Hence, short distances in supergravity correspond to the IR
of the gauge theory, where ${\cal N}=1$ SYM becomes strongly
coupled. This qualitative behavior is pretty much the same as the
${\cal N}=2$ case we discussed in the previous lecture. Still, to make
it more precise we need to pint-point the proper radius/energy
relation, which is missing so far. Before dealing with this important
point, let us turn our attention to the $\theta_{\rm YM}$-angle.

From the identification (\ref{theta1}) it is clear that $U(1)_R$
transformations should be realized as shifts in the angular variable
$\psi$, in the dual supergravity background. The only question is what
is the relation between a rotation in $\psi$ of angle, say, $\alpha$,
and a $U(1)_R$ transformation on the gauge theory side with parameter
$\epsilon$, under which the $\theta_{\rm YM}$-angle  transforms as
$\theta_{\rm YM} - 2 N \epsilon$ (see Insert 4). From eq.~(\ref{theta1}) it follows
that  $\alpha = 2 \epsilon$, which means that under a $U(1)_R$
transformation with parameter $\epsilon$ the dual variable $\psi$
transforms as
\begin{equation}
\label{psie}
\psi \rightarrow \psi + 2 \, \epsilon
\end{equation}
On the other hand eq.~(\ref{theta1}) is also saying that the MN 
solution is describing
the gauge theory in a fixed $\theta_{\rm YM}$-vacuum. This could look
unfortunate at first sight, as it seems to tell us we are not able to
see  anything interesting here. Actually this is not true and the
prediction (\ref{theta1}) is what it should be. Let us see why. We
know that the overall $U(1)$ R-symmetry, enjoyed by ${\cal N}=1$ SYM at
the classical level, is anomalous, and is broken to $\ZZ_{2N}$ by
quantum corrections in the UV. This perturbative symmetry of the 
quantum theory is spontaneously broken to $\ZZ_2$ in the
IR. So, the ``true''  theory itself lives in a fixed (up to $\ZZ_2$
symmetry) $\theta_{\rm YM}$-vacuum. We expect the MN solution to
describe the quantum properties of ${\cal N}=1$ SYM, so it is only the
$\ZZ_2$ invariance left-over in the vacuum we would like our {\it
complete} solution being able to display. As we shall see in a moment,
this symmetry is actually there, and is related to the $\psi$ angle
being fixed to be 0 {\it or} $2 \pi$, see eq.~(\ref{ciclo0}). And in 
the UV we can also see the $\ZZ_{2N}$ symmetry, in fact.

As it can be seen from eqs.~(\ref{mnsol})-(\ref{mnf3}), shifts in
$\psi$ are not symmetries of the MN solution neither isometries of the
metric (this is due to mixed terms coming from the square of
$\omega^{1,2} - A^{1,2}$). However, for $\rho \rightarrow \infty$ we
know that $a(\rho)$ vanishes and so do $A^1$ and $A^2$. So, shifts in
$\psi$ are isometry of the metric at large $\rho$ and the
non-invariant term comes from $C_{(2)}$ only. The point now is that
since  the explicit $\psi$ dependence drops out from the metric, the
relevant $S^2$ obtained by minimizing the corresponding D5-brane
tension  does {\it not} fix $\psi$ anymore. That is to say, what
supergravity tells us about the two-cycle at large $\rho$ is just
eq.~(\ref{ciclo}) rather then eqs.~(\ref{ciclo0})-(\ref{ciclop}). By
computing the integral of $C_{(2)}$ over the cycle (\ref{ciclo}) we
get now
\begin{equation}
\frac{1}{4\pi^2 \alpha'g_s} \int_{S^2}  C_{(2)} =  \frac{N}{2 \pi}
\left( \psi + \psi_0 \right)
\end{equation}
The above integral is allowed to change by integer values so shifts in
$\psi$ such that
\begin{equation}
\psi \rightarrow \psi + 2 \frac{\pi}{N}\, k
\label{psiuv}
\end{equation}
are symmetries of the solution, at large $\rho$. Recalling now
eq.~(\ref{psie}), we see that this corresponds to $U(1)_R$
transformations with parameter $\epsilon = \frac{\pi}{N}\, k$, which
means $\ZZ_{2N}$. Supergravity predicts that at large $\rho$, which
means UV, these should be symmetries of the gauge theory. Hence the symmetry
(\ref{psiuv}) is nothing but the supergravity counterpart of the
non-anomalous $\ZZ_{2N}$ symmetry of the gauge theory! 

This is a nice
result, of course, but we can do more. As already pointed out, the
full solution is not invariant under transformations like
(\ref{psiuv}) as now $a(\rho)$ is there, and is more and more relevant
as we go to short distances (these corresponding to the IR of the dual
gauge theory). Both in the metric and in $C_{(2)}$, see
eq.~(\ref{mnc2}),  $a(\rho)$ multiplies terms proportional to $\cos
\psi$ or $\sin \psi$  which are invariant only for
\begin{equation}
\psi \rightarrow \psi + 2 \pi k \quad {\rm which \;\;means} \quad
\epsilon = k \, \pi
\end{equation}
So this is the only symmetry in $\psi$ enjoyed by the complete
solution. Well, this is nothing but the gravitational counterpart of
the spontaneous breaking of the R-symmetry from $\ZZ_{2N}$ down to
$\ZZ_2$ in the IR. So, the full supergravity solution, which is
expected  to be dual to ${\cal N}=1$ SYM with gauge group $SU(N)$, is
$\ZZ_2$ invariant, as it should be. We also see now what the meaning
of the uncertainness in the exact value of $\psi$ was, in
eq.~(\ref{ciclo0}). Supergravity was just telling us that the vacuum
of ${\cal N}=1$ SYM is $\ZZ_2$ symmetric!

Let us come to the second holographic relation, eq.~(\ref{gym1}),
which involves the gauge coupling. Similarly to the case studied in
the previous lecture, in order to make the correspondence concrete we
should first find out what the radius/energy relation is. Curiously,
the previous analysis of the (apparently unrelated) $\theta_{\rm
YM}$-angle correspondence gives us the answer. What we have just seen
is that $a(\rho)$ is the quantity responsible for the breaking of
the $\ZZ_{2N}$ R-symmetry down to $\ZZ_2$ in the IR. On the gauge
theory side this breaking is accompanied by the phenomenon of the
gaugino condensation, i.e. the operator $\langle\lambda^2 \rangle$
acquires a vacuum expectation value in the IR, $\langle\lambda^2
\rangle = \Lambda^3$.  Something similar happens to $a(\rho)$ which is
sensibly different from zero only at short distances, while it vanishes
at large distances. In other words, the function $a(\rho)$ plays the
same r\^ole in supergravity  has the gaugino condensate in the gauge
theory. It is then natural to conjecture that $a(\rho)$ is the
supergravity dual of the gaugino condensate
\begin{equation}
\label{cond0}
\langle\lambda^2 \rangle \longleftrightarrow a(\rho)
\end{equation}
The gaugino condensate has dimension three while the function
$a(\rho)$ is of course dimensionless so the precise identification
reads
\begin{equation}
\label{cond1}
\frac{\Lambda^3}{\mu^3} \,= \, \frac{2 \rho}{\sinh 2 \rho}
\end{equation}
where $\mu$ is the subtraction scale where the gauge theory is defined
and we have substituted for $a(\rho)$ its explicit form. Notice  that
as it was the case in the previous lecture, where we identified the
radial coordinate of space-time with the scalar field  of the ${\cal
N}=2$ vector multiplet, we have here an identification between a
supergravity field and a protected operator of the gauge theory
(i.e. an operator whose dimension does not get changed by quantum
corrections). This is necessary, as supergravity fields do not change
their physical dimension with $\rho$, of course. The above
identification is also remarkable since it says that the field which
de-singularizes the solution, $a(\rho)$, gets associated with the
non-trivial IR dynamics of the dual gauge theory. This is something we
have already experienced, when discussing the enhan\c{c}on mechanism
for theories with 8 supercharges.

From eq.~(\ref{cond1}) we can extract the desired radius/energy
relation  which we can then use in the holographic expression
(\ref{gym1}) and get an exact expression for the running of the  gauge
coupling. To invert eq.~(\ref{cond1}) is actually a difficult  task,
unfortunately. Instead, it is rather simple to get the
$\beta$-function. We can write
\begin{equation} 
\beta(g_{\rm YM}) = \frac{\partial g_{\rm YM}}{\partial \ln
(\mu/\Lambda)} = \frac{\partial g_{\rm YM}}{\partial
\rho}\frac{\partial \rho}{\partial \ln (\mu/\Lambda)}
\end{equation} 
and compute the two derivative contributions from eqs.~(\ref{gym1}) and
(\ref{cond1}), respectively. In doing so, let us first disregard
the exponential corrections, which are sub-leading at large $\rho$,
this being the region where perturbative computations make sense in
the gauge theory. Expanding the function $Y(\rho)$ appearing in
eq.~(\ref{gym1}) one gets $Y(\rho) = 4 \,\rho + {\cal O}(e^{-
\rho})$. Similarly we get $a(\rho) = 2 \,\rho \,e^{- 2\rho} + {\cal
O}(e^{- 6\rho})$. From these expressions we easily get
\begin{eqnarray}  
\frac{\partial g_{\rm YM}}{\partial \rho} &=&  \frac{\pi}{N^{1/2}}
\rho^{- 3/2} = -  \frac{N g_{\rm YM}^3}{ 8 \pi^2} \\ \frac{\partial
\rho}{\partial \ln  (\mu/\Lambda)} &=& \frac{3}{2} \, \left(1 -
\frac{1}{2 \rho}\right)^{-1}  = \frac{3}{2} \, \left(1 - \frac{N
g_{\rm YM}^2}{8 \pi^2}\right)^{-1}
\end{eqnarray} 
where in the second step of both equations we have used again
eq.~(\ref{gyminf}). The final result is then
\begin{equation} 
\label{betapert1} 
\beta(g_{\rm YM}) = - 3 \, \frac{N g_{\rm YM}^3}{16 \pi^2}  \left(1 -
\frac{N g_{\rm YM}^2}{8 \pi^2} \right)^{-1}
\end{equation} 
which is the NSVZ $\beta$-function. This means that supergravity
predicts the exact $\beta$-function of ${\cal N}=1$ SYM at all loops!

A couple of remarks are in order at this point.
\begin{itemize}
\item{The first question which naturally arises is why supergravity is
    giving the NSVZ $\beta$-function, which is in fact the
    $\beta$-function obtained in a particular scheme, the
    Pauli-Villars scheme. The only thing we can say in this respect is
    that this depends on the explicit form of the gauge/gravity
    relation (\ref{cond1}) which supergravity does not fix
    uniquely. The geometric considerations leading to the
    identification of the gaugino condensate with the function
    $a(\rho)$ are insensible to a redefinition of the holographic
    relation (\ref{cond1}) by means of an analytic function of the
    gauge coupling. If doing so, one can easily see that the result we
    have obtained, eq.~(\ref{betapert1}), changes. Still, this change 
    enters beyond two loops
    only, meaning that supergravity respects the universality of
    the two-loop coefficient of the $\beta$-function.}
\item{At short distances, which correspond to the IR of the gauge
    theory, one should wonder what the meaning of the $\beta$-function
    is, of course. But let us forget this for the time being and
    compare the $\beta$-function one gets for very large $\rho$ and
    for very small $\rho$. In the first case we have the usual
    one-loop result, $\beta = -3/(16\pi^2) g_{\rm YM}^3 N$. Recalling
    that the dynamically generated scale $\Lambda$ is defined as the
    {\it scale invariant} quantity $\Lambda = \mu \exp[- \int d\,
    g_{\rm YM}/\beta]$  we get
    \begin{eqnarray}
    \label{lpert} 
    \Lambda^3 = \mu^3 \exp\left[- \frac{8 \pi^2}{g_{\rm YM}(\mu)^2
    N}\right]
    \end{eqnarray}
    where $\mu >> \Lambda $.  Expanding eqs.~(\ref{gym1}) and
    (\ref{cond1}) at very small $\rho$ in  powers of $1/\rho$ (we are
    now at low energy), one  obtains instead $\beta = -9/(16\pi^2)
    g_{\rm YM}^3 N$. This   leads to
    \begin{eqnarray} 
    \label{lnpert}
    \Lambda^3 = \mu'^3 \exp\left[- \frac{8 \pi^2}{3 \,g_{\rm
    YM}(\mu')^2 N}\right]
    \end{eqnarray}
    where now $\mu' \geq \Lambda$. The behavior (\ref{lpert}) is
    usually  interpreted as a non-perturbative effect due to
    fractional instantons of charge $1/N$. These should become more
    and more relevant at low energy, where the theory becomes strongly
    coupled. However, the IR estimate (\ref{lnpert}) seems to say
    there is something different there, as the coefficient is
    different of a factor 3. That is to say, if we take
    eqs.~(\ref{gym1})  and (\ref{cond1}) seriously all the way down to
    the IR,  $\langle\lambda^2 \rangle = \Lambda^3$ seems not to have
    a pure fractional  instanton behavior. (I thank P. Olesen for
    pointing this out to me).}
\item{In deriving eq.(\ref{betapert1}) we disregarded, both in
    $Y(\rho)$  and in $a(\rho)$, exponentially suppressed
    corrections. Still, supergravity seems to say they are there, no
    matter how small they are. What is their meaning? From the
    relation between $\rho$ and $g_{\rm YM}$ it is clear that these
    correspond to (unexpected) non-perturbative contributions. In
    order  to compute them one should consider the full expression for
    $Y(\rho)$ and $a(\rho)$ in eq.s~(\ref{gym1}) and (\ref{cond1}) and
    re-derive eq.~(\ref{betapert1}). This can easily be done and one
    can get an exact expression in $\rho$ for the
    $\beta$-function. What is difficult is to re-express the result in
    terms of the gauge coupling. In fact, there is some debate  in the
    literature on  how to treat these non-perturbative corrections
    since it can also be the case they mix with KK degrees of freedom
    not belonging to the gauge theory and which are not decoupled 
    in the supergravity limit (see below). If
    this is the case, without further insights it is very difficult to
    disentangle the two contributions and give a precise mathematical
    recipe on how to re-derive eq.~(\ref{betapert1}). The only thing
    we can say, so far, is that supergravity seems to suggest some
    non-perturbative corrections are there, after all. It is also
    worth pointing out that these corrections would remove the pole of
    the NSVZ $\beta$-function, the physical interpretation of which is
    not completely clear. It would be nice to check this (unexpected)
    prediction by doing some computations in the field theory.}
\end{itemize}
As it was the case for theories with 8 supercharges, also in this case
one can obtain the correct action for the gauge theory instantons. It
is well known that a system of $N$ D3-branes and $k$ D(-1)-branes in
flat space describes the $k$-instanton sector of ${\cal N}=4$ $SU(N)$
SYM in four dimensions. It is also known that D1-branes are instantons
within D5-branes. It is then natural to consider a system of $N$
D5-branes and $k$ (euclidean) D1-branes wrapped on the $S^2$ and see
if the $k$ wrapped D1-branes account for the $k$-instanton
contribution (these are nothing but the analogue  of the fractional
D(-1)-branes we considered in the previous section, of course). The
action of an Euclidean D1-brane (we choose for simplicity $k=1$) is
\begin{equation}
{\cal S}_{D1} =  \frac{T_{1}}{\kappa} \int_{S^2} e^{- \Phi} \sqrt{{\rm
      det} G} - {\rm i}  \frac{T_{1}}{\kappa} \int_{S^2} C_{(2)}
\end{equation}
where all supergravity fields are pulled-back onto the brane world
volume and $T_1/\kappa = 1/(2 \pi g_s \alpha')$. Evaluating the above
action in the background of the MN solution we get after some simple
algebra
\begin{equation}
\label{insgrd}
{\cal S}_{D1} =  \frac{8 \pi^2}{g_{\rm YM}^2} - \, {\rm i}
\;\theta_{\rm YM}
\end{equation}
which is the correct form of the instanton action.

There is one last point that may seem missing. It has to do, again,
with  the $\theta_{\rm YM}$-angle. The MN solution describes  $SU(N)$
${\cal N}=1$  SYM in a fixed vacuum. What about the other $N-1$ vacua?
How to generate  them? From what we have learned so far, it is not
difficult to answer this question. Here it is where the r\^ole of
gauge transformations in the seven-dimensional gauged  supergravity
comes into play. In the seven-dimensional solution
(\ref{mn7d})-(\ref{ga7d}) we are free to make $SU(2)_L$ gauge
transformations
\begin{equation}
\label{gt}
A \to A' \, = \, g^{-1}A \, g + {\rm i} \, g^{-1}d g
\end{equation}
where $g$ is an element of the $SU(2)_L$ group and $A$ is the $SU(2)_L$
gauge  connection. This means that the ten-dimensional solution is not
univocally determined, even if all solutions should be physically
equivalent, of course.  It turns out that different seven dimensional
gauge choices correspond just to different parameterizations of the
relevant ten-dimensional geometry. In particular, if choosing $g =
\exp({\rm i} \epsilon \sigma_3)$ one gets
\begin{equation}
\label{gt1}
A' \;=\; e^{{\rm i} \epsilon \sigma_3} \, A \; e^{- {\rm i} \epsilon
\sigma_3}
\end{equation}
By repeating the reasoning which led to eq.(\ref{ciclo0}) one finds now
\begin{eqnarray} 
\label{ciclo1} 
S^2 &:& \theta_1=-\theta_2 \quad , \quad \phi_1=-\phi_2 \quad , \quad
\psi = 2 \epsilon \;\; \mbox{mod}\;\; 2\pi
\end{eqnarray} 
Re-deriving the expression for the gauge coupling and the $\theta_{\rm
YM}$-angle using this modified cycle, it is easy to see that we are
now describing the {\it same} running but we are studying the theory
in a {\it different} $\theta_{\rm YM}$-vacuum, namely
\begin{equation}
\theta_{\rm YM} = - N \left( \psi_0 + 2 \epsilon\right)
\end{equation}
The integral of the RR two-form potential is indeed in this case
\begin{equation}
\frac{1}{4\pi^2 \alpha'g_s} \int_{S^2}  C_{(2)} =  \frac{N}{2 \pi}
\left( \psi_0 + 2 \epsilon \right)
\end{equation}
For values of $\epsilon$ such that $2 \epsilon =  2 \pi k/N$ with $k=
0,...,N-1$ we are just jumping between one vacuum to another,
describing all the $N$ vacua of the gauge theory! Each vacuum, see
eq.~(\ref{ciclo1}), has its own $\ZZ_2$ symmetry. Summarizing, modifying
the MN solution by means of the gauge transformations (\ref{gt1}), we
can indeed describe all vacua of the SYM theory. Notice that the new
form of the solution seems to suggest that under the transformation
(\ref{gt1}) the gaugino condensate, which we identified with
$a(\rho)$, picks-up a phase of $2 {\rm i} \epsilon$, consistently with
gauge theory expectations. This is a non-trivial confirmation of the
transformation rule (\ref{psie}).

\bigskip
\noindent{\bf Exercise 5 -} {\it Consider the gauge transformation 
generated by  the $SU(2)$ element
\begin{eqnarray}
g ~=~ {\rm e}^{- \frac{\rm i}{2}\theta_1\sigma_1} {\rm e}^{- \frac{\rm
i}{2} \phi_1\sigma_3} \nonumber
\end{eqnarray}
and compute the corresponding gauge connection $A'$ using
eq.~(\ref{gt}). What is the expression for the ten-dimensional metric
(\ref{mnsol}) now? Compute the new  parametrization for the two-cycle,
the corresponding running of the gauge coupling and the $\theta_{\rm
YM}$-angle. Are there any changes in the physics?} Hint: in order to
identify the parametrization of the two-cycle look at the behavior of
$A'$ as $\rho \rightarrow 0$ and recall that the mixing between the
three-sphere one up-lifts the seven-dimensional domain wall solution
and the two-cycle depends on that behavior.      \bigskip

After all these nice findings, let us end this lecture pointing out
some subtleties and open problems hidden in the correspondence we have
investigated. These have to do with the decoupling limit issue we
mentioned in the first lecture. One could wonder whether there is
decoupling between (four-dimensional) gauge degrees of
freedom and gravity 
degrees of freedom within the supergravity regime we have been
considering. As explained in our first lecture it turns out  this is
not the case. This is a general problem one always has to deal with in
non-conformal situations, and is the main obstruction in stating exact
dualities in the regime where explicit computations are
doable. Although we  cannot treat this problem in detail here, there
are a couple of simple remarks  that could help in making what we just
said more concrete.
\begin{itemize}
\item{It is easy to see that the curvature of the MN background,
eq.(\ref{mnsol}), goes like
\begin{equation}
\label{curv}
{\cal R} \sim \frac{1}{\alpha' g_s N}
\end{equation}
Therefore, in order to keep the curvature small in string units and
make the supergravity approximation meaningful, one should take the
limit $g_s N \rightarrow \infty$.}
\item{We already pointed out, see the discussion after
eq.~(\ref{gym0}), that the MN solution predicts a Landau pole at small
$\rho$, suggesting that the gauge coupling diverges in the IR. In
fact, this is not necessarily true. As just noticed, the regime in
which the supergravity approximation is reliable is for large $N$. In
this regime a Landau pole can indeed be present even if the gauge
coupling remains finite at the scale $\Lambda$, since in
eq.~(\ref{gym0}) it is really $g_{\rm YM}^2 N$  which is going to
infinity and not the gauge coupling itself. To discuss the  duality in
the deep IR at finite $N$, one has to go beyond the supergravity
approximation.}
\item{By considering the metric (\ref{mnsol}) at small $\rho$ one sees
there are KK states coming from the three-sphere (and not belonging to
the gauge theory) whose mass spectrum is roughly
\begin{equation}
\label{kk}
M_{KK}^2 \sim \frac{1}{R^2_{S^3}} \sim \frac{1}{\alpha' g_s N}
\end{equation}
At $\rho < 1$ we are probing distances of this same order since $\rho
= r /\sqrt{\alpha' g_s N}$. This means that predictions in the deep IR
of the gauge theory, not taking into account these states, should be
taken with  some care.}
\item{Let us try to be a bit more specific on the point above. As a
common feature  of non-conformal gauge/gravity dualities, it turns out
that the KK states found above cannot be decoupled from gauge theory
states, within the supergravity regime. A way to see this is as follows. 
Using some tools inherited 
from the AdS/CFT
correspondence (as such this should be done with some care and the
results we get should be taken just as qualitative ones), we can give
an estimate of the glueball mass spectrum as predicted from the MN
background. We consider here the simplest possible case. The scalar
glueball $0^{++}$ ($0$ refers to the spin while the two upper indexes
refer to the parity and charge conjugation quantum numbers,
respectively) couples to ${\rm Tr} F^2$, this operator being dual to
the  dilaton field. Hence, using the
AdS/CFT prescription, the masses of such glueballs are in
correspondence to the eigenvalues of the dilaton bulk equation
\begin{equation}
\partial_\mu \left(\sqrt{- {\rm det} G} \,G^{\mu\nu} \partial_\nu
\Phi\right) = 0
\end{equation}
Computing the above equation for small $\rho$, this being the region
where confining effects occur, we can forget about the $S^2$, which
shrink to zero size in the MN solution, and expand the dilaton field
in KK modes over the $S^3$, on which the dilaton in fact does not
depend. This simplifies the above equation considerably (i.e. the
indexes $\mu,\nu$ run just along $0,1,2,3$ and $\rho$) and expanding
$\Phi$ in plane waves, $\Phi = \Phi(\rho) \,e^{{\rm i} k x}$ (where
$x$ stands for $x_0, x_1, x_2, x_3$), so that a mode of momentum $k$
has a mass $m^2 = - k^2$ in four dimensions, one easily finds
\begin{equation}
M_{gb}^2 \sim \frac{1}{\alpha' g_s N}
\end{equation}
which says the glueball masses are of the same order of the KK masses
we computed before, see eq.~(\ref{kk})! This shows (in a qualitative
way, of course) the mixing phenomenon we already mentioned and makes
it more evident why one should go beyond the supergravity
approximation if looking for an exact duality. For instance, it can 
be that {\it part} of the non-perturbative contributions to the 
$\beta$-function we have found are in fact coming from  
these non gauge theory degrees of freedom.}
\item{Using again some tools inherited from the AdS/CFT
    correspondence,  we can also give an estimate of the confining
    string tension. The Nambu-Goto action of a fundamental string in
    the MN background is
\begin{equation}
S = \frac{1}{2 \pi \alpha'} \int dx\,dt \sqrt{f(\rho)^2 + g(\rho)^2
(\frac{d\rho}{dx})^2} = \int dt\, E
\end{equation}
where $f(\rho)^2 = G_{tt} G_{xx}$ and $g(\rho)^2 = G_{tt}
G_{\rho\rho}$. From eq.~(\ref{mnsol}) we get
\begin{equation}
f(\rho)^2 = \frac{\sinh 2\rho}{2 \,e^{h(\rho)}} \quad , \quad
g(\rho)^2 = \alpha' g_s N \, f(\rho)^2
\end{equation}
The area law is a consequence of the fact that $f(\rho)$ has a minimum
at $\rho=0$ so a fundamental string prefers to lie on the
hyper-surface $\rho=0$. From the above action one can read-off the
string tension. The energy reads
\begin{equation}
E = \frac{1}{2 \pi \alpha'} \, f(0) \, L + \mbox{corrections}
\end{equation}
Since $f(0)=1$ one gets for the tension $T_s$ of the confining string
\begin{equation}
T_s = \frac{1}{2\pi\alpha'}
\end{equation}
which equals the fundamental string tension. Recalling eq.~(\ref{kk}),
this means that $M_{KK}^2 \sim T_s/g_s N$ which shows again that in
order to keep the KK states decoupled from the gauge  theory (the
string tension sets the scale for gauge theory states), one should
require $g_s N << 1$, which is the opposite regime for the validity of
the  supergravity approximation. This is a further indication that a
complete string analysis would be needed in order to distil an exact
duality.}
\item{The MN solution has some isometries, as for instance shifts in
$\phi_1$ and $\phi_2$, which do not have a gauge theory
interpretation, as opposite to the case of the ${\rm AdS}_5 \times S^5$
correspondence where the isometries of the solution are in one-to-one
correspondence with the R-symmetry group of the dual gauge theory. The
only global (and broken) symmetry that has a dual interpretation in
the present case is that in $\psi$, which is related to the $U(1)$
R-symmetry of ${\cal N}=1$ SYM. Is this weird abundance of symmetries
related to the fact that to find an exact duality we should go to the
string regime? Are these extra isometries related to symmetries of the
KK spectrum and should not be there in the end? If we believe an exact
duality is there, this should probably be the case. Unfortunately, we
do not have a definitive answer to this problem, yet.}
\end{itemize}
This is all we wanted to say about ${\cal N}=1$ SYM using the MN
dual. In our last lecture we are going to reconsider the whole picture
of non-conformal versions of the gauge/gravity correspondence and
briefly discuss the connections between (some of) the different
approaches that have been used in the literature. Some of the open
problems we have discussed both in this lecture as well as in the
previous one in establishing an exact duality between supergravity (or
better, string theory) and gauge theories, may found easier answers
using different approaches. What follows is just an overview aiming to
let the reader having a rough idea on how alternative paths can be
pursued.

%%%%%%%%%%%%%%%%%%%%%%%%%%%%%%%%%%%%%%%%%%%%%%%%%%%%%%%%%%%%%%%%%%
%%%%%%%%%%%%%%%%%%%%%%%%%%%%%%%%%%%%%%%%%%%%%%%%%%%%%%%%%%%%%%%%%%
\section{Lecture IV - Dualities and connections to other approaches}

This last lecture is a brief overview on the other approaches that
have been used recently to construct four-dimensional supersymmetric
gauge theories  by means of string (or M) theory and which are
related, in a way or another, to the ones we have been
discussed so far. %This  also enables us to have a complementary (or
%better dual) description  of some peculiar features we encountered, as
%for instance the enhan\c{c}on phenomenon. 
What follows is nothing more than a bird-eyes view. That is to say, if
you are looking for explicit computations, skip this, and go directly
to the reference list.

%%%%%%%%%%%%%%%%%%%%%%%%%%%%%%%%%%%%%%%%%%%%%%%%%%%%%%%%%%%%%%%%%%%%%%
\subsection{Branes Suspended Between Branes}

\begin{figure}[ht]
\begin{center}
{\includegraphics{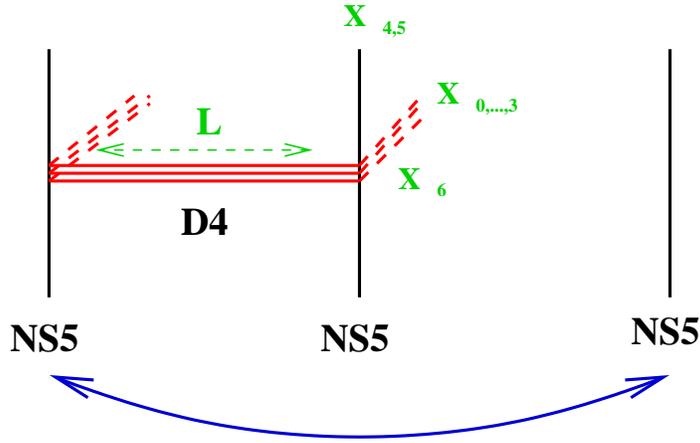}}
\caption{\small A configuration of D4-branes stretched between two
  parallel NS5-branes. The $x_6$ direction is compact, this meaning
  that the first and the third NS5-brane in the picture are
  identified. One can have as well D4-branes stretched from the second
  NS5-brane two the third. These would couple to the NS5-brane world
  volume fields with equal strength as the former, but with opposite sign.}
\label{ns5}
\end{center}
\end{figure}

Let us suppose to be in ten-dimensional flat space and to have two
parallel NS5-branes extending along directions $x_0,...,x_5$ but
which  are at a finite distance $L$ in one of the transverse
directions, say $x_6$ . Suppose this direction is compact, with radius
$R$. Let us now stretch $N$ parallel D4-branes between the two
NS5-branes, with world volume directions along $x_0,..., x_3$ and
$x_6$. At weak string coupling the NS5-branes are much heavier than
the D4-branes (there is a factor $1/g_s$ between the tension of
NS5-branes and D-branes) and therefore at low energy, much lower than
the inverse of $L$, the effective theory of this system is simply  the
gauge theory living on the non-compact part of the D4-brane world
volume, which is actually four-dimensional. This configuration is
depicted in figure \ref{ns5}.

Perform now a T-duality along the compact direction along
which the D4-branes are stretched. What happens? This is something
interesting. It turns out that the T-dual configuration is a bunch of
$N$ fractional D3-branes at the singularity of a $\IC_2/\ZZ_2$
orbifold, the configuration we have been studied in our second
lecture! This can be proved by implementing T-duality rules correctly,
but we do not do it here. Let us just say what the crucial point is:
the T-dual of NS-branes in flat space are orbifold-like
singularities. Starting from a configuration of NS5-branes as that
depicted in figure \ref{ns5} (forget for a while about the D4-branes)
and making a T-duality along $x_6$ one ends up in the orbifold
$\IC_2/\ZZ_2$ and no branes left. On the other hand, it is rather easy
to understand that under T-duality D4-branes become D3-branes. The
fact that these D4-branes are peculiar, i.e. they have a finite
extension along the direction the T-duality is performed, makes the
resulting D3-branes being equally peculiar, i.e. fractional branes.

Everything we learned in our second lecture can be translated into
this branes-suspended-between-branes picture language. Geometrically,
at least, this can be convenient. To start with, it is clear what the
twisted fields really are: they are the world volume fields of the
NS5-branes. This is why they have six-dimensional dynamics, only. The
$B_{(2)}$-flux along the vanishing two-cycle, which is so relevant in
the gauge/gravity dictionary, is related to the distance $L$ between
the NS5-branes. This should not be surprising, too. Under a T-duality
we know that $B_{(2)}$ components and metric components get exchanged
and this is what happens here. In order for string theory to be
well-behaved we know that the background value of the $B_{(2)}$-flux should be
equal to $1/2$, in string units (see eq.~(\ref{bIbg}) and the
subsequent discussion). This implies that in the T-dual configuration,
figure \ref{ns5}, $L$ should be just half the length of the compact
direction $x_6$. The enhan\c{c}on mechanism has a simple geometrical
interpretation in this T-dual set-up. Starting from the configuration
depicted in figure \ref{ns5} and switching-on interactions,  the
D4-branes tend to pull the NS5-branes, bending them. The final stable
configuration, once interactions are take into account, is then
described by NS5-branes which are bent along the $x_6$ direction,
i.e. $L$ is a function of $\rho$. At a distance $\rho = \rho_E$
(recall  $\rho$ is the radial distance in the $x_4,x_5$ plane which is
transverse to the D4-branes but lyes on the world volume of the
NS5-branes) $L \rightarrow 0$ and the branes touch. The new light
fields come from world volume field supermultiplets becoming massless 
and charged under
both the NS5-branes.  Recall that the world volume
theory of NS5-branes of type IIA is described by (2,0) theory and we
have scalars and self-dual two-forms (plus fermions) as light fields,
so we cannot speak of an enhancement of gauge symmetry in this case,
consistently with what we found in the fractional D3-brane
scenario. At the enhan\c{c}on the solution should be described by a
1/2 BPS $A_1$ (2,0) theory ($A_1$ is the first element of the $A$
series of Dynkin diagrams and is related to the fact that we have two
NS5-branes in this case). But this is hard to do explicitly, in practice.

We have learned that there are two different types of fractional D3-branes
on the orbifold $\IC_2/\ZZ_2$ and we described here the T-dual of just
one type. What about the other? This is just a D4-brane stretched from
the second NS5-branes to the third one (recall that $x_6$ is
periodic). In this T-dual picture it is easy to see the opposite
coupling to the twisted fields the two types of fractional branes
have: since the corresponding D4-branes end on opposite sides of the
NS5-branes, the coupling has of course opposite sign. A bound state of
the two different D4-branes corresponds to a D4-brane which is not
tied anymore to the NS5-branes (this can be seen as the coupling to
the NS5-branes world volume fields cancel). This brane does not end on
the NS5-branes and therefore can move freely in the transverse space.
This is nothing but the T-dual description of a regular D3-brane, of
course.  Note how in this picture it is straightforward to understand
why fractional  branes of the $\IC_2/\ZZ_2$ orbifold have half the
tension (and charge) of  regular ones: the tension of (both type of)
fractional D3-branes is $T^f_3 =  T_4 \,L$ (recall $L$ is half the
length of the compact direction), $T_4$  being the tension of the
D4-brane. The tension of a regular D3-brane is  instead $T_3 = 2 \,T_4
L$, which is twice $T^f_3$. The T-dual of the fractional 
D(-1)-branes, which we
learned are related to the instantons of the gauge theory, are now
Euclidean D0-branes stretched between the NS5-branes (it is known that
D0-branes are instantons within D4-branes, in fact).

All what we have been saying above is general. One can consider $k$
parallel NS5-branes and in this case, after T-duality, one ends up
with the orbifold $\IC_2/\ZZ_k$, the rank of the orbifold being
related to the number of NS5-branes. Similarly, one can start from a
type IIB configuration of $N$ D3-branes stretched between a couple of
NS5-branes. Under T-duality this is the system of fractional D2-branes
on $\IC_2/\ZZ_2$ we discussed at the end of our second lecture which
couples non-trivially to the twisted vector field $A_{(1)}$ and which
describes three-dimensional SYM with eight supercharges. In this case
we really have an enhancement of gauge symmetry at the enhan\c{c}on,
see (\ref{encd2}). Indeed, the six-dimensional theory living on the
NS5-branes of type IIB is $(1,1)$ theory  (this is the low energy
theory of the D1-branes living on them) and  is made of vectors and
scalars (plus fermions). Once the NS5-branes touch, the $U(1)$ gauge
symmetry related to the vector field the D3-branes couple to,  gets
enhanced to $SU(2)$. This is exactly the same phenomenon one
encounters when putting usual D-branes on top of each other: the gauge
theory describing their dynamics at low energy gets enhanced.

One can also construct configurations preserving four supercharges,
only,  by just tilting the NS5-branes. This breaks further
supersymmetry and  represent the T-dual version of fractional branes
on singular version  (orbifolds or conifolds) of CY three-folds.

One can build-up a zoology of gauge theories playing with these basic
tools, and this approach has been widely used in the past years to
engineer the most different kind of gauge theories in terms of brane
intersections. Here, we just pointed out what was useful in order to
get the connection with the fractional brane construction. 

In fact,
there is another piece of information that might be interesting for
us. Start again from the type IIA configuration of figure \ref{ns5}
and perform now a T-duality along a transverse direction which is
transverse {\it both} to the NS5-branes and to the D4-branes. In this
case one does not generate an orbifold anymore but the corresponding
smooth ALE space (the Eguchi-Hanson space, in this case). The
D4-branes become D5-branes (now the T-duality has been performed
transverse to their world volume so this is again reasonable) wrapped
on a topologically non-trivial two-cycle inside the ALE space. This is
nothing but the (${\cal N}=2$ version of the) configuration we studied
in the last lecture: a wrapped brane on a CY manifold! So, we see that
through the brane-suspended-between-branes  picture we make contact
between fractional and wrapped branes. This also show, once more, that
fractional branes can be seen just as a particular kind of wrapped
branes (similar conclusions hold for the ${\cal N}=1$ case).

%%%%%%%%%%%%%%%%%%%%%%%%%%%%%%%%%%%%%%%%%%%%%%%%%%%%%%%%%%%%%%%%%%%%%%
\subsection{M5-branes Wrapped On Riemann Surfaces}

The branes-suspended-between-branes picture is also an intermediate
step for embedding fractional branes into M-theory. Indeed, the
configuration of figure \ref{ns5} can be lifted-up in eleven
dimensions. It turns out that the all system of NS5-branes and
D4-branes becomes a single M5-brane wrapped on a complicated  (but
smooth) Riemann surfaces $\Sigma$. The singular points of the
configuration  of figure \ref{ns5}, i.e. the points on the NS5-brane
world volumes where the D4-branes end, get smoothed out by the
D4-brane becoming an M5, as they get extended in the eleventh
direction, $x_{10}$.

The remarkable point is that the condition for preserving ${\cal N}=2$
supersymmetry restricts the embedding of the M5-branes world volume,
and the function describing this embedding are precisely the
corresponding Seiberg-Witten curves (the instantons are automatically
taken care of as D0-branes are simply KK momentum modes in M-theory)!
So the M-theory way to realize ${\cal N}=2$ SYM is where the
appearance of the quantum moduli space of the theory shows-up more
naturally, and is therefore in principle the best place where to study
the quantum property of supersymmetric gauge theories in terms of
supergravity duals. On the other hand, we know little about the very
structure of M-theory and is therefore difficult to pursue such a
program.

\begin{figure}[ht]
\begin{center}
{\includegraphics{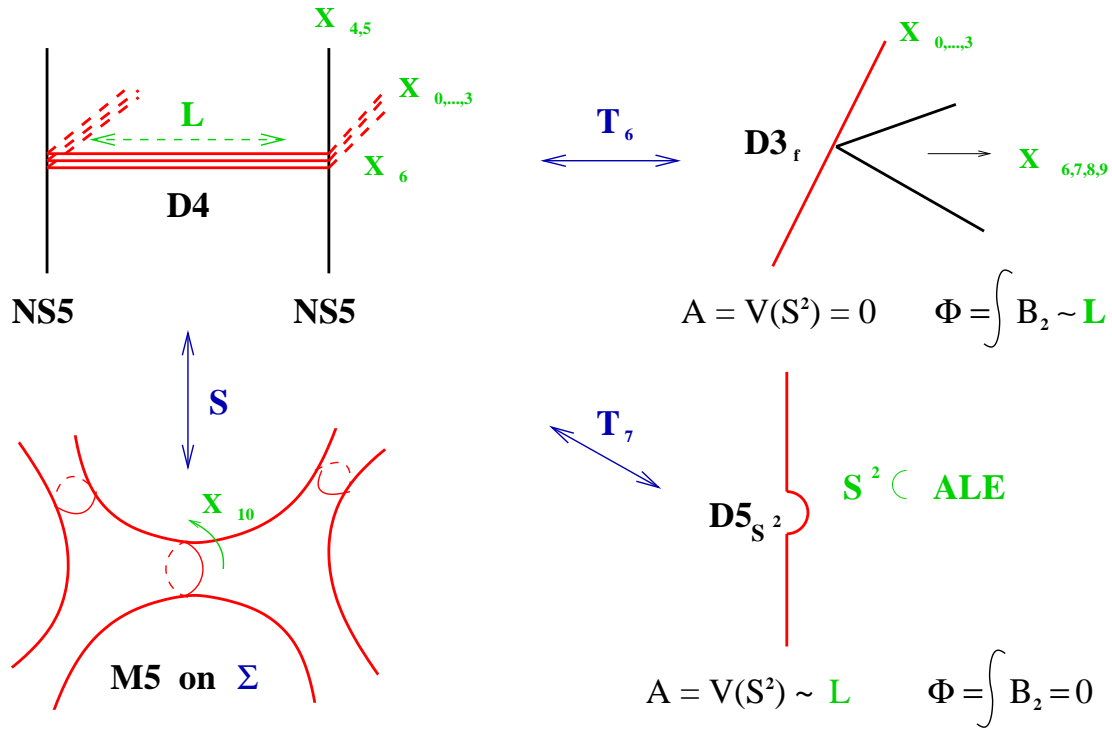}}
\caption{\small The ${\cal N}=2$ duality web. On the left hand side up
  we have D4-branes stretched between two NS5-branes. By performing a
  T-duality along $x_6$ we get fractional D3-branes on the orbifold
  $\IC_2/\ZZ_2$. The distance $L$ between the NS5-branes translates
  into the (background) value of the $B_{(2)}$ flux. Performing a
  T-duality along $x_7$, instead, we get D5-branes wrapped on a
  two-sphere inside an $A_1$ ALE space. The distance $L$ between the
  NS5-branes becomes now the (background) value of the volume of the
  two-sphere. Finally, performing a S-duality (left hand side down) we
  end up in M-theory with a M5-brane wrapped on a Riemann surface. All
  these different configurations describe the same physics at low
  energy: four-dimensional ${\cal N}=2$ SYM.}
\label{dual1}
\end{center}
\end{figure}

Putting together all what we have been learning so far, we can extract
a duality-web, which we summarize in figure
\ref{dual1}.

All what we have been saying, both in the
branes-suspended-between-branes context as well as in the present one,
can be extended to cases with four supercharges (i.e. ${\cal N}=1$ in
four dimensions) and a similar duality web as the one of figure
\ref{dual1} can be drawn. As already noticed, by tilting or rotating
the NS5-brane with respect to each other one can construct T-duals of
fractional branes on ${\cal N}=1$ conifold and orbifolds respectively,
and the M-theory picture of that correspond to M5-branes wrapped on
suitable supersymmetric two-cycles.

%%%%%%%%%%%%%%%%%%%%%%%%%%%%%%%%%%%%%%%%%%%%%%%%%%%%%%%%%%%%%%%%%%%%%
\subsection{Fractional Branes on Conifolds}

Another notable example of a supergravity dual of ${\cal N}=1$ SYM
theory was obtained by considering D-branes on the conifold, i.e. a CY
three-fold whose topology is that of a cone on $T^{1,1}$ (recall that
$T^{1,1} \sim S^2 \times S^3$). This is a singular manifold since at
the apex both the two-sphere and the three-sphere shrink to zero
size. By considering a bound state of $N$ D3-branes and $M$ D5-branes
wrapped on the two-sphere (which are nothing but fractional D3-branes,
recall lecture two), Klebanov and Strassler (KS)  were able to find
the corresponding smooth supergravity solution. This is expected to be
dual, in the same sense of what we have been doing in the  previous
lectures, to the gauge theory living on the world volume of these
bound state of D-branes. This is an ${\cal N}=1$ SYM theory  with
gauge group $SU(N+M) \times SU(N)$, matter in the bi-fundamental and a
quartic superpotential. There is a lot interesting physics in here,
but at first sight we do not see how pure ${\cal N}=1$  SYM could come
about. In fact, it turns out that through a chain of Seiberg dualities
dubbed "duality cascade" the above theory flows to pure $SU(M)$ SYM in
the IR. Indeed, studying the KS solution in the corresponding dual
region, i.e. near the tip of the cone, most of the results we obtained
in the last lecture about ${\cal N}=1$ SYM using the MN solution, can
be similarly  obtained here.

There are a couple of things that may be worth pointing out, though.

As in the IR both the MN and KS solution are expected to be dual to
the same theory, they could not look so different,
geometrically. Indeed, they do not. As it is the case for the MN
solution,  the KS solution realizes a geometric transition, too. The
metric, in the deep IR, is nothing but that of a deformed conifold,
similarly to what we found for the MN solution (in this case this
statement is even more precise as the topology of the six-dimensional
transverse space is that of a cone over $T^{1,1}$ for {\it any} value
of $\rho$). Starting from a singular manifold, with a shrinking
two-cycle at the apex, and putting fractional branes there, one ends
up with a smooth space, where a three-cycle has blown-up. At $\rho =
0$ the metric is exactly equal to eq.~(\ref{met0}). As it was the case
for the MN solution, the deformation of the conifold is related to the
IR physics of the gauge theory, more precisely to the phenomenon of
gaugino condensation.

Another interesting feature of the KS model is that it can be obtained
as a deformation of a ${\cal N}=2$ model. Let us start by considering
$N$ regular D3 branes on the orbifold $\IC^2/\ZZ_2$. The gauge theory
is a superconformal ${\cal N}=2$ theory with gauge group $SU(N)\times
SU(N)$ and the correspoding supergravity dual has an ${\rm
AdS}_5\times S^5/\ZZ_2$ geometry.  Adding a relevant perturbation to
the superpotential of the ${\cal N}=2$ theory and integrating out the
adjoint scalars, one breaks ${\cal N}=2$ to ${\cal N}=1$ and a quartic
superpotential is generated. From the geometric point of view this
operation corresponds to blow-up the fixed circle of the $S^5$, ending
up with the conifold on $T^{(1,1)}$. Adding M fractional D3 branes,
one can repeat the same reasoning and ends up with the non-conformal
${\cal N}=1$ SYM theory we described above.

%%%%%%%%%%%%%%%%%%%%%%%%%%%%%%%%%%%%%%%%%%%%%%%%%%%%%%%%%%%%%%%%%%%%%%
\subsection{Dualities As Geometric Transitions}

The gauge/gravity dualities we have been discussed so far can be
rephrased in a more geometrical language. This is sometime called the
gauge/geometry correspondence and its formulation is mainly due to
work of Vafa with different collaborators in the last few years.

The original duality was obtained embedding in string theory (in fact,
type IIA string theory) a previously discovered large $N$ duality
between Chern-Simon theory and topological closed string theory. Upon
mirror symmetry it was then possible to obtain a similar duality in
type IIB string theory. These example were further generalized by
subsequent works and it turns out a large class of ${\cal N}=1$ SYM
theory dualities can be casted in this geometric way.

In what follows we refer to the type IIB version of the duality since
this is where comparison with our findings are straightforward. The
general idea is to engineer an ${\cal N}=1$ supersymmetric gauge
theory  by means of D5-branes wrapped on a topologically non-trivial
two-cycle  of some CY space. In the corresponding dual closed string
realization, the geometry goes through a geometric transition: the
two-cycle shrinks to zero size and a three-cycle blows-up. The
geometry is completely smooth, no singularities are there, i.e. the
D-branes have disappeared. They get replaced by NSNS and RR 
fluxes.  Most of the gauge theory quantities, as most notably the
effective superpotential, can be  obtained by integrating these
fluxes, as well as other geometric quantities, on suitable cycles
inside the CY.

Another feature of the ${\cal N}=1$ theories considered in this
construction is that they can all be seen as deformations of ${\cal
N}=2$ SYM. One starts with $N$ D5-branes wrapped on a supersymmetric
two-cycle in a CY two-fold, say a (non-compact) $K3$ manifold. The
geometry of the ten-dimensional space-time is then $\IR^{1,3}\times K3
\times \IC$ and the gauge theory living on the D5-branes is
four-dimensional ${\cal N}=2$ gauge theory (at energies where we can
forget KK excitations on the two-cycle). Suppose now to add a scalar
superpotential $W(\phi)$ ($\phi$ being the complex scalar field of the
${\cal N}=2$ vector multiplet) which breaks ${\cal N}=2$ to ${\cal
N}=1$. From the target space geometry point of view, it turns out this
corresponds to make the direct product $K3 \times \IC$ a non-trivial
fibration, and the space a proper Calabi-Yau three-fold with $SU(3)$
holonomy, ending up with the configuration described above.  This
means all these ${\cal N}=1$ theories are strict relatives of the
${\cal N}=2$ theory, and this is something which makes the study of
their properties more manageable.

All what we have been saying should alert the reader: there is more
than one thing  pointing to the configurations we have discussed, the
MN and the KS models. Strictly speaking none of them is really a Vafa
model (although the KS solution is pretty near to it), but both of
them do realize a geometric transition, as  we have already discussed:
both the MN and the KS brane configurations modify the original CY
space into a deformed one, and branes get replaced by fluxes (in fact,
it is quite a general feature of all these gauge/gravity duals to
realize a geometric transition). Moreover, the gaugino condensate is
related to the deformation parameter, this also being the case for
Vafa models. There are also some differences, especially in the MN
solution. For one thing, we have a dilaton there, which is absent in
the Vafa model,  and we do not know the exact expression for the
equations giving  the superpotential for cases with running
dilaton. Moreover, in the  MN solution we do not have NSNS three-form
flux but only RR one. On the other hand, the KS solution has much in
common with the general picture just discussed. It realizes a
geometric transition. It has non-trivial RR and NSNS three-form
fluxes. It has trivial dilaton. It has a superpotential, which can be
seen as generated deforming a ${\cal N}=2$ SYM theory, this
corresponding, from the geometry point of view, to start from
$\IC_2/\ZZ_2$ and partially resolve the orbifold singularity into a CY
three-fold (a conifold in this case). On the other hand, the starting
point of the KS configuration is different, i.e. a singular conifold
and not a resolved one. Moreover, the UV completion of the KS solution
is a chain of Seiberg dualities and the gauge theory  remains
four-dimensional all way long (this is typical for any kind of
fractional brane as the cycles they are wrapped on are geometrically
vanishing). In Vafa models, instead, at very high energy one start
seeing the KK states on the two-cycle the brane are wrapped on,
therefore far in the UV the theory becomes six-dimensional (this same
thing happens for the MN solution). In fact, all these theories have
in general different UV completions but are equivalent in the IR.

We cannot discuss these interesting issues further, here. Still, it
should be already surprising what we have learnt so far. The lesson is
that all these apparently different approaches are related to each
other: fractional branes on orbifolds, fractional branes on conifolds,
M5-branes wrapped on Riemann surfaces, D-branes wrapped on
supersymmetric cycles of CY spaces, geometric transitions, etc... seem
to be part of a unique picture. Each approach has of course its
advantages and disadvantages, but having an understanding and a
control on all of them could considerably increase the chances of
being able to solve, one day, the still open problems lying on the
carpet. This would not only mean to eventually distil an exact duality
for non-conformal supersymmetric (and may be non-supersymmetric)
Yang-Mills theories, but also to have a chance to understand more
deeply the ultimate  structure of the underlying microscopic theory.

Actually, before we end, there is one more topic we would like to
discuss. And this  hides some more (and conceptually remarkable)
surprises.

%%%%%%%%%%%%%%%%%%%%%%%%%%%%%%%%%%%%%%%%%%%%%%%%%%%%%%%%%%%%%%%%%%%%%%
\subsection{M-theory On $G_2$-holonomy Manifolds}

The study of compactifications of Heterotic string on CY three-folds
turned out to be a very useful way to study ${\cal N}=1$
supersymmetric  theories in four dimensions by means of string
theory. There is an M-theory analogue of that and amounts to consider
M-theory on $G_2$-holonomy manifolds. These are seven-dimensional
manifolds with $G_2$ holonomy. They preserve 1/8  supercharges and
therefore, when considering M-theory propagating on these spaces, one
obtains four-dimensional ${\cal N}=1$ supersymmetric theories at low
energy.

An obstruction to find phenomenologically interesting four-dimensional
models out of M-theory on $G_2$-holonomy manifolds is that one cannot
get chiral fermions in four dimensions if the manifold is smooth. This
problem can be overcome taking a singular $G_2$-manifold  where some
three-cycle has shrunk to zero size. M-theory behaves smoothly on
these spaces since the effective volume gets complexified as we have a
three-form potential in the spectrum (this has the same r\^ole played
by the NSNS two-form potential when studying ten-dimensional string
theory on singular spaces). In this case, at very low energy, one can
show the theory is described solely by gauge degrees of freedom near
the singularity (i.e. gravitational modes are decoupled) and one can
indeed get chiral matter. The nature of the singularity dictates the
structure of the four-dimensional gauge theory (gauge group, matter
content, etc...) so in principle one can study a plethora of ${\cal
N}=1$ gauge theories by means of different $G_2$-holonomy spaces. Note
that branes are not present in this construction.

So far, so good. But where is the connection with what we have been
discussing before? It turns out the connection is there, and is rather
profound. It has been shown that the gauge/geometry duality that Vafa
originally studied in type IIA theory, given in terms of D6-branes
wrapped on a three-cycle of the deformed conifold which are dual to
the resolved conifold with fluxes,  can be up-lifted in M-theory. What
one ends up with is just M-theory on two different points in the
moduli space of a $G_2$-manifold! Moreover no branes are present on
either sides of the correspondence (this is not  too surprising
though, as D6-branes are purely gravitational objects in M-theory, so,
once the embedding is performed, it is natural ending up with
something where branes do not appear). The D6-brane/deformed conifold
(gauge theory) side of the duality corresponds to a singular
$G_2$-manifold while the fluxes/resolved conifold (gravity) side
corresponds to a smooth $G_2$-manifold. These two manifolds are
related by a so-called flop transition. So, the duality between the
${\cal N}=1$ SYM theory and supergravity gets reformulated as a purely
geometric transition when studying M-theory propagating on a
$G_2$-holonomy manifold.

The remarkable thing is that it turns out the moduli space  is
smoothly connected, i.e. there is no phase transition at the quantum
level in going from the smooth to the singular points mentioned above.
This means that the equivalence between the two descriptions (the
gauge theory description and the supergravity description) is
guaranteed as this is the ``same'' physics from the M-theory point of
view. This may suggest the idea that the gauge/gravity duality could
be really proved in M-theory. Once again, M-theory seems to be the
right place where to look for answers to  fundamental
questions. However, in order to make this more concrete, we need a
deeper understanding of M-theory of what we presently have.

%%%%%%%%%%%%%%%%%%%%%%%%%%%%%%%%%%%%%%%%%%%%%%%%%%%%%%%%%%%%%%%%%%
%%%%%%%%%%%%%%%%%%%%%%%%%%%%%%%%%%%%%%%%%%%%%%%%%%%%%%%%%%%%%%%%%%
\section{A (Biased And Incomplete) Guided Tour Through The Literature}

Let me finally give some references on the material presented in these
lectures. The literature on the subject is endless and I cannot
provide an exhaustive list here. I just mention some of the papers I
think could help the reader in deepening the issues I discussed.
Essentially, this amounts to list those papers I found useful myself
in preparing the material presented in this course. Often, these
papers do not even coincide with the original works. This is why what
follows  is biased and incomplete.

A comprehensive review on the AdS/CFT correspondence is
\cite{Aharony:1999ti} while an updated discussion of the  conceptual
problems in finding exact dualities for non-conformal  theories can be
found in \cite{Aharony:2002up}, which is a recent review on similar
topics as those presented in these  lectures.
 
\vskip 8pt {\it - Fractional Branes}

The first work where the supergravity solution for fractional branes
was presented and the corresponding gauge dual discussed is
\cite{Bertolini:2000dk} (see also \cite{Polchinski:2000mx}). This
dealt with the case of fractional D3-branes on the most simple
orbifold, $\IC_2/\ZZ_2$. The corresponding compact case was discussed
in  \cite{Frau:2000gk} while in \cite{Billo:2001vg} the analysis was
extended to orbifolds of the full ADE series. Addition of D7-branes
was considered in \cite{Grana:2001xn} and \cite{Bertolini:2001qa},
this giving ${\cal N}=2$ SYM coupled to fundamental matter. Cases with
${\cal N}=1$ supersymmetry were discussed in \cite{Bertolini:2001gg}
and extended to include fundamental matter in
\cite{Marotta:2002gc}. Finally, in \cite{Bertolini:2002xu} a 
complete treatment on the correspondence between the
$U(1)_R$ anomaly and the supergravity duals was given (see  also
\cite{Klebanov:2002gr} for the related case of fractional branes on
conifolds). The work
\cite{Bertolini:2001gq} is a (partial) review which has some
overlap with our second lecture. I say partial because it includes
theories with 8 supercharges, only, and misses some findings which
were achieved later. Still, it can be rather useful as various
computational aspects (the derivation of the string spectrum, the
solution of the equations of motion, the condition of supersymmetry
preservation, etc...) are given in great detail.

Fractional branes were introduced long ago in \cite{Douglas:1996xg}
and \cite{Diaconescu:1997br} (see also the more recent and very useful
\cite{Billo:2000yb}). Looking to the citation list of these papers
one can find much more fractional branes literature. Let me just
mention the two original papers dealing with strings and D-branes on
orbifolds, \cite{Douglas:1996sw} and \cite{Johnson:1996py}, discussing
orbifolds of the A and of the DE series, respectively, and
\cite{Klebanov:1999rd}, where first investigations on the r\^ole
played by fractional branes in the gauge/gravity correspondence were
done.

The enhan\c{c}on phenomenon was discovered in \cite{Johnson:1999qt}
and discussed much further in many contexts. To our purpose, let us
just mention three papers. In \cite{Johnson:2001wm} and
\cite{Merlatti:2001gd} (for the case of wrapped and fractional branes,
respectively) the excision criteria, outlined in
\cite{Johnson:1999qt}, was systematically studied and proved to be
consistent at the pure supergravity level, while
\cite{Wijnholt:2001us} is the only paper, to our acknowledge, where an
honest supergravity attempt to solve the enhan\c{c}on, by including
the new light degrees of freedom, was pursued. This was done  for the
case of three-dimensional  supersymmetric gauge theory,
though. Finally, in the nice work \cite{Petrini:2001fk} 
the relation between the enhan\c{c}on, the running of the
five-form flux and the Seiberg-Witten curve as predicted by the dual
SYM theory was discussed, showing agreement between supergravity, 
the enhan\c{c}on
picture and the field theory expectations (this was done exploiting
the $T\times S$-dual picture of M5-branes wrapped on Riemann surfaces,
where Seiberg-Witten curves naturally arise).

Fractional branes on conifolds is a subject that was pioneered by
I. Klebanov with a number of collaborators during the last few
years. The original paper, containing the smooth supergravity
solution, is \cite{Klebanov:2000hb} , which elaborated on some
previous works, \cite{Klebanov:2000nc}, \cite{Gubser:1998fp} and
\cite{Klebanov:1998hh} (another interesting paper discussing the basis
of D-branes on conifolds is \cite{Morrison:1998cs}). A recent,
complete and very nice review on this subject is \cite{Herzog:2002ih},
where much more references can be found. The complete ${\cal N}=1$
$\beta$-function for the case of fractional branes on the conifold was
found in \cite{Imeroni:2002me}, after the above mentioned review
appeared on the archive.

\vskip 8pt {\it - Wrapped Branes}

Let us start mentioning those papers discussing (and refining) the
proposal originally made by Maldacena and Nu\~nez which is relevant to
study pure ${\cal N}=1$ SYM in four dimensions. The first paper is of
course \cite{Maldacena:2000yy}, where the supergravity solution was
found (see also the preceding work \cite{Maldacena:2000mw} for a
discussion of the gauged supergravity approach to curved branes
supergravity solutions). This solution was obtained up-lifting in ten
dimensions a previously found non-abelian monopole solution in four
dimensions \cite{Chamseddine:1997nm}.

In \cite{Apreda:2001qb} it was first proposed
the identification between the supergravity field $a(\rho)$ and the
gaugino condensate and some AdS/CFT arguments were given in favor of
this identification. In \cite{DiVecchia:2002ks} this identification
was used to extract the radius/energy relation and the exact
$\beta$-function was found (see however \cite{Olesen:2002nh}).
Finally, in \cite{Bertolini:2002yr}, a refinement on the gauge/gravity
dictionary was provided, pint-pointing the correct two-cycle to be
used in the correspondence and solving a number of problems present in
previous works. The third lecture of this course (which also includes some
unpublished material) aims to be a self-contained and updated
treatment of the MN solution and its gauge dual.

A useful reference making some interesting qualitative remarks on
${\cal N}=1$ duals is \cite{Loewy:2001pq}. The ${\cal N}=2$ analogue
of the MN solution, discussing D5-branes wrapped on a two-cycle inside
a K3 manifold (this is the first twist we discussed) was provided in
\cite{Gauntlett:2001ps} and \cite{Bigazzi:2001aj}. Some recent remarks on 
radius/energy relations for ${\cal N}=2$ and ${\cal N}=1$ can be 
found in \cite{Wang:2002es} and \cite{Wang:2002ka}, respectively.

The case of
three-dimensional supersymmetric gauge theory with 8 supercharges was
discussed in \cite{DiVecchia:2001uc} and that of 4 supercharges in
\cite{Maldacena:2001pb} (see also \cite{Gauntlett:2001ur}). Other 
examples of wrapped brane solutions and
their gauge duals can be found for instance in
\cite{Edelstein:2001pu}, \cite{Gomis:2001aa} and many others.

Some useful technical details about the geometry of $T^{1,1}$
manifolds (coordinate transformation, identification of cycles,
etc...) can be found, for instance, in \cite{Minasian:1999tt},
\cite{Papadopoulos:2000gj}  and \cite{Gubser:1998fp}, this being
relevant for studying both the MN as well as the KS
solutions. Finally, in \cite{Cvetic:2000dm} the up-lift formul\ae$\,$
we used have been derived and their consistency checked. For works 
discussing general methods for obtaining smooth supergravity duals of 
non-maximally supersymmetric gauge theories, see for instance 
\cite{Cvetic:2000mh}.

\vskip 8pt 
{\it - Other Approaches}

Let me end this brief tour giving some references on the other
approaches we discussed in the last lecture (this will be even more
incomplete and aims just to open a little crack on a big world).

The study of four-dimensional supersymmetric gauge theories using
configuration of branes suspended between branes was initiated by
Witten \cite{Witten:1997sc} and  \cite{Witten:1997ep}, elaborating on
some previous work \cite{Hanany:1996ie}. Some nice works where
extensions are considered and also relations to fractional branes both on
orbifolds and on conifolds are investigated are for instance 
\cite{Andreas:1998hh}, \cite{Dasgupta:1998su} and
\cite{Dasgupta:1999wx}. They can be a good starting point for studying
these things. Looking to the citation list of these works, much more
papers can be found.

The same paper of Witten \cite{Witten:1997sc} is where these
NS5/D-branes configurations was lifted up in M-theory and the study of
M5 wrapped on Riemann surfaces was initiated. Let me also mention the
works  \cite{Brinne:2000fh} and \cite{Brinne:2000nf} where concrete
attempts to work out the corresponding (eleven-dimensional)
supergravity/gauge theory duality, for ${\cal N}=2$ and
${\cal N}=1$ cases respectively, were done.

The use $G_2$-holonomy manifolds to study four-dimensional SYM
theories was somehow initiated in \cite{Acharya:1998pm} and received a
boost in \cite{Atiyah:2000zz} and \cite{Acharya:2000gb}, where it was
shown the connection with the geometric transition picture (or more
generally with the D-brane engineer/geometry-with-fluxes duality of
ten-dimensional string theory). In the work \cite{Atiyah:2001qf} it
was shown the smooth interpolation between the different vacua (where
SYM and supergravity live at low energy, respectively) of M-theory on
$G_2$-holonomy manifolds. This is a crucial property in order to
re-interpret the gauge/gravity (geometry) duality of ten-dimensional
superstring theory in purely geometric terms in M-theory. There have
been many papers discussing various aspects of M-theory on
$G_2$-holonomy manifolds in the last few years. Two reviews,
straightening complementary aspects of this subject, are
\cite{Acharya:gu} and \cite{Gubser:2002mz}, and much more references
can be found there.

Finally, let us come to geometric transitions. This idea and in
particular how gauge/gravity dualities can be thought of from a more
geometrical point of view is due to work done by Vafa with
different collaborators during the last few years. The first paper,
where the type IIA superstring version of the conjecture appeared
(this is where the up-lift from the pure topological
Chern-Simon/topological string duality was done) is
\cite{Vafa:2000wi}, which also describes the mirror type IIB version
of it. There  have been many papers after this first. Let me just
mention a couple of them, \cite{Cachazo:2001jy} and
\cite{Cachazo:2002pr},  which generalize the duality discussed in
\cite{Vafa:2000wi} to a large  class of matter coupled ${\cal N}=1$
SYM theories.

%%%%%%%%%%%%%%%%%%%%%%%%%%%%%%%%%%%%%%%%%%%%%%%%%%%%%%%%%%%%%%%%%%%%5
\subsection*{Acknowledgments}
I thank the Elementary Particle Theory sector of SISSA/ISAS for
inviting  me to give these lectures and for financial support, and the
people attending  the lectures for their questions and comments, some
of which have been  incorporated in this written version. I would also
like to thank A. Lerda, P. Merlatti  and R. Russo for useful comments
on the manuscript. Finally, I would like to express my gratitude to
all the people I discussed and/or collaborated with on the topics of
these lectures in the last three years, and which are too numerous to
be listed here. This work was supported by an EC Marie Curie
Individual Fellowship under contract number HPMF-CT-2000-00847.

%\newpage

\addcontentsline{toc}{section}{References}

%%%%%%%%%%%%%%%%%%%%%%%%%%%%%%%%%%%%%%%%%%%%%%%%%%%%%%%%%%%%%%%%%%%%%%%


\begin{thebibliography}{99} 

%\cite{Aharony:1999ti}
\bibitem{Aharony:1999ti} O.~Aharony, S.~S.~Gubser, J.~M.~Maldacena,
H.~Ooguri and Y.~Oz, \emph{Large N field theories, string theory and
gravity},  Phys.\ Rept.\  {\bf 323} (2000) 183, {\tt hep-th/9905111}.
%%CITATION = HEP-TH 9905111;%%
%\cite{Aharony:2002up}
\bibitem{Aharony:2002up} O.~Aharony, \emph{The non-AdS/non-CFT 
correspondence, or three different paths to QCD}, 
Cargese 2002, Progress in string, field and particle theory 3-24 
{\tt hep-th/0212193}. 
%%CITATION = HEP-TH 0212193;%%
\vskip 10pt
\hskip -15pt {\it - Fractional Branes}

%\cite{Bertolini:2000dk}
\bibitem{Bertolini:2000dk} M.~Bertolini, P.~Di Vecchia, M.~Frau,
A.~Lerda, R.~Marotta and I.~Pesando, \emph{Fractional D-branes and
their gauge duals},  JHEP {\bf 0102} (2001) 014, {\tt hep-th/0011077}.
%%CITATION = HEP-TH 0011077;%%

%\cite{Polchinski:2000mx}
\bibitem{Polchinski:2000mx} J.~Polchinski, \emph{${\cal N} = 2$ gauge-gravity
duals},  Int.\ J.\ Mod.\ Phys.\ A {\bf 16} (2001) 707, {\tt
hep-th/0011193}.
%%CITATION = HEP-TH 0011193;%%

%\cite{Frau:2000gk}
\bibitem{Frau:2000gk}
M.~Frau, A.~Liccardo and R.~Musto, 
\emph{The geometry of fractional branes}, 
Nucl.\ Phys.\ B {\bf 602} (2001) 39, {\tt hep-th/0012035}.
%%CITATION = HEP-TH 0012035;%%

%\cite{Billo:2001vg}
\bibitem{Billo:2001vg} M.~Billo, L.~Gallot and A.~Liccardo,
\emph{Classical geometry and gauge duals for fractional branes on  ALE
orbifolds},  Nucl.\ Phys.\ B {\bf 614} (2001) 254, {\tt
hep-th/0105258}.
%%CITATION = HEP-TH 0105258;%%

%\cite{Grana:2001xn}
\bibitem{Grana:2001xn} M.~Grana and J.~Polchinski, \emph{Gauge /
gravity duals with holomorphic dilaton},  Phys.\ Rev.\ D {\bf 65}
(2002) 126005, {\tt hep-th/0106014}.
%%CITATION = HEP-TH 0106014;%%

%\cite{Bertolini:2001qa}
\bibitem{Bertolini:2001qa} M.~Bertolini, P.~Di Vecchia, M.~Frau,
A.~Lerda and R.~Marotta, \emph{${\cal N} = 2$ gauge theories on systems of
fractional D3/D7 branes},  Nucl.\ Phys.\ B {\bf 621} (2002) 157,  {\tt
hep-th/0107057}.
%%CITATION = HEP-TH 0107057;%%

%\cite{Bertolini:2001gg}
\bibitem{Bertolini:2001gg} M.~Bertolini, P.~Di Vecchia, G.~Ferretti
and R.~Marotta, \emph{Fractional branes and ${\cal N} = 1$  gauge theories},
Nucl.\ Phys.\ B {\bf 630} (2002) 222,  {\tt hep-th/0112187}.
%%CITATION = HEP-TH 0112187;%%

%\cite{Marotta:2002gc}
\bibitem{Marotta:2002gc} R.~Marotta, F.~Nicodemi, R.~Pettorino,
F.~Pezzella and F.~Sannino, \emph{${\cal N} = 1$  matter from fractional
branes},  JHEP {\bf 0209} (2002) 010 {\tt hep-th/0208153}.
%%CITATION = HEP-TH 0208153;%%

%\cite{Bertolini:2002xu}
\bibitem{Bertolini:2002xu} M.~Bertolini, P.~Di Vecchia, M.~Frau,
A.~Lerda and R.~Marotta, \emph{More anomalies from fractional branes},
Phys.\ Lett.\ B {\bf 540} (2002) 104, {\tt hep-th/0202195}.
%%CITATION = HEP-TH 0202195;%%

%\cite{Klebanov:2002gr}
\bibitem{Klebanov:2002gr} I.~R.~Klebanov, P.~Ouyang and E.~Witten,
\emph{A gravity dual of the chiral anomaly},  Phys.\ Rev.\ D {\bf 65}
(2002) 105007, {\tt hep-th/0202056}.
%%CITATION = HEP-TH 0202056;%%

%\cite{Bertolini:2001gq}
\bibitem{Bertolini:2001gq} M.~Bertolini, P.~Di Vecchia and R.~Marotta,
\emph{${\cal N} = 2$  four-dimensional gauge theories from fractional branes},
{\tt hep-th/0112195}.
%%CITATION = HEP-TH 0112195;%%

%\cite{Douglas:1996xg}
\bibitem{Douglas:1996xg} M.~R.~Douglas, \emph{Enhanced gauge symmetry
in M(atrix) theory},  JHEP {\bf 9707} (1997) 004, {\tt hep-th/9612126}.
%%CITATION = HEP-TH 9612126;%%

%\cite{Diaconescu:1997br}
\bibitem{Diaconescu:1997br} D.~E.~Diaconescu, M.~R.~Douglas and
J.~Gomis, \emph{Fractional branes and wrapped branes},  JHEP {\bf
9802} (1998) 013, {\tt hep-th/9712230}.
%%CITATION = HEP-TH 9712230;%%

%\cite{Billo:2000yb}
\bibitem{Billo:2000yb} M.~Billo, B.~Craps and F.~Roose, \emph{Orbifold
boundary states from Cardy's condition},  JHEP {\bf 0101} (2001) 038,
{\tt hep-th/0011060}.
%%CITATION = HEP-TH 0011060;%%

%\cite{Douglas:1996sw}
\bibitem{Douglas:1996sw} M.~R.~Douglas and G.~W.~Moore,
\emph{D-branes, Quivers, and ALE Instantons}, {\tt hep-th/9603167}.
%%CITATION = HEP-TH 9603167;%%

%\cite{Johnson:1996py}
\bibitem{Johnson:1996py} C.~V.~Johnson and R.~C.~Myers, \emph{Aspects
of type IIB theory on ALE spaces},  Phys.\ Rev.\ D {\bf 55} (1997)
6382, {\tt hep-th/9610140}.
%%CITATION = HEP-TH 9610140;%%

%\cite{Klebanov:1999rd}
\bibitem{Klebanov:1999rd} I.~R.~Klebanov and N.~A.~Nekrasov,
\emph{Gravity duals of fractional branes and logarithmic RG flow},
Nucl.\ Phys.\ B {\bf 574} (2000) 263, {\tt hep-th/9911096}.
%%CITATION = HEP-TH 9911096;%%

%\cite{Johnson:1999qt}
\bibitem{Johnson:1999qt} C.~V.~Johnson, A.~W.~Peet and J.~Polchinski,
\emph{Gauge theory and the excision of repulson singularities}, Phys.\
Rev.\ D {\bf 61} (2000) 086001, {\tt hep-th/9911161}.
%%CITATION = HEP-TH 9911161;%%

%\cite{Johnson:2001wm}
\bibitem{Johnson:2001wm} C.~V.~Johnson, R.~C.~Myers, A.~W.~Peet and
S.~F.~Ross, \emph{The enhancon and the consistency of excision},
Phys.\ Rev.\ D {\bf 64} (2001) 106001, {\tt hep-th/0105077}.
%%CITATION = HEP-TH 0105077;%%

%\cite{Merlatti:2001gd}
\bibitem{Merlatti:2001gd}
P.~Merlatti, 
\emph{The enhancon mechanism for fractional branes}, 
Nucl.\ Phys.\ B {\bf 624} (2002) 200, {\tt hep-th/0108016}.
%%CITATION = HEP-TH 0108016;%%

%\cite{Wijnholt:2001us}
\bibitem{Wijnholt:2001us} M.~Wijnholt and S.~Zhukov, \emph{Inside an
enhancon: Monopoles and dual Yang-Mills theory},  Nucl.\ Phys.\ B {\bf
639} (2002) 343, {\tt hep-th/0110109}.
%%CITATION = HEP-TH 0110109;%%

%\cite{Petrini:2001fk}
\bibitem{Petrini:2001fk} M.~Petrini, R.~Russo and A.~Zaffaroni,
\emph{${\cal N} = 2$  gauge theories and systems with fractional
  branes}, Nucl.\
Phys.\ B {\bf 608} (2001) 145, {\tt hep-th/0104026}.
%%CITATION = HEP-TH 0104026;%%

%\cite{Klebanov:2000hb} 
\bibitem{Klebanov:2000hb} I.R. Klebanov and M.J. Strassler,
\emph{Supergravity and a confining gauge theory: Duality cascades and
$\chi_{SB}$-resolution of naked singularities}, JHEP {\bf 08} (2000)
052, {\tt hep-th/0007191}.
%%CITATION = HEP-TH 0007191;%%" 

%\cite{Klebanov:2000nc}
\bibitem{Klebanov:2000nc} I.~R.~Klebanov and A.~A.~Tseytlin,
\emph{Gravity duals of supersymmetric SU(N) x SU(N+M) gauge theories},
Nucl.\ Phys.\ B {\bf 578} (2000) 123, {\tt hep-th/0002159}.
%%CITATION = HEP-TH 0002159;%%

%\cite{Gubser:1998fp}
\bibitem{Gubser:1998fp} S.~S.~Gubser and I.~R.~Klebanov, \emph{Baryons
and domain walls in an ${\cal N} = 1$ superconformal gauge theory}, Phys.\
Rev.\ D {\bf 58} (1998) 125025, {\tt hep-th/9808075}.
%%CITATION = HEP-TH 9808075;%%

%\cite{Klebanov:1998hh}
\bibitem{Klebanov:1998hh} I.~R.~Klebanov and E.~Witten,
\emph{Superconformal field theory on threebranes at a Calabi-Yau
singularity},  Nucl.\ Phys.\ B {\bf 536} (1998) 199, {\tt
hep-th/9807080}.
%%CITATION = HEP-TH 9807080;%%

%\cite{Morrison:1998cs}
\bibitem{Morrison:1998cs} D.~R.~Morrison and M.~R.~Plesser,
\emph{Non-spherical horizons. I},  Adv.\ Theor.\ Math.\ Phys.\  {\bf
3} (1999) 1, {\tt hep-th/9810201}.
%%CITATION = HEP-TH 9810201;%%

%\cite{Herzog:2002ih}
\bibitem{Herzog:2002ih} C.~P.~Herzog, I.~R.~Klebanov and P.~Ouyang,
\emph{D-branes on the conifold and ${\cal N} = 1$  gauge / gravity dualities},
{\tt hep-th/0205100}.
%%CITATION = HEP-TH 0205100;%%

%\cite{Imeroni:2002me}
\bibitem{Imeroni:2002me} E.~Imeroni, \emph{On the ${\cal N} = 1$ beta-function
from the conifold},  Phys.\ Lett.\ B {\bf 541} (2002) 189, {\tt
hep-th/0205216}.
%%CITATION = HEP-TH 0205216;%%

\vskip 10pt
\hskip -15pt {\it - Wrapped Branes}

%\cite{Maldacena:2000yy}
\bibitem{Maldacena:2000yy}  J. Maldacena and C. Nu\~nez, \emph{Toward
The Large $N$ Limit Of ${\cal N} = 1$  Super Yang Mills}, Phys. Rev. Lett.
{\bf 86} (2001) 588, {\tt hep-th/0008001}.
%%CITATION = HEP-TH 0008001;%% 
 
%\cite{Maldacena:2000mw}
\bibitem{Maldacena:2000mw} J.~M.~Maldacena and C.~Nunez,
\emph{Supergravity description of field theories on curved manifolds
and a no  go theorem},  Int.\ J.\ Mod.\ Phys.\ A {\bf 16} (2001) 822,
{\tt hep-th/0007018}.
%%CITATION = HEP-TH 0007018;%%

%\cite{Chamseddine:1997nm}
\bibitem{Chamseddine:1997nm}
A.~H.~Chamseddine and M.~S.~Volkov, 
\emph{Non-Abelian BPS monopoles in N = 4 gauged supergravity}, 
Phys.\ Rev.\ Lett.\  {\bf 79} (1997) 3343, {\tt hep-th/9707176}.
%%CITATION = HEP-TH 9707176;%%

%
\bibitem{Apreda:2001qb} R. Apreda, F. Bigazzi, A.L. Cotrone,
M. Petrini, A. Zaffaroni, \emph{Some comments on ${\cal N}=1$ gauge
theories  from wrapped branes}, Phys. Lett. B {\bf 536} (2002) 161,
{\tt  hep-th/0112236}.
%%CITATION =  HEP-TH 0112236;%% 
 
%\cite{DiVecchia:2002ks} 
\bibitem{DiVecchia:2002ks} P.~Di Vecchia, A.~Lerda and P.~Merlatti,
\emph{${\cal N} = 1$ and ${\cal N} = 2$ super Yang-Mills theories from
wrapped branes}, Nucl. Phys. B {\bf 624} (2002) 200, {\tt hep-th/0205204}.
%%CITATION = HEP-TH 0205204;%% 

%\cite{Olesen:2002nh} 
\bibitem{Olesen:2002nh} P.~Olesen and F.~Sannino, \emph{${\cal N} = 1$
super  Yang-Mills from supergravity: The UV-IR connection}, {\tt
hep-th/0207039}.
%%CITATION = HEP-TH 0207039;%% 

%\cite{Bertolini:2002yr} 
\bibitem{Bertolini:2002yr} M. Bertolini and P. Merlatti,   \emph{A
note on the dual of ${\cal N} = 1$ super Yang-Mills theory},  
Phys. Lett. B {\bf 556} (2003) 80 {\tt hep-th/0211142}.
%%CITATION = HEP-TH 0211142;%%

%\cite{Loewy:2001pq} 
\bibitem{Loewy:2001pq}  A. Loewy and J. Sonnenschein, \emph{On The
Holographic Duals Of ${\cal N}=1$ Gauge Theories}, JHEP {\bf 0108}
(2001) 007, {\tt hep-th/0103163}.
%%CITATION = HEP-TH 0103163;%% 
 
%\cite{Gauntlett:2001ps}
\bibitem{Gauntlett:2001ps} J.~P.~Gauntlett, N.~Kim, D.~Martelli and
D.~Waldram, \emph{Wrapped fivebranes and ${\cal N} = 2$ super 
Yang-Mills theory}, Phys.\ Rev.\ D {\bf 64} (2001) 106008, 
{\tt hep-th/0106117}.
%%CITATION = HEP-TH 0106117;%%

%\cite{Bigazzi:2001aj}
\bibitem{Bigazzi:2001aj} F.~Bigazzi, A.~L.~Cotrone and A.~Zaffaroni,
\emph{${\cal N} = 2$ gauge theories from wrapped five-branes},  
Phys.\ Lett.\ B {\bf 519} (2001) 269, {\tt hep-th/0106160}.
%%CITATION = HEP-TH 0106160;%%

%\cite{Wang:2002es}
\bibitem{Wang:2002es} X.~J.~Wang and S.~Hu,
\emph{Gauge / gravity duality, Green functions of N = 2 SYM and 
radial/energy scale relation}, 
JHEP {\bf 0210} (2002) 005, {\tt hep-th/0207145}.
%%CITATION = HEP-TH 0207145;%%

%\cite{Wang:2002ka}
\bibitem{Wang:2002ka} X.~J.~Wang and S.~Hu,
\emph{Green functions of N = 1 SYM and radial / energy-scale
  relation}, {\tt hep-th/0210041}.
%%CITATION = HEP-TH 0210041;%%

%\cite{DiVecchia:2001uc}
\bibitem{DiVecchia:2001uc} P.~Di Vecchia, H.~Enger, E.~Imeroni and
E.~Lozano-Tellechea,
\emph{Gauge theories from wrapped and fractional branes}, 
Nucl.\ Phys.\ B {\bf 631} (2002) 95, {\tt hep-th/0112126}.
%%CITATION = HEP-TH 0112126;%%

%\cite{Maldacena:2001pb}
\bibitem{Maldacena:2001pb}
J.~M.~Maldacena and H.~Nastase,
\emph{The supergravity dual of a theory with dynamical supersymmetry 
breaking}, JHEP {\bf 0109} (2001) 024, {\tt hep-th/0105049}.
%%CITATION = HEP-TH 0105049;%%

%\cite{Gauntlett:2001ur}
\bibitem{Gauntlett:2001ur} J.~Gauntlett, N.~Kim, D.~Martelli and
D.~Waldram, \emph{Fivebranes wrapped on SLAG three-cycles and related
geometry},  JHEP {\bf 0111} (2001) 018, {\tt hep-th/0110034}.
%%CITATION = HEP-TH 0110034;%%

%\cite{Edelstein:2001pu}
\bibitem{Edelstein:2001pu} J.~D.~Edelstein and C.~Nunez, \emph{D6
branes and M-theory geometrical transitions from gauged supergravity},
JHEP {\bf 0104} (2001) 028, {\tt hep-th/0103167}.
%%CITATION = HEP-TH 0103167;%%

%\cite{Gomis:2001aa}
\bibitem{Gomis:2001aa} J.~Gomis and J.~G.~Russo, 
\emph{D = 2+1 ${\cal N} = 2$ Yang-Mills theory from wrapped branes}, 
JHEP {\bf 0110} (2001) 028, {\tt hep-th/0109177}.
%%CITATION = HEP-TH 0109177;%%

%\cite{Minasian:1999tt} 
\bibitem{Minasian:1999tt} R.~Minasian and D.~Tsimpis, \emph{On the
geometry of non-trivially embedded branes},  Nucl.\ Phys.\ B {\bf 572}
(2000) 499, {\tt hep-th/9911042}.
%%CITATION = HEP-TH 9911042;%% 

%\cite{Papadopoulos:2000gj} 
\bibitem{Papadopoulos:2000gj}  G. Papadopoulos and A.A. Tseytlin,
\emph{Complex geometry of conifolds and 5-brane wrapped on 2-sphere},
Class. Quant. Grav. {\bf 18} (2001) 1333, {\tt hep-th/0012034}.
%%CITATION = HEP-TH 0012034;%%" 

%\cite{Cvetic:2000dm} 
\bibitem{Cvetic:2000dm}  M. Cvetic, H. Lu and C.N. Pope,
\emph{Consistent Kaluza-Klein sphere reductions}, Phys. Rev. {\bf D62}
(2000) 064028, {\tt hep-th/0003286}.
%%CITATION = HEP-TH 0003286;%%" 

%\cite{Cvetic:2000mh}
\bibitem{Cvetic:2000mh} 
M.~Cvetic, H.~Lu and C.~N.~Pope,
\emph{Brane resolution through transgression}, 
Nucl.\ Phys.\ B {\bf 600}, 103 (2001), {\tt hep-th/0011023}.
%%CITATION = HEP-TH 0011023;%%

\vskip 10pt
\hskip -15pt {\it - Other Approaches}

%\cite{Witten:1997sc}
\bibitem{Witten:1997sc}
E.~Witten,
\emph{Solutions of four-dimensional field theories via M-theory}, 
Nucl.\ Phys.\ B {\bf 500} (1997) 3, {\tt hep-th/9703166}.
%%CITATION = HEP-TH 9703166;%%

%\cite{Witten:1997ep}
\bibitem{Witten:1997ep}
E.~Witten,
\emph{Branes and the dynamics of {QCD}}, 
Nucl.\ Phys.\ B {\bf 507} (1997) 658, {\tt hep-th/9706109}.
%%CITATION = HEP-TH 9706109;%%

%\cite{Hanany:1996ie}
\bibitem{Hanany:1996ie}
A.~Hanany and E.~Witten,
\emph{Type IIB superstrings, BPS monopoles, and three-dimensional
  gauge  dynamics}, 
Nucl.\ Phys.\ B {\bf 492} (1997) 152, {\tt hep-th/9611230}.
%%CITATION = HEP-TH 9611230;%%

%\cite{Andreas:1998hh}
\bibitem{Andreas:1998hh}
B.~Andreas, G.~Curio and D.~Lust,
\emph{The Neveu-Schwarz five-brane and its dual geometries}, 
JHEP {\bf 9810} (1998) 022, {\tt hep-th/9807008}.
%%CITATION = HEP-TH 9807008;%%

%\cite{Dasgupta:1998su}
\bibitem{Dasgupta:1998su}
K.~Dasgupta and S.~Mukhi,
\emph{Brane constructions, conifolds and M-theory}, 
Nucl.\ Phys.\ B {\bf 551} (1999) 204, {\tt hep-th/9811139}.
%%CITATION = HEP-TH 9811139;%%

%\cite{Dasgupta:1999wx}
\bibitem{Dasgupta:1999wx}
K.~Dasgupta and S.~Mukhi,
\emph{Brane constructions, fractional branes and anti-de Sitter 
domain walls}, 
JHEP {\bf 9907} (1999) 008, {\tt hep-th/9904131}.
%%CITATION = HEP-TH 9904131;%%

%\cite{Brinne:2000fh}
\bibitem{Brinne:2000fh}
B.~Brinne, A.~Fayyazuddin, S.~Mukhopadhyay and D.~J.~Smith,
\emph{Supergravity M5-branes wrapped on Riemann surfaces and 
their QFT duals}, 
JHEP {\bf 0012} (2000) 013, {\tt hep-th/0009047}.
%%CITATION = HEP-TH 0009047;%%

%\cite{Brinne:2000nf}
\bibitem{Brinne:2000nf}
B.~Brinne, A.~Fayyazuddin, T.~Z.~Husain and D.~J.~Smith,
\emph{${\cal N} = 1$ M5-brane geometries}, 
JHEP {\bf 0103} (2001) 052, {\tt hep-th/0012194}.
%%CITATION = HEP-TH 0012194;%%


%\cite{Acharya:1998pm}
\bibitem{Acharya:1998pm} B.~S.~Acharya,
\emph{M theory, Joyce orbifolds and super Yang-Mills}, 
Adv.\ Theor.\ Math.\ Phys.\  {\bf 3} (1999) 227, {\tt hep-th/9812205}.
%%CITATION = HEP-TH 9812205;%%

%\cite{Atiyah:2000zz}
\bibitem{Atiyah:2000zz} M.~Atiyah, J.~M.~Maldacena and C.~Vafa,
\emph{An M-theory flop as a large N duality}, 
J.\ Math.\ Phys.\  {\bf 42} (2001) 3209, {\tt hep-th/0011256}.
%%CITATION = HEP-TH 0011256;%%

%\cite{Acharya:2000gb}
\bibitem{Acharya:2000gb} B.~S.~Acharya, \emph{On realising ${\cal N} = 1$ super
Yang-Mills in M theory}, {\tt hep-th/0011089}.
%%CITATION = HEP-TH 0011089;%%

%\cite{Atiyah:2001qf}
\bibitem{Atiyah:2001qf}
M.~Atiyah and E.~Witten, 
\emph{M-theory dynamics on a manifold of G(2) holonomy}, 
Adv.\ Theor.\ Math.\ Phys.\  {\bf 6} (2003) 1, {\tt hep-th/0107177}.
%%CITATION = HEP-TH 0107177;%%

%\cite{Acharya:gu}
\bibitem{Acharya:gu}
B.~S.~Acharya,
\emph{M Theory, G(2)-Manifolds And Four-Dimensional Physics}, 
Class.\ Quant.\ Grav.\  {\bf 19} (2002) 5619.
%%CITATION = CQGRD,19,5619;%%

%\cite{Gubser:2002mz}
\bibitem{Gubser:2002mz}
S.~S.~Gubser, 
\emph{Special holonomy in string theory and M-theory}, 
{\tt hep-th/0201114}.
%%CITATION = HEP-TH 0201114;%%

%\cite{Vafa:2000wi}
\bibitem{Vafa:2000wi} C. Vafa, \emph{Superstrings and topological
strings at large N}, J. Math. Phys. {\bf 42} (2001), 2798, {\tt
hep-th/0008142}.
%%CITATION = HEP-TH 0008142;%%" 
 
%\cite{Cachazo:2001jy}
\bibitem{Cachazo:2001jy} F. Cachazo, K.A. Intriligator and C. Vafa,
\emph{A large N duality via a geometric transition}, Nucl. Phys. {\bf
B603} (2001) 3, {\tt hep-th/0103067}.
%%CITATION = HEP-TH 0103067;%%" 

%\cite{Cachazo:2002pr}
\bibitem{Cachazo:2002pr}
F.~Cachazo and C.~Vafa,
\emph{${\cal N} = 1$ and ${\cal N} = 2$ geometry from fluxes}, 
{\tt hep-th/0206017}.
%%CITATION = HEP-TH 0206017;%%

\end{thebibliography}
\end{document}